\documentclass[a4paper,11pt]{article}

\usepackage{jheppub} 

\pdfoutput=1

\usepackage{amsmath,amssymb,bm,graphicx}
\usepackage[dvipsnames]{xcolor}
\usepackage{textcomp,gensymb}
\usepackage{subcaption}
\usepackage{array}
\usepackage{tikz}
\usepackage[normalem]{ulem}

\allowdisplaybreaks
\newcommand{\eps}{\epsilon}
\newcommand{\taubar}{\bar\tau}
\newcommand{\Tau}{\mathcal{T}}
\def\rd{\mathrm{d}}
\newcommand{\nn}{\nonumber}

\newcommand{\bfT}{{\mathbf T}}
\newcommand{\gSzero}{\ensuremath{\gamma_0^S}}
\newcommand{\gSone}{\ensuremath{\gamma_1^S}}
\newcommand{\gSk}{\ensuremath{\gamma_k^S}}

\newcommand{\I}[1]{\textcolor{ForestGreen}{#1}}
\newcommand{\IIa}[1]{\textcolor{YellowGreen}{#1}}
\newcommand{\IIb}[1]{\textcolor{YellowOrange}{#1}}
\newcommand{\IIIa}[1]{\textcolor{Red}{#1}}
\newcommand{\IIIb}[1]{\textcolor{Cyan}{#1}}

\title{
The NNLO soft function for N-jettiness in hadronic collisions}

\author[a]{Guido Bell}
\author[b]{Bahman Dehnadi}
\author[a]{Tobias Mohrmann}
\author[c]{Rudi Rahn}

\affiliation[a]{Theoretische Physik 1, Center for Particle Physics Siegen, Universit\"at Siegen, Walter-Flex-Strasse 3, 57068 Siegen, Germany}
\affiliation[b]{Deutsches Elektronen-Synchrotron DESY, Notkestr.  85, 22607 Hamburg, Germany}
\affiliation[c]{Department of Physics and Astronomy, University of Manchester, Manchester, M13 9PL, United Kingdom}

\emailAdd{bell@physik.uni-siegen.de}
\emailAdd{bahman.dehnadi@desy.de}
\emailAdd{rudi.rahn@manchester.ac.uk}

\abstract{We compute the N-jettiness soft function in hadronic collisions to next-to-next-to-leading order (NNLO) in the strong-coupling expansion. Our calculation is based on an extension of the {\tt SoftSERVE} framework to soft functions that involve an arbitrary number of lightlike Wilson lines. We present numerical results for 1-jettiness and 2-jettiness, and illustrate that our formalism carries over to a generic number of jets by calculating a few benchmark points for 3-jettiness. We also perform a detailed analytic study of the asymptotic behaviour of the soft-function coefficients at the edges of phase space, where one of the jets becomes collinear to another jet or beam direction, and comment on previous calculations of the N-jettiness soft function.}

\arxivnumber{2312.11626}

\begin{document}


\maketitle

\section{Introduction}
\label{sec:intro}

Over the past decade the N-jettiness event shape has emerged as a prominent reference observable for precision studies at the Large Hadron Collider (LHC). Whereas N-jettiness was originally proposed as a resolution variable for discriminating events with different jet multi\-plicities~\cite{Stewart:2010tn}, it nowadays finds wide applications in collider phenomenology, as it is used e.g.~as a slicing parameter for higher-order perturbative computations~\cite{Boughezal:2015dva,Gaunt:2015pea}, for combining fixed-order calculations with a parton shower within the {\tt GENEVA} framework~\cite{Alioli:2015toa} or for certain jet substructure studies~\cite{Thaler:2010tr}.

The key advantage of N-jettiness in comparison to other resolution variables that are based on a jet algorithm lies in its global nature, which significantly simplifies the logarithmic structure of the differential distribution at higher orders. The starting point for a resummation of these logarithms, which arise in the limit in which the N-jettiness variable $\Tau_N$ becomes small, is a factorisation theorem that can be written in the schematic form~\cite{Stewart:2010tn,Stewart:2009yx}
\begin{align}
\label{eq:fact}
\frac{d\sigma}{d\Tau_N}&=\sum_{i,j,\{k_n\}} 
B_i \otimes B_j \otimes \,\prod_{n=1}^N \,J_{k_n} 
\otimes\, \text{tr}\big[H_{ij\to\{k_n\}} * S_{ij\to\{k_n\}}\big] \,+\, \mathcal{O}(\Tau_N)\,.
\end{align}
Here the sum runs over all partonic channels and the final state typically includes colour-neutral (signal) particles on top of the indicated colour-charged partons $\{k_n\}$. The beam functions $B_{i,j}$ capture the collinear radiation into the beam directions, as do the N jet functions $J_{k_n}$ for the final-state collinear emissions. The process-dependent hard functions $H_{ij\to\{k_n\}}$ encode the virtual corrections to the underlying Born process, and they are correlated with the soft functions $S_{ij\to\{k_n\}}$ in colour space, as reflected by the colour trace in \eqref{eq:fact}. The \mbox{symbol $\otimes$} furthermore indicates a convolution with respect to the observable, and $*$ the integration over the directions of the hard partons.

The ingredients in the factorisation theorem have been studied extensively in the past years. Specifically, the N-jettiness beam  functions are currently known to next-to-next-to-next-to-leading order (N$^3$LO) in perturbation theory~\cite{Stewart:2010qs,Berger:2010xi,Gaunt:2014xga,Gaunt:2014cfa,Ebert:2020unb,Baranowski:2022vcn}, and the quark and gluon jet functions are also available at this accuracy~\cite{Bauer:2003pi,Bosch:2004th,Becher:2006qw,Bruser:2018rad,Becher:2009th,Becher:2010pd,Banerjee:2018ozf}. While a general procedure for calculating the N-jettiness soft function at NLO was developed in~\cite{Jouttenus:2011wh}, current NNLO calculations have focused  specifically on 0-jettiness~\cite{Kelley:2011ng,Monni:2011gb,Hornig:2011iu}, 1-jettiness~\cite{Boughezal:2015eha,Campbell:2017hsw} and 2-jettiness~\cite{Bell:2018mkk,Jin:2019dho}. For 0-jettiness there currently exists an on-going effort to extend the calculation to N$^3$LO accuracy, and partial results in this direction have been published in~\cite{Baranowski:2022khd,Chen:2020dpk}. Studies of 0-jettiness in the context of massive particle production relevant for top-quark observables were considered in~\cite{Li:2018tsq,Alioli:2021ggd}.

Whenever the N-jettiness variable is used as a slicing parameter for fixed-order calculations, it is crucial to control power-suppressed terms of $\mathcal{O}(\Tau_N)$ to improve the numerical accuracy and stability of the method. Power corrections to the N-jettiness distribution have therefore attracted considerable interest in the past, and the dominant corrections at next-to-leading power have been studied for 0-jettiness~\cite{Moult:2016fqy,Boughezal:2016zws,Moult:2017jsg,Boughezal:2018mvf,Ebert:2018lzn} and 1-jettiness~\cite{Boughezal:2019ggi}. In this context it was also pointed out that the precise definition of the N-jettiness variable matters for controlling the size of power-suppressed terms at large rapidities~\cite{Moult:2016fqy}.

The above factorisation theorem is based on the assumption that Glauber gluons cancel in the N-jettiness distribution. This view has been challenged in~\cite{Gaunt:2014ska,Zeng:2015iba,Rothstein:2016bsq}, where it was argued that two-Glauber-exchange diagrams may give a non-vanishing contribution to the 0-jettiness distribution starting at $\mathcal{O}(\alpha_s^4)$ or next-to-next-to-next-to-next-to-leading logarithmic (N$^4$LL) accuracy in a logarithmic counting. It was later pointed out, however, that these diagrams cancel for generic single-scale observables, and that factorisation may only be violated through the interplay of Glauber modes with soft radiation~\cite{Schwartz:2018obd}. In addition to this, factorisation might be violated through effects involving active collinear partons, potentially similar to the breakdown of factorisation for transverse-momentum-dependent observables~\cite{Rogers:2010dm}. In any case such factorisation-violating effects may only appear at perturbative orders beyond what we will consider in this paper.

In this work we present a systematic method for calculating the N-jettiness soft function \emph{for an arbitrary number of jets} at NNLO accuracy. Our method is based on the {\tt SoftSERVE} \mbox{framework~\cite{Bell:2018vaa,Bell:2018oqa,Bell:2020yzz}}, which in its current version can be used to compute soft functions that are defined in terms of two back-to-back lightlike Wilson lines. In contrast, the current calculation involves (N+2) lightlike directions in non-back-to-back kinematics, and for N $\geq2$ it includes non-trivial three-particle correlations in addition to the usual dipole contributions.

Our calculation goes beyond previous studies of the NNLO N-jettiness soft function~\cite{Kelley:2011ng,Monni:2011gb,Hornig:2011iu,Boughezal:2015eha,Campbell:2017hsw,Bell:2018mkk,Jin:2019dho} in several respects:
\begin{itemize}
\item   
While we mostly disregard the 0-jettiness case, which some of us already studied in~\cite{Bell:2018oqa}, we recompute the 1-jettiness soft function and perform a detailed comparison to the results of~\cite{Campbell:2017hsw}. Whereas the authors of~\cite{Campbell:2017hsw} provided simple-to-use fit functions for the coefficients of the renormalised soft function, they did not publish the corresponding correlation matrices, which inflates the theory uncertainties somewhat. In contrast, we provide the raw data of our calculation in an ancillary file attached to the electronic version of our paper.
\item
We already presented results for the 2-jettiness soft function in~\cite{Bell:2018mkk} (see also~\cite{Jin:2019dho}), but these results were limited to a specific kinematic configuration in which the two jets are back-to-back in the hadronic center-of-mass frame. In the current paper we sample the entire phase space of the hard emitters in terms of multi-dimensional grids consisting of about $30,\!000$ points (compared to around $20$ that were provided in \cite{Bell:2018mkk,Jin:2019dho}). While we display lower-dimensional projections of these grids for illustration in later sections, the full set of data points is included in the ancillary electronic file.
\item
In order to illustrate that our method applies to a generic number of jets, we for the first time present numbers for the 3-jettiness soft function for a few phase-space points.
\item
As can readily be seen from the plots in later sections, the coefficients of the renormalised soft function tend to grow logarithmically towards the edges of phase space, i.e.~in the limit in which the number of jets is reduced by one unit. We perform a detailed method-of-regions~\cite{Beneke:1997zp} analysis to understand this logarithmic growth analytically, which allows us to extrapolate the multidimensional grids in these kinematic regions in a controlled manner. This can become relevant, for instance, when the soft function is evaluated in a highly boosted frame.
\end{itemize}
This paper is organised as follows. In Sec.~\ref{sec:calculation} we discuss the technical aspects of the calculation, which includes the definition of the N-jettiness observable and the method we use for computing the bare soft function to NNLO. While we largely follow the {\tt SoftSERVE} strategy, we emphasise the differences that arise in the N-jet case with non-back-to-back kinematics and we discuss the renormalisation of the soft function in this section. In Sec.~\ref{sec:results} we present numerical results for 1-jettiness, 2-jettiness and 3-jettiness, while Sec.~\ref{sec:method-of-regions} is devoted to an analytic method-of-regions analysis of the N to (N-1)-jet transition. We conclude in Sec.~\ref{sec:conclusion} and present further technical details of our analysis in the appendix.

\section{Technical aspects of the calculation}
\label{sec:calculation}

\subsection{Definition of the soft function}

The N-jettiness event shape is defined as~\cite{Stewart:2010tn}
\begin{align}
\mathcal{T}_{N}= \sum_{m} \,\min_i \bigg\{\frac{2q_i \cdot k_m}{Q_i}\bigg\}\,,
\end{align}
where $q_i^\mu$ are massless reference momenta associated with the beams ($i=1,2$ ) and N signal jets ($i=3,\ldots, N+2$). In the context of Soft-Collinear Effective Theory~\cite{Bauer:2000ew,Bauer:2000yr,Bauer:2001yt,Beneke:2002ph}, these trace the orientation of the collinear partons produced in the hard partonic interaction, which constitute the initial- and final-state jets. Below we will also refer to these emitters as ``hard'' or ``high-energetic''. For the calculation of the soft function, we assume that these reference momenta are fixed, whereas the momenta of the emitted soft partons $k_m^\mu$ will be integrated over. Moreover, the $Q_i$ denote arbitrary normalisation factors. Specifically, we write
\begin{align}
\label{eq:jet-ref}
q_i^\mu = \omega_i\, n_i^\mu \equiv \omega_i (1,\vec{n}_i)\,.
\end{align}
where $\omega_i$ are the corresponding energies and $\vec{n}_i$ are distinct unit vectors along the beam/jet directions. By choosing $Q_i= 2 \omega_i$, the above definition turns into
\begin{align}
\label{eq:n-jettiness}
\mathcal{T}_{N}= \sum_{m} \,\min_i\big\{n_i \cdot k_m\big\}\,,
\end{align}
and the soft function depends only on the kinematics of the hard process via the variables $n_{ij}\equiv n_i \cdot n_j = 1 - \vec{n}_i \cdot \vec{n}_j$. In fact, this choice refers to the standard definition of the N-jettiness in the hadronic center-of-mass frame, while the soft function for other choices $Q_i\propto \omega_i$ can be obtained by a longitudinal boost along the beam direction.

The N-jettiness soft function is defined as a correlation function of Wilson lines,
\begin{align}
\label{eq:N-jet-soft-func}
S(\tau, \mu) = \sum_X \; 
\mathcal{M}(\tau;\lbrace k_{m} \rbrace)
\langle 0 | \bar T \big(S_{n_1}^\dagger S_{n_2}^\dagger\cdots S_{n_{N+2}}^\dagger\big)| X \rangle
\langle X | T\big( S_{n_1} S_{n_2}\cdots S_{n_{N+2}} \big) | 0 \rangle \,,
\end{align}
where the sum represents the phase space of the emitted partons with \mbox{momenta $\{k_m\}$}. The Wilson lines $S_{n_i}$ extend along the lightlike directions $n_i^\mu$, and they depend on the colour representation of the associated hard partons. The soft function by itself is a matrix in colour space, and we make use of the colour-space notation of~\cite{Catani:1996vz} throughout this article. The term $\mathcal{M}(\tau;\lbrace k_{m} \rbrace)$ finally specifies that the N-jettiness is measured on the soft radiation. Following~\cite{Bell:2018vaa,Bell:2018oqa,Bell:2020yzz}, we assume that the measurement function is formulated in Laplace space, with $\tau$ being the Laplace-conjugate variable to $\mathcal{T}_{N}$. Specifically, we write 
\begin{align}
\mathcal{M}(\tau;\lbrace k_{m} \rbrace) = 
\exp\bigg(-\tau\, \sum_{m} \,\min_i\big\{n_i \cdot k_m\big\}\bigg)\,.
\label{eq:measurement}
\end{align}
The major advantage of working in Laplace space is that the expressions contain ordinary functions rather than distributions, and that the renormalisation of the soft function simplifies considerably as we will see below. 

We are interested in computing the N-jettiness soft function to NNLO in perturbation theory. To this end, we write the perturbative expansion of the bare soft function as
\begin{align}
\label{eq:Sbare:pertexpansion}
S_0(\tau) = 1 &+ \left(\frac{Z_\alpha\alpha_s}{4\pi}\right) \,
(\mu^2 \taubar^2)^\eps \, S^{(1)}(\eps) 
+ \left(\frac{Z_\alpha\alpha_s}{4\pi}\right)^2 
(\mu^2 \taubar^2)^{2\eps} \,S^{(2)}(\eps) 
+ \mathcal{O}(\alpha_s^3)\,,
\end{align}
where $\eps=(4-d)/2$ is the dimensional regulator, $\taubar = \tau e^{\gamma_E}$, and $\alpha_{s}$ is the strong-coupling constant renormalised in the $\overline{\text{MS}}$-scheme, which is related to the bare coupling $\alpha_s^0$ via $Z_\alpha \alpha_s\,\mu^{2\eps}=e^{-\eps\gamma_E}(4\pi)^\eps \alpha_s^0$ with $Z_\alpha = 1-\beta_0\alpha_s/(4\pi\eps)$ and $\beta_0 = 11/3\, C_A - 4/3\,T_F n_f$. In this notation, the first term represents the unit operator in colour space, whereas the NLO coefficient $S^{(1)}(\eps)$ and the NNLO coefficient $S^{(2)}(\eps)$ have a non-trivial colour structure, and they implicitly depend on the kinematic variables $n_{ij}$.

\subsection{Phase-space parametrisations}
\label{sec:parametrisations}

Before entering the technical details of the calculation, we specify the variables that we use to parametrise the phase space of the soft emissions. The parametrisations are a generalisation of the ones that some of us introduced for dijet soft functions in back-to-back kinematics in~\cite{Bell:2018vaa,Bell:2018oqa,Bell:2020yzz}. We first address the single-emission case, subsequently turn to the double real-emission phase space, and finally discuss the form of the measurement function for a generic N-jet observable. Throughout this section, we assume that the dipole that emits the soft radiation is spanned by the directions $n_i^\mu$ and $n_j^\mu$. We will see later that the same type of parametrisations can be used for the three-particle correlations as well.

\paragraph{Single emission:} 
We start with the Sudakov decomposition of the momentum $k^\mu$ of the emitted soft parton in terms of the dipole directions $n_i^\mu$ and $n_j^\mu$,
\begin{align}
k^\mu &=\,  k_j \,\frac{n_i^\mu}{n_{ij}} +\, k_i \,\frac{n_j^\mu}{n_{ij}} + k_\perp^\mu\,,
\end{align}
where $k_\perp^\mu$ with $k_\perp \cdot n_i = k_\perp \cdot n_j =0$ parametrises the transverse space with respect to the dipole directions, and we introduced the short-hand notation $k_i = k \cdot n_i$ etc. In the non-back-to-back case, the vector $k_\perp^\mu$ has a non-zero temporal component, but it always satisfies $k_\perp^2<0$. It is therefore possible to find a reference frame in which the dipole directions are back-to-back. This is achieved by a boost with velocity $\vec{\beta}= -\frac12 (\vec{n}_i + \vec{n}_j)$ and boost factor $\gamma=(1-\vec{\beta}^2)^{-1/2}=\sqrt{2/n_{ij}}$ (see~\cite{Kasemets:2015uus} for an explicit construction of such a boost).

In the boosted back-to-back frame, in which we denote all four-vectors with a prime, the transverse vector $k^{\prime \,\mu}_\perp$ is then purely spatial. One may therefore decompose this vector in the usual Cartesian basis $\vec{e}_{\perp_n}$ for the spatial components,
\begin{align}
\label{eq:kperp}
k^{\prime \,\mu}_\perp &= - (k' \cdot e'_{\perp_1}) e_{\perp_1}^{\prime \,\mu} - (k' \cdot e'_{\perp_2}) e_{\perp_2}^{\prime \,\mu} - \ldots\,,
\end{align}
where the first two directions refer to the physical transverse space in four space-time dimensions, while there are $(d-4)$ additional directions in $d$ dimensions, hiding in the dots. Notice that we have 
$e_{\perp_n}^{\prime \,\mu}=(0,\vec{e}_{\perp_n}$) with 
$e'_{\perp_n} \cdot e'_{\perp_m}=-\delta_{nm}$ in four-vector notation. In contrast to the symmetric back-to-back case, there exists a preferred direction in the transverse space for non-back-to-back kinematics that is singled out by the boost vector $\vec{\beta}$. We may then use this direction to parametrise the first vector $e_{\perp_1}^{\prime \,\mu}$, and for convenience we choose $e_{\perp_1}^{\prime \,\mu}=(0,-\vec{n}_\beta)$ with $\vec{n}_\beta=\vec{\beta}/|\vec{\beta}|$. We furthermore construct the remaining vector in the physical transverse space via $e_{\perp_2}^{\prime \,\mu}=\big(0,(\vec{n}_i\times\vec{n}_j)/|\vec{n}_i\times\vec{n}_j|\big)$, while we leave the orientation of the additional $(d-4)$ directions unspecified for the moment.

In the original non-back-to-back frame the vector $k_\perp^\mu$ has a similar decomposition as in \eqref{eq:kperp}, and the corresponding basis vectors transform into
\begin{align}
e_{\perp_1}^{\mu} = \Big(\gamma |\vec{\beta}|,-\gamma \,\vec{n}_\beta\Big)\,,
\qquad\qquad
e_{\perp_2}^{\mu} = \bigg(0,\frac{\vec{n}_i\times\vec{n}_j}{|\vec{n}_i\times\vec{n}_j|}\bigg)\,,
\end{align}
where as argued before the first vector now has a non-zero temporal component. It is easy to verify that these vectors satisfy $e_{\perp_1} \cdot n_i = e_{\perp_2} \cdot n_i =0$, and similar for their projections onto $n_j^\mu$, along with $e_{\perp_1} \cdot e_{\perp_2} = 0$. Instead of parametrising the transverse space in Cartesian coordinates, we will use $(d-2)$-dimensional spherical coordinates that we introduce as follows,
\begin{align}
\label{eq:spherical-coordinates}
k \cdot e_{\perp_1} 
&=k' \cdot e'_{\perp_1} 
= - k_T \,c_{k_1}\,,
& &\qquad\;\vdots
\\
k \cdot e_{\perp_2} 
&=k' \cdot e'_{\perp_2} 
= - k_T \,s_{k_1} \,c_{k_2}\,,\quad
& 
k \cdot e_{\perp_{d-3}} 
&=k' \cdot e'_{\perp_{d-3}}
= - k_T \, s_{k_1} \, \ldots \, s_{k_{d-4}} \, c_{k_{d-3}}\,,
\nonumber\\[-0.5em]
&\qquad\;\vdots
&
k \cdot e_{\perp_{d-2}} 
&=k' \cdot e'_{\perp_{d-2}}
= - k_T \, s_{k_1} \, \ldots \, s_{k_{d-4}} \, s_{k_{d-3}}\,,
\nonumber
\end{align}
where $c_{k_i}=\cos\theta_{k_i}$, $s_{k_i}=\sin\theta_{k_i}$ 
and $k_\perp^2=-k_T^2$. This parametrisation is obviously Lorentz-invariant, and from the construction it should be clear that the interpretation of the variables $k_T$ and the angles $\theta_{k_i}$ refers to the boosted back-to-back frame, although we omitted the primes on these variables to simplify the notation. 

We are now in the position to generalise the phase-space parametrisation from~\cite{Bell:2018vaa,Bell:2018oqa,Bell:2020yzz} to non-back-to-back dipoles. Specifically, we use in the single-emission case
\begin{align}
y_k = \frac{k_i}{k_j}\,, \qquad\qquad 
k_T = \sqrt{\frac{2k_ik_j}{n_{ij}}}\,, \qquad\qquad
t_{k_i} = \frac{1-\cos\theta_{k_i}}{2}\,,
\end{align}
where $y_k$ is a measure of the rapidity of the emitted parton with respect to the dipole directions $\vec{n}_i$ and $\vec{n}_j$, and we have used the on-shell condition to relate the transverse component $k_T$ to the longitudinal projections $k_i$ and $k_j$. According to \eqref{eq:measurement}, the N-jettiness measurement requires us to determine the minimal projection of the momentum $k^\mu$ onto all lightlike vectors $n_X^\mu$ with $X\in\{1,\ldots,N+2\}$,
\begin{align}
\mathcal{M}_1(\tau;k) = 
\exp\Big(-\tau\, \min_X \,k_X\Big)\,,
\label{eq:M1:Njettiness}
\end{align}
where 
\begin{align}
k_X = k \cdot n_X = k_T \, \sqrt{\frac{n_{ij}}{2}} \, \sqrt{y_k} \;
Z_X(y_k, \theta_{k_1}, \theta_{k_2})\,,
\label{eq:single-emission:kX}
\end{align}
with
\begin{align}
Z_X(y_k, \theta_{k_1}, \theta_{k_2}) &= \frac{k_X}{k_i} =  
 \frac{n_{jX}}{n_{ij}} + \frac{n_{iX}}{n_{ij}} \frac{1}{y_k}  
+ \sqrt{\frac{2}{n_{ij}}}\, \frac{1}{\sqrt{y_k}} 
\Bigl( n_{X\perp_1} \cos\theta_{k_1} + 
n_{X\perp_2} \sin\theta_{k_1} \cos\theta_{k_2} \Bigr)\,,
\label{eq:single-emission:ZX}
\end{align}
and $n_{X\perp_m}= n_X \cdot e_{\perp_m}$. The latter factors take a particularly simple form for a projection onto the dipole directions, $Z_i(y_k, \theta_{k_1}, \theta_{k_2})= 1$ and $Z_j(y_k, \theta_{k_1}, \theta_{k_2})=1/y_k$, while the projection onto an arbitrary direction depends at most on two angles $\theta_{k_1}$ and $\theta_{k_2}$, since the vectors $n_X^\mu$ live in the physical four-dimensional subspace. In comparison to the  construction from~\mbox{\cite{Bell:2018vaa,Bell:2018oqa,Bell:2020yzz}}, we thus observe that the second angle arises here as a consequence of the boost vector $\vec{\beta}$, which is absent in the back-to-back case. 

\paragraph{Two emissions:} 
For the double real-emission contribution, we similarly aim at generalising the phase-space parametrisations from~\cite{Bell:2018vaa,Bell:2018oqa,Bell:2020yzz} to non-back-to-back dipoles. For the variables that depend on the longitudinal projections of the momenta $k^\mu$ and $l^\mu$ of the emitted soft partons, this generalisation reads
\begin{align}
\label{eq:M2:variables}
p_T =  \sqrt{\frac{2}{n_{ij}}(k_{i} + l_{i})(k_{j} + l_{j})} \,, \quad\quad
y &= \frac{k_{i} + l_{i}}{k_{j} + l_{j}} \,, 
\quad\quad
a = \sqrt{\frac{k_{j}l_{i}}{k_{i}l_{j}}} \,,
\quad\quad
b =\sqrt{\frac{k_{i}k_{j}}{l_{i}l_{j}}} \,.
\end{align}
In the spirit of the correlated-emission parametrisation from~\cite{Bell:2018vaa,Bell:2018oqa}, we thus employ two variables $p_T$ and $y$ that depend on the collective light-cone components of the two emissions, and two further variables $a$ and $b$ that depend on the relative ones. 

The parametrisation of the $(d-2)$-dimensional transverse space is, on the other hand, more involved in the double real-emission case. Following the construction from above, we have
\begin{align}
k_\perp \cdot n_X &= k_T 
\Big( n_{X\perp_1} \cos\theta_{k_1} + 
n_{X\perp_2} \sin\theta_{k_1} \cos\theta_{k_2} \Big)\,,
\nonumber\\
l_\perp \cdot n_X &= l_T 
\Big( n_{X\perp_1} \cos\theta_{l_1} + 
n_{X\perp_2} \sin\theta_{l_1} \cos\theta_{l_2} \Big)\,,
\label{eq:kl-projections}
\end{align}
where the angles $\theta_{l_1}$ and $\theta_{l_2}$ are introduced in analogy to \eqref{eq:spherical-coordinates}, and their interpretation therefore also refers to the boosted back-to-back frame. As the soft matrix element depends on the angle $\theta_{kl}$ between the vectors $\vec{k}_\perp^{\,\prime}$ and $\vec{l}_\perp^{\,\prime}$, we want to trade some of these angles for $\theta_{kl}$, similar to the strategy that was applied for dijet soft functions in~\cite{Bell:2018oqa}. The explicit construction of this variable transformation is presented in~\ref{app:angles}, where we show that the phase-space integrals depend at most on five non-trivial angles $\{\theta_{kl},\theta_{5},\theta_{6},\theta_{l_1},\theta_{l_2}\}$ in the double real-emission case, in comparison to the two angles $\{\theta_{k_1},\theta_{k_2}\}$ that we found for a single emission. Here $\theta_{kl}$ is the desired angle with
\begin{align}
k_\perp \cdot l_\perp &= - k_T \,l_T \cos \theta_{kl}\,,
\end{align} 
whereas $\theta_{5}$ and $\theta_{6}$ are auxiliary angles that appear in the construction (we adopt the terminology used in~\cite{Bell:2018oqa}). 

In terms of these variables, the projection of the four-vector $k_\perp^\mu$ onto the (N+2) lightlike vectors $n_X^\mu$ 
takes the form
\begin{align}
\label{eq:kperpdotnX}
k_\perp \cdot n_X &= k_T \Big( n_{X\perp_1} \, \lambda_3(\theta_{kl},\theta_5,\theta_{l_1}) + n_{X\perp_2} \, \lambda_4(\theta_{kl},\theta_5,\theta_6,\theta_{l_1},\theta_{l_2}) \Big)\,,
\end{align}
with 
\begin{align}
\lambda_3(\theta_{kl},\theta_5,\theta_{l_1})  &=
\cos \theta_{kl} \cos \theta_{l_1}-\sin \theta_{kl}\cos \theta_5 \sin \theta_{l_1}\,,
\nn  \\
\lambda_4(\theta_{kl},\theta_5,\theta_6,\theta_{l_1},\theta_{l_2}) &=  
\cos \theta_{kl} \sin \theta_{l_1} \cos \theta_{l_2} 
\nn \\ 
&\quad+ \sin \theta_{kl} \,\big(\cos \theta_5 \cos \theta_{l_1} \cos \theta_{l_2} - \sin \theta_5\cos \theta_6 \sin \theta_{l_2}\big)\,,
\label{eq:lambda34}
\end{align}
while the projection of $l_\perp^\mu$ in the second relation of \eqref{eq:kl-projections} remains unchanged, or is equivalently given by
\begin{align}
\label{eq:lperpdotnX}
l_\perp \cdot n_X &= l_T \Big( n_{X\perp_1} \, \lambda_3(0,0,\theta_{l_1}) + n_{X\perp_2} \, \lambda_4(0,0,0,\theta_{l_1},\theta_{l_2}) \Big)\,.
\end{align}
Coming back to the N-jettiness measurement function, we then start in the double real-emission case from the representation
\begin{align}
\mathcal{M}_2(\tau;k,l) = 
\exp\Big(-\tau\, \big(\min_X \,k_X
+ \min_{X} \,l_{X} \big)\Big)\,,
\label{eq:M2:Njettiness}
\end{align}
where the minimisation is performed with respect to the lightlike vectors $n_X^\mu$ independently in each term. In analogy to \eqref{eq:single-emission:kX}, we then write the relevant projections of the momenta $k^\mu$ and $l^\mu$ in the form
\begin{align}
k_X &= k \cdot n_X = p_T \, \sqrt{\frac{n_{ij}}{2}} \, \sqrt{y} 
\; \frac{b}{a+b} \;
Z_{X,k}(a,b,y, \{\theta\})\,,
\nonumber\\
l_X &= l \cdot n_X = p_T \, \sqrt{\frac{n_{ij}}{2}} \, \sqrt{y} 
\;\frac{a}{a+b}\;
Z_{X,l}(a,b,y, \{\theta\})\,,
\label{eq:kXlX}
\end{align}
where $\{\theta\} = \{\theta_{kl},\theta_{5},\theta_{6},\theta_{l_1},\theta_{l_2}\}$ is a short-hand notation for the angular variables, and
\begin{align}
Z_{X,k}(a,b,y, \{\theta\}) = \frac{k_X}{k_i} &=  
 \frac{n_{jX}}{n_{ij}} + \frac{n_{iX}}{n_{ij}} \, 
\frac{a(a+b)}{1+a b} \, \frac{1}{y} 
+ \sqrt{\frac{2}{n_{ij}}}\sqrt{\frac{a(a+b)}{1+a b}}\, 
\frac{1}{\sqrt{y}} 
\nonumber\\
&\qquad \times \Big( n_{X\perp_1} \, \lambda_3(\theta_{kl},\theta_5,\theta_{l_1}) + n_{X\perp_2} \, \lambda_4(\theta_{kl},\theta_5,\theta_6,\theta_{l_1},\theta_{l_2}) \Big)\,,
\nonumber\\
Z_{X,l}(a,b,y, \{\theta\}) = \frac{l_X}{l_i} &=  
 \frac{n_{jX}}{n_{ij}} + \frac{n_{iX}}{n_{ij}} \, 
\frac{a+b}{a(1+a b)} \, \frac{1}{y} 
+ \sqrt{\frac{2}{n_{ij}}}\sqrt{\frac{a+b}{a(1+a b)}}\, 
\frac{1}{\sqrt{y}} 
\nonumber\\
&\qquad \times \Big( n_{X\perp_1} \, \cos \theta_{l_1} + 
n_{X\perp_2} \, \sin \theta_{l_1} \cos \theta_{l_2}\Big)\,.
\end{align}
Once again, the latter factors take a particularly simple form for a projection onto the dipole directions $n_i^\mu$ and $n_j^\mu$, since their scalar products with the basis vectors $e_{\perp_1}^\mu$ and $e_{\perp_2}^\mu$ vanish by construction.

\paragraph{Generic N-jet observables:} 
While we focus on the N-jettiness soft function in this article, the method we present for computing N-jet soft functions at hadron colliders is general, and it can hence be applied to a much broader class of observables. The {\tt SoftSERVE} approach~\mbox{\cite{Bell:2018vaa,Bell:2018oqa,Bell:2020yzz}} distinguishes, moreover, between observables that respect the non-Abelian exponentiation (NAE) theorem~\cite{Gatheral:1983cz,Frenkel:1984pz} and those that violate it. The exponential structure of the soft matrix elements in terms of connected webs, in particular, relates the uncorrelated-emission contribution to the lower-order coefficients for the former, while it is unrelated and needs to be calculated explicitly for the latter.  In the present work, we assume that the considered N-jet observables comply with NAE, which obviously is the case for N-jettiness, since the two-emission measure\-ment function in \eqref{eq:M2:Njettiness} trivially factorises into a product of two single-emission functions in Laplace space. A generalisation of the N-jet approach for NAE-violating observables would be needed, for instance, for soft functions that involve the action of a jet algorithm or a grooming procedure. For convenience, we furthermore restrict our attention to \mbox{SCET-1} observables in this article, for which all divergences are regularised by the dimensional \mbox{regulator $\eps$}.

Adopting the notation from~\cite{Bell:2018vaa,Bell:2018oqa} (see also~\cite{Bell:2018mkk}), the single-emission measurement function for a generic N-jet observable in Laplace space is then expressed as
\begin{align}
\mathcal{M}_1(\tau; k) = \exp\biggl\{-\tau\, \Big(\frac{n_{ij}}{2}\Big)^{n/2} \,k_{T}\bigg( 
y_k^{n/2}\, f_A(y_k,\theta_{k_1},\theta_{k_2}) \,\Theta_1 + 
y_k^{-n/2}\, f_B(1/y_k,\theta_{k_1},\theta_{k_2}) \,\Theta_2
\biggr)\bigg\},
\label{eq:M1}
\end{align}
where we split the phase space into two contributions, indicated by $\Theta_1=\theta(1-y_k)=1-\Theta_2$, in order to disentangle the collinear divergences that arise in the limits \mbox{$y_k\to0$} and $y_k\to\infty$. In the approach from~\cite{Bell:2018vaa,Bell:2018oqa}, these contributions are related by the \mbox{$n^\mu\leftrightarrow\bar n^\mu$} symmetry for back-to-back dipoles, which obviously does not hold in the N-jet case. As a consequence, twice as many ingredients are needed to compute N-jet soft functions than for the ones considered in~\cite{Bell:2018vaa,Bell:2018oqa}. 

Closer inspection of \eqref{eq:M1} reveals that a generic N-jet observable is char\-acterised by a parameter $n$ and two functions $f_{A,B}(y_k,\theta_{k_1},\theta_{k_2})$ in our framework. The parameter $n$ is related to the scaling of the momentum modes in the effective theory~\cite{Bell:2018oqa}, and it can in practise be determined by requiring that the functions $f_{A,B}(y_k,\theta_{k_1},\theta_{k_2})$ are finite and non-zero in the limit $y_k\to0$. We furthermore assume that the Laplace variable $\tau$ has the dimension $1/$mass, which fixes the linear dependence on $k_T$, while $(n_{ij}/2)^{n/2}$  is factored out for convenience. In this notation, the N-jettiness case is recovered for $n=1$, along with
\begin{align}
f_{A}(y_k,\theta_{k_1},\theta_{k_2})=\min_X \,Z_X(y_k,\theta_{k_1},\theta_{k_2}),  \qquad
f_{B}(y_k,\theta_{k_1},\theta_{k_2})=1/y_k\, f_{A}(1/y_k,\theta_{k_1},\theta_{k_2}), 
\end{align}
and it is easy to verify that these functions are finite as $y_k\to 0$.

Similarly, we write the double-emission measurement function for a generic N-jet observable in Laplace space in the form
\begin{align}
\mathcal{M}_2(\tau; k,l) &= \exp\bigg\{-\tau\, \Big(\frac{n_{ij}}{2}\Big)^{n/2}
 \,p_{T}
\bigg( 
y^{n/2}\, F_A\big(a,b,y,\{\theta\}\big) \,\Theta_A
+ y^{n/2} \, F_B\big(a,1/b,y,\{\theta\}\big) \,\Theta_B\nn\\
&\qquad\qquad
+ 
y^{-n/2}\, F_C\big(a,b,1/y,\{\theta\}\big) \,\Theta_C
+ y^{-n/2}\, F_D\big(a,1/b,1/y,\{\theta\}\big) \,\Theta_D
\bigg)\bigg\},
\label{eq:M2}
\end{align}
where the phase space is now split into four contributions according to 
\begin{align}
\Theta_A&=\theta(1-a)\,\theta(1-b)\,\theta(1-y)\,,
& 
\Theta_B&=\theta(1-a)\,\theta(b-1)\,\theta(1-y)\,,
\nonumber\\
\Theta_C&=\theta(1-a)\,\theta(1-b)\,\theta(y-1)\,,
& 
\Theta_D&=\theta(1-a)\,\theta(b-1)\,\theta(y-1)\,,
\end{align}
and the various functions $F_i\big(a,b,y,\{\theta\}\big)$ with $i\in\{A,B,C,D\}$ are again assumed to be finite and non-zero in the limit $y\to 0$. As will be explained in later sections, the additional contributions with $a \geq 1$ can be mapped onto these regions using the symmetry under the exchange of the momenta $k^\mu$ and $l^\mu$. Compared to~\cite{Bell:2018vaa,Bell:2018oqa}, which contains  two functions for the double real-emission contribution, the number is thus again doubled here because of the lacking \mbox{$n^\mu\leftrightarrow\bar n^\mu$} symmetry. For the N-jettiness soft function, the required ingredients are then
\begin{align}\label{eq:2loopF}
F_A\big(a,b,y,\{\theta\}\big)=
\frac{b}{a+b}\; \min_X \, Z_{X,k}(a,b,y, \{\theta\})
+\frac{a}{a+b}\; \min_X \, Z_{X,l}(a,b,y, \{\theta\})\,,
\end{align}
along with 
$F_B\big(a,b,y,\{\theta\}\big) = F_A\big(a,1/b,y,\{\theta\}\big)$,
$F_C\big(a,b,y,\{\theta\}\big) = 1/y\, F_A\big(a,b,1/y,\{\theta\}\big)$, as well as $F_D\big(a,b,y,\{\theta\}\big) = 1/y\,F_A\big(a,1/b,1/y,\{\theta\}\big)$.

\subsection{NLO calculation}
\label{sec:NLO}

While the NLO calculation of the N-jettiness soft function was already presented in~\cite{Jouttenus:2011wh}, we briefly review it here using the {\tt SoftSERVE} strategy for a generic N-jet observable that is characterised by the measurement function $\mathcal{M}_1(\tau; k)$ in  \eqref{eq:M1}. As mentioned earlier, we restrict our attention to SCET-1 observables here for which all divergences show up as poles in the dimensional regulator $\eps=(4-d)/2$. In this regularisation scheme the purely virtual corrections are scaleless and vanish, and we are thus left with the real-emission contribution at this order. As the reference vectors $n_i^\mu$ are lightlike, the real-emission diagrams that connect the same Wilson line also vanish, and the NLO coefficient $S^{(1)}(\eps)$ defined in \eqref{eq:Sbare:pertexpansion} becomes a sum over dipole contributions,
\begin{equation}
\label{eq:S1}
S^{(1)}(\eps) = \sum_{i\neq j} \; \bfT_{i} \cdot \bfT_{j} \;
\Big( \frac{n_{ij}}{2} \Big)^{n\eps} \, S_{ij}^{(1)}(\eps)\,,
\end{equation}
where $\bfT_i$ and $\bfT_j$ are the colour generators of the $i$-th and $j$-th hard parton with reference vectors $n_i^\mu$ and $n_j^\mu$, respectively. In this notation, the dipole contributions can be calculated from
\begin{equation}
S_{ij}^{(1)}(\eps)  = -\frac{(2^{2-n}\, n_{ij}^{n}\,\pi e^{\gamma_E} \tau^2)^{-\eps}}{(2\pi)^{d-1}} \,
\int d^{d}k \;\, \delta(k^{2}) \,\theta(k^{0}) 
\,\mathcal{M}_1(\tau; k) \, |\mathcal{A}^R_{ij}(k)|^{2} \,,
\end{equation}
where $|\mathcal{A}^R_{ij}(k)|^{2}=32 \pi^2 (n_{ij}/2/k_i/k_j)$ is proportional to the squared tree-level matrix element.

In order to evaluate this expression for a generic N-jet observable, we express the Lorentz-invariant phase-space measure in terms of the variables that we introduced in the previous section. To do so, we adopt the frame in which the reference vectors are back-to-back with 
$n_i^{\prime\,\mu} = 1/\gamma\,(1,\vec{n}_i^\prime)$, $n_j^{\prime\,\mu} = 1/\gamma \,(1,\vec{n}_j')$ and
$\vec{n}_j' = - \vec{n}_i'$. We recall that the transverse vector $k^{\prime \,\mu}_\perp$ is purely spatial in this frame, and the phase-space measure then translates into
\begin{align}
& \int d^{d}k' \; \delta(k^{\prime\,2}) \,\theta(k^{\prime\,0}) 
=\frac{1}{n_{ij}} \,\int dk_i^\prime \; dk_j^\prime\; 
d^{d-2}k_\perp^\prime \;\, 
\delta\Big(\frac{2k_i^\prime k_j^\prime}{n_{ij}} + k_\perp^{\prime\,2}\Big) \,
\theta\Big(\frac{k_i^\prime+k_j^\prime}{\gamma n_{ij}}\Big) 
\nonumber\\
&\qquad
=\frac{\pi^{-\eps}}{2\,\Gamma(-\eps)}\;
\int_0^\infty \!dk_T \; k_T^{1-2\eps}\;
\int_0^\infty \frac{dy_k}{y_k} \; 
\int_{-1}^1 \!dc_{k_1} \,s_{k_1}^{-1-2\eps} \;
\int_{-1}^1 \!dc_{k_2} \,s_{k_2}^{-2-2\eps} \,,
\end{align}
where we recall that $c_{k_i}=\cos\theta_{k_i}$ and $s_{k_i}=\sin\theta_{k_i}$.
After inserting the explicit form \eqref{eq:M1} of the measurement function and integrating over the dimensionful variable $k_T$, this yields 
\begin{align}
S_{ij}^{(1)}(\eps)  &= -\frac{2\,e^{-\gamma_E\eps}}{\pi}\,
\frac{\Gamma(-2\eps)}{\Gamma(-\eps)}\;
\int_0^1 \! dy_k \;\, y_k^{-1+n\eps}\;
\int_{-1}^1 \!dc_{k_1} \,s_{k_1}^{-1-2\eps} \;
\int_{-1}^1 \!dc_{k_2} \,s_{k_2}^{-2-2\eps}
\nonumber\\
&\qquad
\times\;
\Big\{f_A(y_k,\theta_{k_1},\theta_{k_2})^{2\eps}
+ f_B(y_k,\theta_{k_1},\theta_{k_2})^{2\eps}\Big\}\,,
\label{eq:NLO:intermediate}
\end{align}
where we mapped the contribution from region B to the unit interval using $y_k\to 1/y_k$. In this representation the singularities are completely factorised and regularised; the collinear divergence associated with the limit $y_k\to0$ produces a pole in $\eps$ for SCET-1 soft functions with $n\neq0$, while the prefactor $\Gamma(-2\epsilon)$ captures the soft singularity from $k_T\to 0$. The integration over the angle $\theta_{k_2}$ produces, moreover, a spurious divergence, which is compensated by the prefactor $1/\Gamma(-\epsilon)= \mathcal{O}(\epsilon)$. Despite being unphysical, we need to expose this singularity properly in our numerical approach, and to do so we follow the strategy from~\cite{Bell:2018oqa}, which consists in mapping $c_{k_2}\to 1-2t_{k_2}$, splitting the resulting integration at $t_{k_2}=1/2$, and rescaling the individual contributions via $t_{k_2} \rightarrow t_{k_2}'/2$ and $t_{k_2} \rightarrow 1- t_{k_2}'/2$. As a result, we obtain the following master formula for the computation of the NLO dipole contributions
\begin{align}
S_{ij}^{(1)}(\eps)  &= -\frac{4\,e^{-\gamma_E\eps}}{\pi}\,
\frac{\Gamma(-2\eps)}{\Gamma(-\eps)}\;
\int_0^1 \!dy_k \;\, y_k^{-1+n\eps}\;
\int_{0}^1 \!dt_{k_1} \,(4 t_{k_1} \bar t_{k_1})^{-1/2-\eps} 
\nonumber\\
&\qquad
\times\; 
\int_{0}^1 \!dt'_{k_2} \,(t'_{k_2}(2-t'_{k_2}))^{-1-\eps} \;
\tilde f(y_k,\theta_{k_1},\theta_{k_2})\,,
\label{eq:NLO:master}
\end{align}
where 
\begin{align}
\tilde f(y_k,\theta_{k_1},\theta_{k_2})&= 
\Big[\mathcal{O}_1(\theta_{k_1},\theta_{k_2}) + \mathcal{O}_2(\theta_{k_1},\theta_{k_2})\Big]\;
\Big\{f_A(y_k,\theta_{k_1},\theta_{k_2})^{2\eps}
+ f_B(y_k,\theta_{k_1},\theta_{k_2})^{2\eps}\Big\}\,,
\end{align}
and $\mathcal{O}_{1,2}(\theta_{k_1},\theta_{k_2})$ are replacement operators that act on a test function $g(\theta_{k_1},\theta_{k_2})$ as
\begin{align}
\mathcal{O}_1(\theta_{k_1},\theta_{k_2})\; g(\theta_{k_1},\theta_{k_2})&=
g(\theta_{k_1},\theta_{k_2})\,\Big| \scriptsize\begin{array}{l} c_{k_1} \to 1-2 t_{k_1}\\[-0.2em]
	c_{k_2} \to 1- t'_{k_2}\end{array}\,,
\nonumber\\[0em]
\mathcal{O}_2(\theta_{k_1},\theta_{k_2})\; g(\theta_{k_1},\theta_{k_2})&=
g(\theta_{k_1},\theta_{k_2})\,\Big| \scriptsize\begin{array}{l} c_{k_1} \to 1-2 t_{k_1}\\[-0.2em]
	c_{k_2} \to t'_{k_2}-1\end{array}\,.
\label{eq:O12}
\end{align}
Our result in \eqref{eq:NLO:master} generalises the corresponding expression for dijet soft functions in~\cite{Bell:2018oqa}, where we assumed that the two lightlike directions are back-to-back and that the observable is symmetric under $n^\mu\leftrightarrow\bar n^\mu$ exchange, to cases that involve multiple lightlike directions and arbitrary kinematics.

\subsection{NNLO calculation}
\label{sec:NNLO}

At NNLO the coefficient $S^{(2)}(\eps)$ defined in \eqref{eq:Sbare:pertexpansion} consists of three contributions
\begin{equation}
S^{(2)}(\eps)= S^{(2,{\rm RV})}(\eps) +{S}^{(2,q\bar{q})}(\eps) +{S}^{(2,gg)}(\eps)\,,
\label{eq:NNLOcontributions}
\end{equation}
where the first term describes the mixed real-virtual (RV) corrections, and the latter two capture the corrections from the emission of a soft quark-antiquark pair ($q\bar q$) or two soft gluons ($gg$). Whereas the evaluation of the double real-emission corrections represents the most challenging part of the calculation, the purely virtual contributions are again scaleless and vanish in the applied regularisation scheme. In this section we will derive master formulae for all of the above corrections for a generic N-jet observable that is described by the measurement functions $\mathcal{M}_1(\tau; k)$ in  \eqref{eq:M1} and $\mathcal{M}_2(\tau; k,l)$ in  \eqref{eq:M2}.

\paragraph{Real-virtual corrections:} 
The mixed real-virtual contribution  consists of two-particle and three-particle correlations~\cite{Catani:2000pi},
\begin{align}
\label{eq:NNLO:RV}
S^{(2,{\rm RV})}(\eps) &= C_A \,\sum_{i\neq j} \; \bfT_{i} \cdot \bfT_{j} \;
\Big( \frac{n_{ij}}{2} \Big)^{2n\eps} \, S_{ij}^{(2,{\rm Re})}(\eps)
\nonumber\\
&\quad
+  \sum_{i\neq l\neq j} (\lambda_{il} - \lambda_{ip}-\lambda_{lp})\,
f_{ABC}\; \bfT_{i}^A \;\bfT_{l}^B \;\bfT_{j}^C  \;
\Big( \frac{n_{ij}}{2} \Big)^{2n\eps} \,
S_{ilj}^{(2,{\rm Im})}(\eps)\,,
\end{align}
where the indices $i$, $j$ and $l$ refer to the hard partons associated with the respective Wilson lines and $p$ represents the emitted soft gluon. Similar to the NLO structure in \eqref{eq:S1}, the first term consists of a sum of dipole contributions, whereas the second term describes three-parton correlations which we will refer to as  tripole contributions. The dipoles are, moreover, associated with the real part of the underlying loop integral, whereas the tripoles arise from its imaginary part, which is process dependent. This process dependence is reflected by the  $\lambda_{AB}$ factors in \eqref{eq:NNLO:RV}, which take the values $\lambda_{AB}=1$ if the partons $A$ and $B$ are both incoming or outgoing, and $\lambda_{AB}=0$ otherwise. In processes with three hard partons, the sum of the tripole contributions vanishes because of colour conservation~\cite{Catani:2000pi}, but for general processes with four or more partons it leads to a non-trivial correction. 
\bigskip

We first consider the dipole contributions, which can be calculated from~\cite{Catani:2000pi} 
\begin{equation}
S_{ij}^{(2,{\rm Re})}(\eps)  = \frac{(2^{2-n}\, n_{ij}^{n}\,\pi e^{\gamma_E} \tau^2)^{-2\eps}}{(2\pi)^{d-1}} \,
\int d^{d}k \;\, \delta(k^{2}) \,\theta(k^{0}) 
\,\mathcal{M}_1(\tau; k) \, |\mathcal{A}^{\rm Re}_{ij}(k)|^{2} \,,
\end{equation}
with
\begin{equation}
|\mathcal{A}^{\rm Re}_{ij}(k)|^{2} =  64 \pi^2 \cos(\pi\eps) \, \frac{(4\pi)^\eps}{\eps^2} \,
\frac{\Gamma^3(1-\eps)\Gamma^2(1+ \eps)}{\Gamma(1-2 \eps)}\;
\Big(\frac{n_{ij}}{2k_{i}k_{j} }\Big)^{1+\eps}\,.
\end{equation}
As the structure of the soft matrix element is similar to the NLO real emission matrix element,  $|\mathcal{A}^R_{ij}(k)|^{2}\propto  (n_{ij}/2/k_i/k_j)$, we can proceed along the same steps as in the previous section to derive the respective master formula. The result reads
\begin{align}
\label{eq:NNLO:RV:dipoles:master}
S_{ij}^{(2,{\rm Re})}(\eps)  &=  \frac{8\,e^{-2\gamma_E\eps}}{\pi}\, \frac{1}{\eps^2} \,
\frac{\Gamma^3(1-\eps)\Gamma^2(1+ \eps)\Gamma(-4\eps)\cos(\pi\eps) }{\Gamma(1-2 \eps)\Gamma(-\eps)}
\;\int_0^1 \! dy_k \;\, y_k^{-1+2n\eps}
\nonumber\\
&\qquad
\times\;
\int_{0}^1 \!dt_{k_1} \,(4 t_{k_1} \bar t_{k_1})^{-1/2-\eps} \;
\int_{0}^1 \!dt'_{k_2} \,(t'_{k_2}(2-t'_{k_2}))^{-1-\eps} \;
\hat f(y_k,\theta_{k_1},\theta_{k_2})\,,
\end{align}
where 
\begin{align}
\hat f(y_k,\theta_{k_1},\theta_{k_2})&= 
\Big[\mathcal{O}_1(\theta_{k_1},\theta_{k_2}) + \mathcal{O}_2(\theta_{k_1},\theta_{k_2})\Big]\;
\Big\{f_A(y_k,\theta_{k_1},\theta_{k_2})^{4\eps}
+ f_B(y_k,\theta_{k_1},\theta_{k_2})^{4\eps}\Big\}\,,
\end{align}
and $\mathcal{O}_{1,2}(\theta_{k_1},\theta_{k_2})$ are the replacement operators that we introduced in \eqref{eq:O12}. This expression generalises the corresponding result that was given for back-to-back dipoles in~\cite{Bell:2018oqa}.
\bigskip

We next turn to the tripole contributions, which are given by~\cite{Catani:2000pi}
\begin{equation}
S_{ilj}^{(2,{\rm Im})}(\eps)  = \frac{(2^{2-n}\, n_{ij}^{n}\,\pi e^{\gamma_E} \tau^2)^{-2\eps}}{(2\pi)^{d-1}} \,
\int d^{d}k \;\, \delta(k^{2}) \,\theta(k^{0}) 
\,\mathcal{M}_1(\tau; k) \, |\mathcal{A}^{\rm Im}_{ilj}(k)|^{2} \,,
\end{equation}
with
\begin{equation}
|\mathcal{A}^{\rm Im}_{ilj}(k)|^{2} =  128 \pi^2 \sin(\pi\eps) \, \frac{(4\pi)^\eps}{\eps^2} \,
\frac{\Gamma^3(1-\eps)\Gamma^2(1+ \eps)}{\Gamma(1-2 \eps)}\;
\Big(\frac{n_{ij}}{2k_{i}k_{j} }\Big)\Big(\frac{n_{il}}{2k_{i}k_{l} }\Big)^{\eps}\,.
\label{eq:tripoles:matrixelement}
\end{equation}
Note that the tripoles are not symmetric under the exchange of any pair of indices, since the directions $n_i^\mu$ and $n_l^\mu$ enter the loop diagram interfering with an emission from the direction $n_j^\mu$, whereas the directions $n_i^\mu$ and $n_j^\mu$ have been singled out by the  Sudakov decomposition. The singularity structure of the soft matrix element is then again controlled by the factor $(n_{ij}/2/k_i/k_j)$, and we can therefore apply the same phase-space parametrisation as in the dipole case. The calculation therefore again proceeds along the steps outlined above, and we can similarly derive the corresponding master formula for the tripole contributions,
\begin{align}
\label{eq:NNLO:RV:tripoles:master}
S_{ilj}^{(2,{\rm Im})}(\eps)  &=  \frac{16\,e^{-2\gamma_E\eps}}{\pi}\, \frac{1}{\eps^2} \,
\frac{\Gamma^3(1-\eps)\Gamma^2(1+ \eps)\Gamma(-4\eps)\sin(\pi\eps) }{\Gamma(1-2 \eps)\Gamma(-\eps)}\;
\\
&\quad
\times\,
\int_{0}^1 \!dt_{k_1} \,(4 t_{k_1} \bar t_{k_1})^{-1/2-\eps}
\int_{0}^1 \!dt'_{k_2} \,(t'_{k_2}(2-t'_{k_2}))^{-1-\eps} \;
\int_0^1 \!dy_k \;\, 
\nn\\[0.2em]
&\quad
\times\,
\Big\{y_k^{-1+2n\eps}\; \bar f_A(y_k,\theta_{k_1},\theta_{k_2})
+ y_k^{-1+(2n+1)\eps}\; \bar f_B(y_k,\theta_{k_1},\theta_{k_2})\Big\}\,,
\nonumber
\end{align}
where now
\begin{align}
\bar f_A(y_k,\theta_{k_1},\theta_{k_2})&= 
\Big[\mathcal{O}_1(\theta_{k_1},\theta_{k_2}) + \mathcal{O}_2(\theta_{k_1},\theta_{k_2})\Big]\;
\bigg( \frac{n_{il}}{n_{ij} \,y_k\, Z_l(y_k,\theta_{k_1},\theta_{k_2})} \bigg)^\eps\,
	f_A(y_k,\theta_{k_1},\theta_{k_2})^{4\eps}\,,
\nonumber\\
\bar f_B(y_k,\theta_{k_1},\theta_{k_2})&= 
\Big[\mathcal{O}_1(\theta_{k_1},\theta_{k_2}) + \mathcal{O}_2(\theta_{k_1},\theta_{k_2})\Big]\;
\bigg( \frac{n_{il}}{n_{ij} \,Z_l(1/y_k,\theta_{k_1},\theta_{k_2})} \bigg)^\eps\,
f_B(y_k,\theta_{k_1},\theta_{k_2})^{4\eps}\,,
\end{align}
with $\mathcal{O}_{1,2}(\theta_{k_1},\theta_{k_2})$ from \eqref{eq:O12}. In this expression the term in the round parenthesis reflects the last factor from \eqref{eq:tripoles:matrixelement}, which was written in a form in which the combinations $y_k\, Z_l(y_k,\theta_{k_1},\theta_{k_2})$ and $Z_l(1/y_k,\theta_{k_1},\theta_{k_2})$ are both finite and non-zero in the limit $y_k\to 0$. This asymmetry between the two regions A and B is also the reason that different powers of $y_k$ have been factored out in \eqref{eq:NNLO:RV:tripoles:master} to properly extract the associated collinear divergences.

\paragraph{Emission of quark-antiquark pair:} We next consider the correction from the emission of a soft quark-antiquark pair, which can be obtained from~\cite{Catani:1999ss}
\begin{equation}
{S}^{(2,q\bar{q})}(\eps) =  T_F n_f \,\sum_{i\neq j} \; \bfT_{i} \cdot \bfT_{j} \;
\Big( \frac{n_{ij}}{2} \Big)^{2n\eps} \, S_{ij}^{(2,q\bar{q})}(\eps)
\end{equation}
with 
\begin{align}
S_{ij}^{(2,q\bar{q})}(\eps)  = -\frac{(2^{2-n}\, n_{ij}^{n}\,\pi e^{\gamma_E} \tau^2)^{-2\eps}}{(2\pi)^{2d-2}} \!
\int \!d^{d}k \; \delta(k^{2}) \,\theta(k^{0}) \!
\int \!d^{d}l \; \delta(l^{2}) \,\theta(l^{0}) \,
\mathcal{M}_2(\tau; k,l) \, |\mathcal{A}^{q\bar{q}}_{ij}(k,l)|^{2} ,
\end{align}
and 
\begin{equation}
|\mathcal{A}^{q\bar{q}}_{ij}(k,l)|^{2} =  256 \pi^{4} \; 
\frac{n_{ij} \,(k_{i} + l_{i}) \, (k_{j} + l_{j}) \, (k\cdot l) - (k_{i}l_{j}-k_{j}l_{i})^{2}}{(k_{i} + l_{i})^{2}\,(k_{j} + l_{j})^{2}\,(k\cdot{l})^{2}} \,.
\label{eq:NNLO:RR:qq:matrixelement}
\end{equation}
In this case we express the Lorentz-invariant phase-space measures in terms of the variables $\{p_T,y,a,b\}$ from \eqref{eq:M2:variables}, along with the corresponding angular variables $\{\theta_{kl},\theta_{5},\theta_{6},\theta_{l_1},\theta_{l_2}\}$ that we introduced in~\ref{app:angles}. Working once more in the frame in which the reference vectors are back-to-back, the phase-space measures then translate into
\begin{align}
& \int d^{d}k' \; \delta(k^{\prime\,2}) \,\theta(k^{\prime\,0}) \,
\int d^{d}l' \; \delta(l^{\prime\,2}) \,\theta(l^{\prime\,0}) 
\nonumber\\
&\quad
=\frac{\Omega_{d-4}\Omega_{d-5}}{4n_{ij}^2} \,
\int_0^\infty \!dk_i^\prime \,\int_0^\infty \!dk_j^\prime \,
\int_0^\infty \!dl_i^\prime \,\int_0^\infty \!dl_j^\prime \;
\bigg(\frac{2k_i^\prime k_j^\prime}{n_{ij}}\bigg)^{-\eps} \,
\bigg(\frac{2l_i^\prime l_j^\prime}{n_{ij}}\bigg)^{-\eps} \,
\nonumber\\
&\quad\qquad\times\;
\int_{-1}^1 \!d c_{l_1} \, s_{l_1}^{d-5}  \int_{-1}^1 \!d c_{l_2} \, s_{l_2}^{d-6} 
\int_{-1}^1 \!d c_{kl} \, s_{kl}^{d-5}  \int_{-1}^1 \!d c_{5} \, s_{5}^{d-6}  \int_{-1}^1 \!d c_{6} \, s_{6}^{d-7}
\nonumber\\
&\quad
=\frac{(2\pi)^{-1-2\eps}}{4\,\Gamma(-1-2\eps)}\,
\int_0^\infty \!\!dp_T \; p_T^{3-4\eps}\,
\int_0^\infty \frac{dy}{y} \, 
\int_0^\infty \!\!da\, 
\int_0^\infty \!\!db \; 
(a b)^{1-2\eps} \,(a+b)^{-2+2\eps} \,(1+a b)^{-2+2\eps}
\nonumber\\
&\quad\qquad\times\;
\int_{-1}^1 \!d c_{l_1} \, s_{l_1}^{d-5}  \int_{-1}^1 \!d c_{l_2} \, s_{l_2}^{d-6} 
\int_{-1}^1 \!d c_{kl} \, s_{kl}^{d-5}  \int_{-1}^1 \!d c_{5} \, s_{5}^{d-6}  \int_{-1}^1 \!d c_{6} \, s_{6}^{d-7}\,.
\end{align}
While the integration over the dimensionful variable $p_T$ will again be performed analytical\-ly, we find it convenient to map the remaining integrations onto the unit hypercube. To do so, we exploit the fact that the measurement function cannot distinguish between the two partons, and it is therefore symmetric under the exchange of the momenta $k^\mu$ and $l^\mu$. As can readily be read off from \eqref{eq:NNLO:RR:qq:matrixelement}, the matrix element for the emission of a quark-antiquark pair also enjoys this symmetry, which can then be used to reduce the number of integrations on the unit hypercube. More specifically, our phase-space variables transform under $k^\mu\leftrightarrow l^\mu$ exchange as
\begin{equation}
p_T \rightarrow p_T\,, \qquad 
y \rightarrow y\,, \qquad 
a \rightarrow \frac{1}{a}\,, \qquad 
b \rightarrow \frac{1}{b}\,, \qquad 
\theta_{k_i} \rightarrow \theta_{l_i}\,, \qquad 
\theta_{l_i} \rightarrow \theta_{k_i}\,,
\end{equation}
which allows us, for instance, to map the integration domain with $a\geq1$ onto the one with $a\leq1$. The construction is, in fact, analogous to the one discussed in Sec.~3.3 of~\cite{Bell:2018oqa}, except that the constraints from the \mbox{$n^\mu\leftrightarrow\bar n^\mu$} symmetry do not apply here. In particular, the method exploits the fact that there exists a freedom in defining the angular variables, which was discussed in detail in~\cite{Bell:2018oqa}. Once this symmetry has been used to restrict the integration domain to $a\leq1$, we insert the explicit form \eqref{eq:M2} of the measurement function and integrate over the variable $p_T$. 
This yields the intermediate result
\begin{align}
\label{eq:RR:qqbar:intermediate}
& S_{ij}^{(2,q\bar{q})}(\eps)  = -\frac{4^{1-\eps}\,e^{-2\gamma_E\eps}}{\pi^3}\,
\frac{\Gamma(-4\eps)}{\Gamma(-1-2\eps)}\;
\int_0^1 \! dy \;  y^{-1+2n\eps}
\int_0^1 \! da \; a^{1-2\eps}
\int_0^1 \! db \; b^{-2\eps} 
\\
&\quad\times\;
\int_{-1}^1 \!d c_{l_1} \, s_{l_1}^{d-5}  
\int_{-1}^1 \!d c_{l_2} \, s_{l_2}^{d-6} 
\int_{-1}^1 \!d c_{kl} \, s_{kl}^{d-5}  
\int_{-1}^1 \!d c_{5} \, s_{5}^{d-6}  
\int_{-1}^1 \!d c_{6} \, s_{6}^{d-7}
\nonumber\\
&\quad\times\;
(a+b)^{-2+2\eps} \, (1+ab)^{-2+2\eps}\;
\bigg\{ \frac{(a+b)(1+ab)}{1+a^2-2a c_{kl}}
- \frac{b (1-a^2)^2}{[1+a^2-2a c_{kl}]^2} \bigg\}
\nonumber\\
&\quad\times\;
\Big\{F_A\big(a,b,y,\{\theta\}\big)^{4\eps}
+ F_B\big(a,b,y,\{\theta\}\big)^{4\eps}
+ F_C\big(a,b,y,\{\theta\}\big)^{4\eps}
+ F_D\big(a,b,y,\{\theta\}\big)^{4\eps}\Big\}\,,
\nonumber
\end{align}
which contains three types of physical divergences that are encoded in the pre\-factor $\Gamma(-4\epsilon)$, the limit $y\to 0$, as well as an overlapping divergence that arises in the limit $\{a\to 1, c_{kl}\to 1\}$, which is associated with the configuration in which the two emitted partons become collinear to each other. The angular variables, moreover, now introduce various spurious divergences that need to be isolated properly. To this end, we follow the construction described in the NLO calculation for the variables $c_{l_2}$ and $c_5$ -- which introduces two branches for each of the associated variables $t'_{l_2}$ and $t'_5$ -- while we substitute the remaining ones via $c_{l_1}=1-2t_{l_1}$, $c_{kl}=1-2t_{kl}$ and $c_6=1-2t_6$. 

The integration over the angle $\theta_6$ deserves some explanation, since it does not produce a $1/\eps$ pole in $d=4-2\eps$ dimensions, but it is never\-theless non-integrable in four dimensions. It therefore also requires a workaround for generic N-jet observables, but for the specific N-jettiness soft function it turns out that the integrals depend only trivially on this angle up to the considered $\mathcal{O}(\eps^0)$, and we can therefore perform this integration analytically. The only place where the angle $\theta_6$ actually enters the calculation is via the scalar product $k_\perp \cdot n_X$ in \eqref{eq:kperpdotnX} that appears in the measurement function. This dependence, however, only shows up at $\mathcal{O}(\eps)$  in the last line of \eqref{eq:RR:qqbar:intermediate}, which implies that it needs to appear together with at least one singularity to produce a $\mathcal{O}(\eps^0)$ contribution to the soft function. In other words, as singularities are always accompanied by delta functions, up to the considered $\mathcal{O}(\eps^0)$ term the NNLO soft function does not see the full measurement function, but only certain realisations of it in the singular limits. In \eqref{eq:RR:qqbar:intermediate} these limits are the physical ones with $y\to 0$ and $\{a\to 1, c_{kl}\to 1\}$, as well as the spurious ones with $c_{l_2}\to\pm 1$ and $c_{5}\to\pm 1$. It is, however, easy to see that the angle $\theta_6$ drops out in all of the latter angular limits, since its occurrence is multiplied in \eqref{eq:lambda34} by a factor $s_{kl} \, s_{l_2}\, s_5$. While this is true for generic N-jet observables, the independence in the limit $y\to 0$ is specific to the N-jettiness variable, since the minimisation over all $k_X$ projections always picks one of the dipole directions in this limit, which from the outset are independent of $\theta_6$. We therefore do not have to include a workaround for the $\theta_6$ integration here, but from the discussion it should be clear that if one is interested in computing $\mathcal{O}(\eps)$ corrections to the N-jettiness soft function -- or $\mathcal{O}(\eps^0)$ terms for other N-jet observables -- this treatment needs to be revised.

After mapping the angular integrations onto the unit hypercube, we arrive at the following master formula for the calculation of the quark-antiquark real-emission contribution
\begin{align}
\label{eq:RR:qqbar:master}
& S_{ij}^{(2,q\bar{q})}(\eps)  = \frac{4^{-1-\eps}\,e^{-2\gamma_E\eps}}{\pi^3}\,
\frac{\Gamma(-4\eps)}{\Gamma(-1-2\eps)}
\int_0^1 \!\! dy \;  y^{-1+2n\eps}
\int_0^1 \!\! da \; a^{-2\eps} \!
\int_0^1 \!\! db \; b^{-2\eps} 
\,(a+b)^{2\eps} \, (1+ab)^{2\eps}
\nonumber\\
&\quad\times\;
\int_{0}^1 \!dt_{l_1} \,(4 t_{l_1} \bar t_{l_1})^{-1/2-\eps} 
\int_{0}^1 \!dt'_{l_2} \,(t'_{l_2}(2-t'_{l_2}))^{-1-\eps} \;
\int_{0}^1 \!dt_{kl} \,(4 t_{kl} \bar t_{kl})^{-1/2-\eps} 
\nonumber\\
&\quad\times\;
\int_{0}^1 \!dt'_{5} \,(t'_{5}(2-t'_{5}))^{-1-\eps} \;
\int_{0}^1 \!dt_{6} \,(4 t_{6} \bar t_{6})^{-3/2-\eps} \;\,
k^{(n_f)}(a,b,t_{kl}) \; \tilde F\big(a,b,y,\{\theta\}\big)\,,
\end{align}
where
\begin{align}
&\tilde F\big(a,b,y,\{\theta\}\big)= 
\Big[\mathcal{O}_3(\{\theta\}) + \mathcal{O}_4(\{\theta\}) +
\mathcal{O}_5(\{\theta\}) + \mathcal{O}_6(\{\theta\})\Big]
\\
&\quad\times\;
\Big\{F_A\big(a,b,y,\{\theta\}\big)^{4\eps}
+ F_B\big(a,b,y,\{\theta\}\big)^{4\eps}
+ F_C\big(a,b,y,\{\theta\}\big)^{4\eps}
+ F_D\big(a,b,y,\{\theta\}\big)^{4\eps}\Big\}\,,
\nonumber
\end{align}
and $\mathcal{O}_{3-6}(\{\theta\})$ are NNLO analogues to the NLO replacement operators that act on a test function $h(\{\theta\})$ as
\begin{align}
\mathcal{O}_3(\{\theta\})\; h(\{\theta\})&=
h(\{\theta\})\,\Bigg| \scriptsize
\begin{array}{l} 
c_{l_1} \to 1-2 t_{l_1}\\[-0.2em]
c_{l_2} \to 1-t'_{l_2}\\[-0.2em]
c_{kl} \to 1-2 t_{kl}\\[-0.2em]
c_{5} \to 1-t'_{5}\\[-0.2em]
c_{6} \to 1- 2t_{6}\end{array}\;,
&
\mathcal{O}_4(\{\theta\})\; h(\{\theta\})&=
h(\{\theta\})\,\Bigg| \scriptsize
\begin{array}{l} 
c_{l_1} \to 1-2 t_{l_1}\\[-0.2em]
c_{l_2} \to t'_{l_2}-1\\[-0.2em]
c_{kl} \to 1-2 t_{kl}\\[-0.2em]
c_{5} \to 1-t'_{5}\\[-0.2em]
c_{6} \to 1- 2t_{6}\end{array}\;,
\nonumber\\[0em]
\mathcal{O}_5(\{\theta\})\; h(\{\theta\})&=
h(\{\theta\})\,\Bigg| \scriptsize
\begin{array}{l} 
c_{l_1} \to 1-2 t_{l_1}\\[-0.2em]
c_{l_2} \to 1-t'_{l_2}\\[-0.2em]
c_{kl} \to 1-2 t_{kl}\\[-0.2em]
c_{5} \to t'_{5}-1\\[-0.2em]
c_{6} \to 1- 2t_{6}\end{array}\;,
&
\mathcal{O}_6(\{\theta\})\; h(\{\theta\})&=
h(\{\theta\})\,\Bigg| \scriptsize
\begin{array}{l} 
c_{l_1} \to 1-2 t_{l_1}\\[-0.2em]
c_{l_2} \to t'_{l_2}-1\\[-0.2em]
c_{kl} \to 1-2 t_{kl}\\[-0.2em]
c_{5} \to t'_{5}-1\\[-0.2em]
c_{6} \to 1- 2t_{6}\end{array}\;.
\label{eq:O3456}
\end{align}
For latter convenience, we also introduced the integration kernel
\begin{align}
k^{(n_f)}(a,b,t_{kl}) &=
\frac{128a}{(a+b)^2(1+a b)^2} \;\bigg\{
\frac{b (1-a^2)^2}{[(1-a)^2+4a t_{kl}]^2}
- \frac{(a+b)(1+ab)}{(1-a)^2+4a t_{kl}} \bigg\}\,,
\end{align}
which contains the overlapping collinear divergence in the limit $\{a\to 1, t_{kl}\to 0\}$. Following~\cite{Bell:2018oqa}, we resolve this singularity by an additional substitution, namely $a=1-u(1-v)$ and $t_{kl}=u^2 v/(1-u(1-v))$, which maps the overlapping divergence onto $u\to0$. Our result in \eqref{eq:RR:qqbar:master} can again be compared to the corresponding expression for back-to-back dipoles in~\cite{Bell:2018oqa}.

\paragraph{Emission of two gluons:} The calculation of the last piece in \eqref{eq:NNLOcontributions} proceeds similarly. Here the starting point is~\cite{Catani:1999ss}
\begin{align}
{S}^{(2,gg)}(\eps) &=  C_A \,\sum_{i\neq j} \; \bfT_{i} \cdot \bfT_{j} \;
\Big( \frac{n_{ij}}{2} \Big)^{2n\eps} \, S_{ij}^{(2,gg)}(\eps)
\nonumber\\
& \quad
+ \frac{1}{4} \, \sum_{i \neq j} \,\sum_{k \neq l} \;
\big\{\bfT_{i} \cdot \bfT_{j}\,,\,\bfT_{k} \cdot \bfT_{l}\big\} 
\Big( \frac{n_{ij}}{2} \Big)^{n \eps}
\Big( \frac{n_{kl}}{2} \Big)^{n \eps}
S_{ij}^{(1)}(\eps)  \,S_{kl}^{(1)}(\eps)\,,
\end{align}
where the last term involves three-particle and four-particle correlations, which are however trivially related to the NLO dipoles $S_{ij}^{(1)}(\eps)$ as long as the observable respects NAE. The only non-trivial correction for this class of observables is therefore again of dipole type, which can be calculated from
\begin{equation}
S_{ij}^{(2,gg)}(\eps)  = -\frac{(2^{2-n}\, n_{ij}^{n}\,\pi e^{\gamma_E} \tau^2)^{-2\eps}}{(2\pi)^{2d-2}} \!
\int \!d^{d}k \; \delta(k^{2}) \,\theta(k^{0}) \!
\int \!d^{d}l \; \delta(l^{2}) \,\theta(l^{0}) \,
\mathcal{M}_2(\tau; k,l)  |\mathcal{A}^{gg}_{ij}(k,l)|^{2} ,
\end{equation}
with~\cite{Catani:1999ss}
\begin{align}  
\big|\mathcal{A}^{gg}_{ij}(k,l)\big|^2 & =
 64 \pi^{4} \,\bigg\{ 
 \frac{2(1-\eps)\,(k_{i}l_{j}-k_{j}l_{i})^{2}}{(k_{i} + l_{i})^{2}\,(k_{j} + l_{j})^{2}\,(k\cdot{l})^{2}}
- \frac{4n_{ij}}{ (k_{i}+ l_{i})\,(k_{j}+ l_{j})\, (k\cdot l) } 
\nonumber \\
&\qquad\quad\quad 
- \frac{n_{ij}^2}{k_i\, k_j\, l_i\,l_j} \bigg(2-\frac{ k_i l_j + k_j l_i}{(k_i + l_i)\, (k_j +l_j)} \bigg)\bigg( 1 - \frac{k_i l_j + k_j l_i}{n_{ij}\,(k\cdot l)} \bigg) \bigg\}\,.
\label{eq:NNLO:RR:gg:matrixelement}
\end{align}
As the matrix element is again symmetric under $k^\mu\leftrightarrow l^\mu$ exchange, the calculation then follows along the same lines as for the quark-antiquark correction, and it leads to a similar master formula for the two-gluon real-emission contribution $S _{ij}^{(2,gg)}(\eps)$, which takes the form \eqref{eq:RR:qqbar:master}, except that the integration kernel $k^{(n_f)}(a,b,t_{kl})$ is replaced by
\begin{align}
k^{(C_A)}(a,b,t_{kl}) &=
-\frac{32}{ab(a+b)^2(1+a b)^2} \;\bigg\{
\frac{2(1-\eps)a^2b^2(1-a^2)^2}{[(1-a)^2+4a t_{kl}]^2}
- (a+b)(1+ab)
\nonumber\\[0.2em]
&\qquad\times
\bigg[ b(1+a^2)+2a(1+b^2) -\frac{b(1-a^2)^2+2a(1+a^2)(1+b^2)}{(1-a)^2+4a t_{kl}}
\bigg] \bigg\}\,.
\end{align}
In comparison to the quark-antiquark contribution, the two-gluon real-emission correction contains an additional soft singularity that arises in the limit $b\to 0$. This divergence, in particular, challenges our observation from above that the N-jettiness soft function depends only trivially on the angle $\theta_6$ up to the considered $\mathcal{O}(\eps^0)$. From \eqref{eq:2loopF} it is, however, evident that the relevant contribution vanishes in this limit, which is a direct consequence of infrared safety~\cite{Bell:2018oqa}, and therefore our argument from the previous section also carries over to the two-gluon contribution.   

\subsection{Renormalisation}
\label{sec:renormalisation}

Having developed an automated framework for the calculation of bare N-jet soft functions that are defined in SCET-1 and comply with NAE, we now turn to their renormalisation. To this end, we assume that the considered class of soft functions renormalises multiplicatively in Laplace space, which is in particular true for the N-jettiness observable under consideration. In Laplace space the relation between the bare and the renormalised soft function can then be written as
\begin{equation}
\label{eq:Sbare}
S_0(\tau) =  Z_S(\tau, \mu) \; S(\tau, \mu)\; Z_S^{\dagger}(\tau, \mu)\,,
\end{equation}
where all objects are matrices in colour space. The $\overline{\text{MS}}$ renormalised soft function satisfies the renormalisation group (RG) equation
\begin{equation}
\frac{\rd}{\rd \ln\mu}  \; S(\tau,\mu)
= \frac{1}{2} \,\Gamma_S(\tau,\mu) \, S(\tau,\mu) 
+\frac{1}{2} \,S(\tau,\mu)\,  \Gamma_S(\tau,\mu)^{\dagger}
\end{equation}
with anomalous dimension\footnote{We note that the sign of the imaginary part in \eqref{eq:AD} differs from the one in~\cite{Jouttenus:2011wh, Bell:2018mkk},  which can be traced to a typo in Eq.~(21) of~\cite{Jouttenus:2011wh}, which erroneously associates the quoted expression to the anomalous dimension $\hat\gamma_C^\dagger(\mu)$ rather than $\hat\gamma_C(\mu)$ in the notation of that paper. The correct form of the soft anomalous dimension has later been given in \cite{Gaunt:2015pea}.}
\begin{equation}
\label{eq:AD}
\Gamma_S(\tau,\mu)= \frac{1}{n}\, \bigg\{
\sum_{i\neq j} \, \bfT_{i} \cdot \bfT_{j} \; \Gamma_{\rm cusp}(\alpha_s)
\bigg[2 L_{ij} + i  n \pi\, \lambda_{ij}\bigg] + 2 \gamma^S(\alpha_s)\bigg\}\,,
\end{equation}
where $L_{ij}=\ln (\mu \taubar) + \frac{n}{2}\ln (\frac{n_{ij}}{2})$ and the factors $\lambda_{ij}$ have been defined following \eqref{eq:NNLO:RV}. The imaginary part is related to the anomalous dimension of the associated hard function in the factorisation theorem, whose general structure was discussed in~\cite{Becher:2009qa}. The cusp anomalous dimension has a perturbative expansion $\Gamma_{\rm cusp} (\alpha_s)= \sum_{n=0}^{\infty} \Gamma_n \,( \frac{\alpha_s}{4 \pi})^{n+1}$, with leading coefficients $ \Gamma_0 =4$ and $\Gamma_1 = ( 268/9 - 4\pi^2/3) C_A - 80/9\, T_F n_f$. Up to the considered two-loop order, the non-cusp anomalous dimension also has a dipole structure, which we write in the form \mbox{$\gamma^{S}(\alpha_s) = \sum_{i\neq j} \, \bfT_{i} \cdot \bfT_{j} \, \big\{ (\frac{\alpha_s}{4\pi}) \, \gSzero + (\frac{\alpha_s}{4\pi})^2 \, \gSone \big\}$}, and where we assumed that its coefficients $\gSzero$ and $\gSone$ are numbers rather than matrices with indices $(ij)$. Following~\cite{Bell:2018vaa,Bell:2018oqa,Bell:2020yzz}, we find it furthermore convenient to define the anomalous dimensions with a common prefactor $1/n$, where $n$ is related to the asymptotic behaviour of the observable in the soft-collinear limit, see \eqref{eq:M1} and \eqref{eq:M2}, and which for the N-jettiness is simply $n=1$.

In this notation, the two-loop solution of the RG equation takes the form
\begin{align}
\label{eq:RGE:softsolution}
& S(\tau,\mu) = 
1 + \left( \frac{\alpha_s}{4 \pi} \right) 
\sum_{i\neq j} \, \bfT_{i} \cdot \bfT_{j} \,
\bigg( \frac{\Gamma_0}{n} \,L_{ij}^2 
+ \frac{2\gSzero}{n} \,L_{ij} 
+ c_{ij}^{(1)} \bigg) 
+\left( \frac{\alpha_s}{4 \pi} \right)^2
\\
&
\times \bigg\{
\sum_{i\neq j} \, \bfT_{i} \cdot \bfT_{j} \,
\bigg(\frac{2\beta_0\Gamma_0}{3n} \,L_{ij}^3 
+ \bigg( \frac{\Gamma_1}{n} + \frac{2\beta_0 \gSzero}{n}  \bigg) L_{ij}^2 
+ 2 \bigg( \frac{\gSone}{n} +\beta_0 c_{ij}^{(1)} \bigg) L_{ij} 
+ c_{ij}^{(2)} \bigg)
\nn\\
&\hspace{2mm}
 + 2\pi \sum_{i\neq j\neq k} f_{ABC}\; \bfT_{i}^A \;\bfT_{j}^B \;\bfT_{k}^C\;
\bigg( \Gamma_0 \lambda_{ij} \bigg(
\frac{\Gamma_0}{3n} L_{jk}^3 +  \frac{\gSzero}{n} L_{jk}^2 + c_{jk}^{(1)} L_{jk} \bigg)
+ c_{ijk}^{(2)} \bigg)
\nn\\
&\hspace{2mm}
+ \frac{1}{4} \, \sum_{i \neq j} \sum_{k \neq l}
\big\{\bfT_{i} \cdot \bfT_{j}\,,\,\bfT_{k} \cdot \bfT_{l}\big\} 
\bigg( \frac{\Gamma_0}{n} \,L_{ij}^2 + \frac{2\gSzero}{n} \,L_{ij} + c_{ij}^{(1)} \bigg) 
\bigg( \frac{\Gamma_0}{n} \,L_{kl}^2 + \frac{2\gSzero}{n} \,L_{kl} + c_{kl}^{(1)} \bigg)
\bigg\}, \nn
\end{align}
where the tripole contribution in the third line arises from commutators of colour generators, 
e.g.~$\big[ \bfT_{i} \cdot \bfT_{j} \,,\, \bfT_{j} \cdot \bfT_{k} \big] 
= - i f_{ABC} \; \bfT_{i}^A \;\bfT_{j}^B \;\bfT_{k}^C$. For later purposes, it will also be useful to rewrite the tripole contributions for fixed values of $(ijk)$ in the form 
\begin{align}
\label{eq:c2ijknotilde}
&
\Gamma_0 \lambda_{ij} \bigg(
\frac{\Gamma_0}{3n} L_{jk}^3 +  \frac{\gSzero}{n} L_{jk}^2 + c_{jk}^{(1)} L_{jk} \bigg)
+ c_{ijk}^{(2)} 
\nn\\
&\quad
=
 \Gamma_0 \lambda_{ij} \bigg(
\frac{\Gamma_0}{3n} L_\mu^3 +  \Big(\frac{\Gamma_0}{n} \tilde L_{jk} + \frac{\gSzero}{n}\Big) L_\mu^2 
 + \Big( \frac{\Gamma_0}{n} \tilde L_{jk}^2 + \frac{2\gSzero}{n} \tilde L_{jk} 
+  c_{jk}^{(1)} \Big) L_\mu \bigg)
+ \tilde c_{ijk}^{(2)} \,,
\end{align}
where $L_\mu=\ln (\mu \taubar)$ and $\tilde L_{jk}=\frac{n}{2}\ln (\frac{n_{jk}}{2})$. This amounts to shifting kinematic logarithms into the $\mu$-independent coefficients $\tilde c_{ijk}^{(2)} $ such that
\begin{align}
\label{eq:c2ijktilde}
\tilde c_{ijk}^{(2)} 
= c_{ijk}^{(2)}
+  \Gamma_0\lambda_{ij}\bigg(\frac{\Gamma_0}{3n}\tilde{L}_{jk}^3+\frac{\gSzero}{n} \tilde{L}_{jk}^2+ c_{jk}^{(1)} \tilde{L}_{jk} \bigg). 
\end{align}
We remark that there is an important difference between the dipole and tripole contributions in the sense that the dipoles can be renormalised individually for fixed values of $(ij)$, while this is not so for the tripole contribution, which can only be renormalised in the sum over all tripoles, as we will explain below. This then implies that the coefficients $c_{ijk}^{(2)}$ in the third line of \eqref{eq:RGE:softsolution} -- or the alternative coefficients $\tilde c_{ijk}^{(2)}$ in \eqref{eq:c2ijktilde} -- are not individually meaningful, since certain contributions could be added to them that vanish in the sum, without changing the result of the renormalised soft function. 

The counterterm fulfils a simpler RG equation
\begin{equation}
\frac{\rd}{\rd \ln\mu}  \; Z_S(\tau,\mu)
= - \frac{1}{2}  \, Z_S(\tau,\mu) \,\Gamma_S(\tau,\mu)\,,
\end{equation}
and its explicit solution to two-loop order is given by
\begin{align}
\label{eq:counterm}
Z_{S}(\tau,\mu) &= 
1 + \left( \frac{\alpha_s}{4 \pi} \right) 
\sum_{i\neq j} \, \bfT_{i} \cdot \bfT_{j} \,
\bigg( \frac{\Gamma_0}{4n\eps^2}
+ \frac{G^{(0)}_{ij}}{4n\eps} \bigg) 
+\left( \frac{\alpha_s}{4 \pi} \right)^2 
\\
&\quad
\times \bigg\{
\sum_{i\neq j} \, \bfT_{i} \cdot \bfT_{j} \,
\bigg( -\frac{3\beta_0\Gamma_0}{16n\eps^3}
+\frac{\Gamma_1-2\beta_0 G^{(0)}_{ij}}{16n\eps^2}
+ \frac{G^{(1)}_{ij}}{8n\eps} \bigg) 
\nn\\
&\hspace{11mm}
+ \frac{1}{4} \, \sum_{i \neq j} \sum_{k \neq l}
\big\{\bfT_{i} \cdot \bfT_{j}\,,\,\bfT_{k} \cdot \bfT_{l}\big\} 
\bigg( \frac{\Gamma_0}{4n\eps^2}
+ \frac{G^{(0)}_{ij}}{4n\eps} \bigg) 
\bigg( \frac{\Gamma_0}{4n\eps^2}
+ \frac{G^{(0)}_{kl}}{4n\eps} \bigg) 
\bigg\}, \nn
\end{align}
where $G^{(k)}_{ij} = (2  L_{ij}  + i n\pi \, \lambda_{ij})\Gamma_{k} +2\gSk$. Note that there is no tripole contribution in the counterterm, since the commutators $\big[\bfT_{i} \cdot \bfT_{j}\,,\,\bfT_{k} \cdot \bfT_{l}\big]$ multiply a term that vanishes in the sum over all tripoles in this case.

Let us now come back to the peculiarity of the tripole contribution to the renormalised soft function mentioned above. While we have just seen that there exists no tripole term in the counterterm \eqref{eq:counterm} itself, the matrix nature of the relation~\eqref{eq:Sbare} generates an $\mathcal{O}(\alpha_s^2)$ tripole contribution to the renormalised soft function from the interference of the NLO bare soft function and the NLO counterterm via commutators of colour generators. As such a colour structure is anti-hermitean, it can only contribute in the presence of another source of imaginary units -- which the $\text{i} \pi \lambda_{ij}$ term in the anomalous dimension \eqref{eq:AD} (and hence the $1/\epsilon$ coefficient of the bare soft function) provides. In the transition from the bare to the renormalised soft function, this contribution then removes the poles from the bare tripole sum, and it adds, of course, a corresponding finite term. The tripole contribution thus involves two distinct sums: The bare tripoles $S_{ilj}^{(2,{\rm Im})}(\eps)$ in~\eqref{eq:NNLO:RV} are weighed by a factor \mbox{$(\lambda_{il}-\lambda_{ip}-\lambda_{lp})$} originating from the underlying one-loop matrix element, whereas the NLO$\times$NLO cross terms contain a $\lambda_{ij}$ factor originating from the anomalous dimension. As a result, the individual bare tripoles cannot be combined with a corresponding counterterm to remove the respective divergences --  only the sum over all tripoles can be renormalised and is thus free of divergences. This is different from the dipole contributions, which can be renormalised individually for fixed values of $(ij)$.

Using all these expressions, the divergences of the bare soft function can be reconstructed through NNLO via \eqref{eq:Sbare}. Explicitly, we find
\begin{align}
\label{eq:RGE:baresoft}
& S_0(\tau) = 
1 + \Bigl( \frac{Z_\alpha \alpha_s}{4 \pi} \Bigr) 
\sum_{i\neq j} \, \bfT_{i} \cdot \bfT_{j} \,
\biggl( \frac{\Gamma_0}{2n\eps^2}
+ \frac{ \Gamma_{0} L_{ij} + \gSzero}{n\eps} \biggr) 
+\Bigl( \frac{Z_\alpha \alpha_s}{4 \pi} \Bigr)^2 \;
\biggl\{
\sum_{i\neq j} \, \bfT_{i} \cdot \bfT_{j}
\\
&
\times  \,
\biggl( \frac{\beta_0\Gamma_0}{8n\eps^3}
+\frac{\Gamma_1+4\beta_0( \Gamma_{0} L_{ij} + \gSzero)}{8n\eps^2}
+ \frac{ \Gamma_{1} L_{ij} + \gSone +2 \beta_0 (\Gamma_0 L_{ij}^2 + 2 \gamma_0^S L_{ij} + n c_{ij}^{(1)})}{2n\eps}\biggr) 
\nn\\
&
+\sum_{i\neq j\neq k} f_{ABC}\; \bfT_{i}^A \;\bfT_{j}^B \;\bfT_{k}^C\;\, \pi \Gamma_0\lambda_{ij}
\biggl( \frac{ \Gamma_0 L_{jk}}{2n\eps^2}
+ \frac{\Gamma_0 L_{jk}^2 +2\gSzero L_{jk} +n c_{jk}^{(1)}}{n\eps}   \biggr)
+ ~\ldots~
\biggr\} + \mathcal{O}(\eps^0), 
\nn
\end{align}
where the dots refer to the four-particle correlations encoded in 
$\big\{\bfT_{i} \cdot \bfT_{j}\,,\,\bfT_{k} \cdot \bfT_{l}\big\}$ 
that are trivial for NAE-violating observables. For the tripole contribution, on the other hand, we have once more used the property that certain terms -- like the leading $1/\eps^3$ pole --  vanish in the sum over all tripoles, which explains the unusual feature that the dominant pole in the tripole term is multiplied by a kinematic logarithm $L_{jk}$. The very fact that the explicit calculation of the bare soft function that is based on the master formulae from the previous sections matches the divergence structure predicted by the RG equation then provides a strong check of our calculation.

For the N-jettiness soft function, in particular, the soft anomalous dimension is known to the considered two-loop order with~\cite{Stewart:2010qs}
\begin{align}
\gSzero &= 0\,,
\nonumber\\
\gSone &=C_A \,\Big( \frac{404}{27} - \frac{11 \pi^2}{18} - 14 \zeta_3\Big) 
+ T_F n_f \,\Big(-\frac{112}{27} + \frac{2 \pi^2}{9} \Big)\,.
\end{align}
The one-loop matching corrections $c_{ij}^{(1)}$ defined in \eqref{eq:RGE:softsolution} are also known from the analysis in~\cite{Jouttenus:2011wh}, and the quantities of interest are therefore the NNLO dipole corrections $c_{ij}^{(2)}$ and the respective tripole contributions encoded in the coefficients $\tilde{c}_{ijk}^{(2)}$. We recall that the latter only contribute to processes with at least four hard partons, and they are therefore irrelevant for the 1-jettiness soft function.

\section{Results}
\label{sec:results}

\subsection{Numerical implementation}

In the master formulae from the previous section all divergences are factorised into monomials, which can easily be expanded in terms of standard plus distributions, e.g.
\begin{equation}
y_k^{-1+n \epsilon}= \frac{\delta(y_k)}{n \epsilon} + \left[\frac{1}{y_k}\right]_+
+ n \epsilon \left[\frac{\ln y_k}{y_k}\right]_+ + \,\ldots
\label{eq:monomial}
\end{equation}
We are thus in the position to perform the expansion in the dimensional regulator $\eps$ and to compute the respective coefficients in this expansion numerically. For the numerical evaluation we have followed two independent approaches that we briefly describe in this section.

The first approach is based on the public program {\tt pySecDec}~\cite{Borowka:2017idc}, and it uses the {\tt Vegas} routine of the {\tt Cuba} library~\cite{Hahn:2004fe,Hahn:2014fua} for the numerical integrations. While this implementation is rather straightforward, we used a few additional substitutions to remove integrable divergences with the objective of improving the numerical performance. This is similar to the strategy that was applied for dijet soft functions in~\cite{Bell:2018oqa} (for details cf.~Sec.~6 therein). We also note that we used the {\tt pySecDec} approach only for cross checks, and that all numbers published in this paper and the ancillary electronic file were obtained with the second approach, which consists of an extension of our dedicated {\tt C++} program {\tt SoftSERVE}~\cite{Bell:2018vaa,Bell:2018oqa,Bell:2020yzz}. This setup allows us to perform a number of optimisations that are not possible with a multi-purpose program like {\tt pySecDec}. This includes, for instance, simplifications on the integrand level or a proper reduction of the dimensionality of the numerical integrations that is induced by the delta functions related to the implicit phase-space divergences, as shown in \eqref{eq:monomial}. Grouping various integrals of the same dimensionality but with different integration variables can significantly speed up the calculation.

The main difference between the N-jet formalism described in this paper and the one for back-to-back dipoles discussed in~\cite{Bell:2018vaa,Bell:2018oqa,Bell:2020yzz} consists in the complexity of the calculation. For observables like thrust, which is similar to 0-jettiness,  only a single number for a single dipole needs to be computed, which should be contrasted with the 2-jettiness calculation we perform here: Besides there being six independent dipole and 24 tripole contributions, a multi-dimensional space associated with the configuration of the hard emitters --- encoded in the various factors $n_{ij}$ --- must be sampled, which actually presents the main challenge of the calculation. For the 2-jettiness, specifically, we produced multi-dimensional grids consisting of about $30,\!000$ points, which clearly shows that the computation requires some level of automation and parallelisation. Our in-house extension of {\tt SoftSERVE}, which is based on the public version {\tt SoftSERVE~1.0} (available at \href{https://softserve.hepforge.org}{https://softserve.hepforge.org}), therefore also contains a small machinery of scripts to control and combine the various contributions of the calculation.

In this {\tt SoftSERVE} approach the external geometry of the soft function is implemented using the {\tt C++17~constexpr} and variadic function definition capabilities to calculate all Minkow\-ski scalar products at compile time and to have them available as constants at run time. While the results presented in this paper mainly concern the 2-jettiness, the formalism and its numerical implementation are valid for any number of jets N. This will be illustrated below by computing a few benchmark points for the 3-jettiness soft function with exactly the same setup. The main difficulty for N~$>2$ is, of course, again the complexity of the calculation, and in particular the even larger phase space of the hard emitters that needs to be sampled.

Another novelty of the N-jet formalism concerns the angular integrations in the transverse space, which -- as described in Sec.~\ref{sec:parametrisations} -- are parametrised by two (five) angles for one (two) emission(s), as compared to one (three) angle(s) in the setup from~\cite{Bell:2018vaa,Bell:2018oqa,Bell:2020yzz}. While the higher dimensionality of the numerical integrations is obviously more demanding, the integration over the angle $\theta_6$ is particularly tricky, since it involves a factor $(4 t_{6} \bar t_{6})^{-3/2-\eps}$, see \eqref{eq:RR:qqbar:master}. This implies that the $t_6$ integration does not produce a $1/\eps$ pole, but it is nevertheless problematic from a numerical point of view and requires further subtractions. As we saw after \eqref{eq:RR:qqbar:intermediate}, however, the dependence of the N-jettiness soft function on this particular angle is trivial up to the considered $\mathcal{O}(\eps^0)$, and we may therefore integrate this angle out analytically. As a result, our in-house {\tt SoftSERVE} extension contains only one additional integration compared to the public version {\tt SoftSERVE~1.0}. We also remind the reader that we have to compute twice as many ingredients in the N-jet case compared to the setup from~\cite{Bell:2018vaa,Bell:2018oqa,Bell:2020yzz} because of the missing \mbox{$n^\mu\leftrightarrow\bar n^\mu$} symmetry.

The actual computations were performed on a cluster of around two hundred cores, with individual binaries allocated typically between 2 and 4 cores, and running on average ca. 30~min per dipole. A small fraction of integrations (around one per cent of all binaries) failed to converge within a span of several hours, these were recalculated using different integrator settings. A subset of a few dozen of these recalculated points we checked for consistency with the remainder of the grid by comparing against neighbouring points, as well as by comparing to a third run with different integrator settings. The whole procedure of running the integrations for the entire grid and recalculating/checking failed runs was completed over the span of several real-time weeks. The bottleneck in virtually all phase-space calculations proved to be the $C_A$ colour structure, taking up over 80 per cent of the computing time for a given phase-space point. With the settings chosen we then typically achieve sub-percent precision and accuracy.

To summarise, our N-jet {\tt SoftSERVE} extension builds on the publicly available version {\tt SoftSERVE~1.0}, but it contains higher-dimensional angular integrations due to the intricate construction of the transverse space for non-back-to-back dipoles. It furthermore includes some features to control and parallelise the various parts of the calculation. From the numerical perspective, on the other hand, it uses exactly the same integration settings as {\tt SoftSERVE~1.0}, i.e.~the {\tt Divonne} routine of the {\tt Cuba} library with partitioning based on Cuhre cubature rules, main sampling using Korobov random numbers and refinement via subdivision.

All of the results presented in this paper have been cross-checked in several respects. First of all, we saw in Sec.~\ref{sec:renormalisation} that the divergences of the bare N-jettiness soft function are fixed by RG constraints, and we explicitly verified that these conditions are fulfilled within the numerical uncertainties of our calculation. For the finite $\mathcal{O}(\eps^0)$ terms, we used two independent codes that rely on two different numerical integrators as discussed above. Moreover, we compared our 1-jettiness results against existing calculations in the literature, which also provides an indirect check for our 2-jettiness numbers, since we calculated the 1-jettiness soft function as a special case of 2-jettiness as will be explained in the following section.

\subsection{1-jettiness}

We first consider the 1-jettiness soft function that was computed before to NLO in~\cite{Jouttenus:2011wh} and to NNLO in~\cite{Boughezal:2015eha,Campbell:2017hsw}. In this case the external geometry is confined to a plane with two back-to-back beams and one jet direction that we parametrise by the scattering angle $\theta_{13}$. The kinematic invariants thus become $n_{12}=2$, $n_{13}=1-\cos \theta_{13}$ and $n_{23}=1+\cos \theta_{13}=2-n_{13}$. Our results for the 1-jettiness soft function will be displayed as a function of $n_{13}$, and they are based on 24 points in the range $\theta_{13}=\{\pi/25,\ldots, 24\pi/25\}$.

For the 1-jettiness soft function there are three dipole contributions, whereas the tripole sum vanishes because of colour conservation as argued above. We, in fact, already showed the divergences of the bare soft function for the (13)-dipole in~\cite{Bell:2018mkk}, and the divergences of the remaining dipoles can be reconstructed from the ancillary electronic file. In each case we find that our numbers agree with the RG prediction~\eqref{eq:RGE:baresoft} within the numerical uncertainties.

\begin{figure}[t!]
	\centering{
		\includegraphics[width=0.325\textwidth]{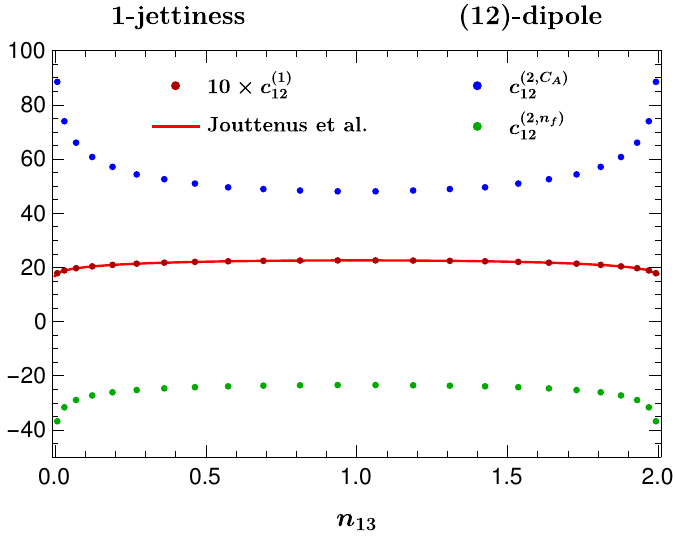}
		\includegraphics[width=0.325\textwidth]{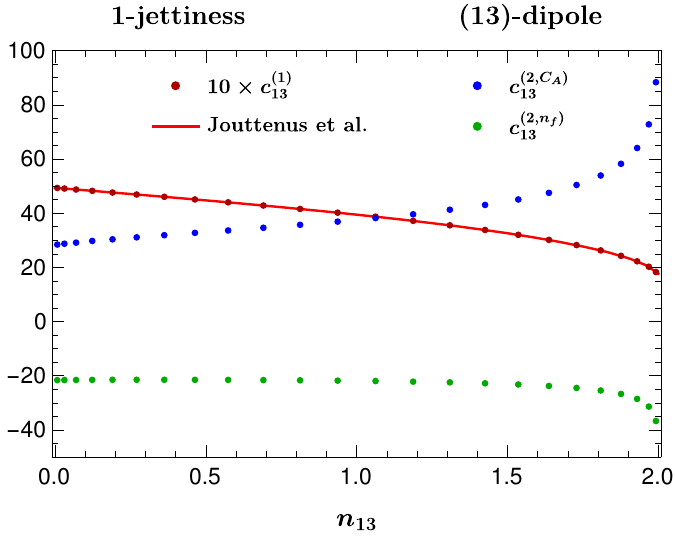}
		\includegraphics[width=0.325\textwidth]{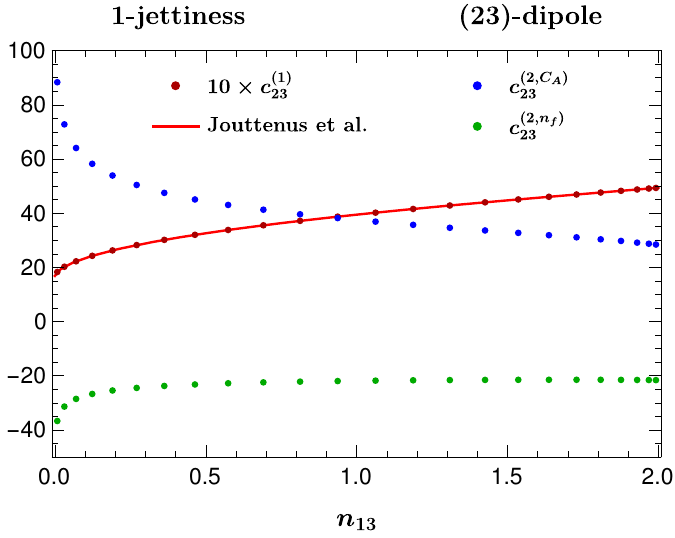}
	}
	\caption{Dipole contributions to the 1-jettiness soft function. The plots show the non-logarithmic terms $c_{ij}^{(1)}$ and $c_{ij}^{(2)}=T_F n_f \,c_{ij}^{(2,n_f)}+ C_A \, c_{ij}^{(2,C_A)}$ of the renormalised soft function~\eqref{eq:RGE:softsolution} for the $(12)$-dipole (left), $(13)$-dipole (middle) and $(23)$-dipole (right) as a function of $n_{13}$. The dots represent our numerical results, and the solid lines were obtained from the integral representations in~\cite{Jouttenus:2011wh}.}
	\label{fig:dipole:finite:1jet}
\end{figure}

Our results for the non-logarithmic terms of the renormalised Laplace-space soft function~\eqref{eq:RGE:softsolution} are displayed in Fig.~\ref{fig:dipole:finite:1jet}. The NLO coefficients $c_{ij}^{(1)}$ (red dots) can be compared to the calculation in~\cite{Jouttenus:2011wh}, which provides integral representations for the NLO dipoles in distribution space. These expressions can readily be transformed to Laplace space and integrated numerically, which yields the solid (red) lines in the plots. For each dipole contribution, our numbers agree perfectly with these predictions.

The NNLO coefficients $c_{ij}^{(2)}=T_F n_f \,c_{ij}^{(2,n_f)}+ C_A \, c_{ij}^{(2,C_A)}$ consist of two colour structures, which are displayed by the green and blue dots in Fig.~\ref{fig:dipole:finite:1jet}, respectively. The plots clearly exhibit a symmetry, which can be traced to the exchange of the two beam directions $(1\leftrightarrow 2$), under which the beam-beam dipole transforms as $c^{(k)}_{12}(n_{13}) = c^{(k)}_{21}(n_{23}) = c^{(k)}_{12}(2 - n_{13})$, whereas the beam-jet dipoles satisfy $c^{(k)}_{13}(n_{13})=c^{(k)}_{23}(n_{23})=c^{(k)}_{23}(2-n_{13})$. We further note that the numerical uncertainties of our predictions are at the subpercent level, and they are therefore not visible on the scale of the plots.

We finally compare our results for the NNLO coefficients $c_{ij}^{(2)}$ to the calculations in~\cite{Boughezal:2015eha,Campbell:2017hsw}. To do so, we adopt the conventions used in~\cite{Campbell:2017hsw}, which denotes the $\mathcal{O}(\eps^0)$-coefficient of the $\delta(\Tau_N)$ term in distribution space by $C_{-1,nab}$ (where ``nab'' refers to the non-abelian part). In this notation, we have
\begin{align}
 C_{-1,nab} &= \sum_{i\neq j} \; \bfT_{i} \cdot \bfT_{j} \; \bigg\{
\Big(\frac83 \tilde L_{ij}^4-\frac{8\pi^2}{3}\tilde L_{ij}^2-\frac{64\zeta_3}{3}\tilde L_{ij}+\frac{2\pi^4}{45}\Big) \,y_4^{ij}\\
&\quad
+ \Big(\frac83 \tilde L_{ij}^3-\frac{4\pi^2}{3}\tilde L_{ij}-\frac{16\zeta_3}{3}\Big) \,y_3^{ij}
+ \Big(2 \tilde L_{ij}^2-\frac{\pi^2}{3}\Big) \,y_2^{ij}
+ \tilde L_{ij}\,y_1^{ij} +\frac{y_0^{ij}}{4}
\nonumber\\
&\quad
-\beta_0 \bigg[
\Big(\frac13 \tilde L_{ij}^3-\frac{\pi^2}{6}\tilde L_{ij}-\frac{2\zeta_3}{3}\Big) \,x_2^{ij}
+ \Big(\frac12 \tilde L_{ij}^2-\frac{\pi^2}{12}\Big) \,x_1^{ij}
+ \frac12 \tilde L_{ij}\,x_0^{ij} +\frac{x_{-1}^{ij}}{4} \bigg] \bigg\},
\nonumber
\end{align}
where $\tilde L_{ij}=\frac{1}{2}\ln (\frac{n_{ij}}{2})$, and the $x_{k}^{ij}$ are the expansion coefficients of the bare NLO dipoles,
\begin{align}
S_{ij}^{(1)}(\eps) =
\frac{x_2^{ij}}{\eps^2} + \frac{x_1^{ij}}{\eps}  + x_0^{ij} + x_{-1}^{ij}\,\eps
+ \mathcal{O}(\eps^2)\,,
\end{align}
whereas the $y_{k}^{ij}$ are the analogous coefficients of the NNLO dipoles defined by
\begin{align}
T_F n_f \, S_{ij}^{(2,q\bar{q})}(\eps)
+ C_A \Big[ S_{ij}^{(2,{\rm Re})}(\eps)
+ S_{ij}^{(2,gg)}(\eps) \Big] =
\frac{y_4^{ij}}{\eps^4} + \frac{y_3^{ij}}{\eps^3}  + \frac{y_2^{ij}}{\eps^2} + \frac{y_1^{ij}}{\eps} + y_0^{ij} + \mathcal{O}(\eps)\,.
\label{eq:def:ycoefficients}
\end{align}
Note that the $y_{k}^{ij}$ coefficients consist of two colour structures.

\begin{figure}[t!]
	\centering{
		\includegraphics[width=0.485\textwidth]{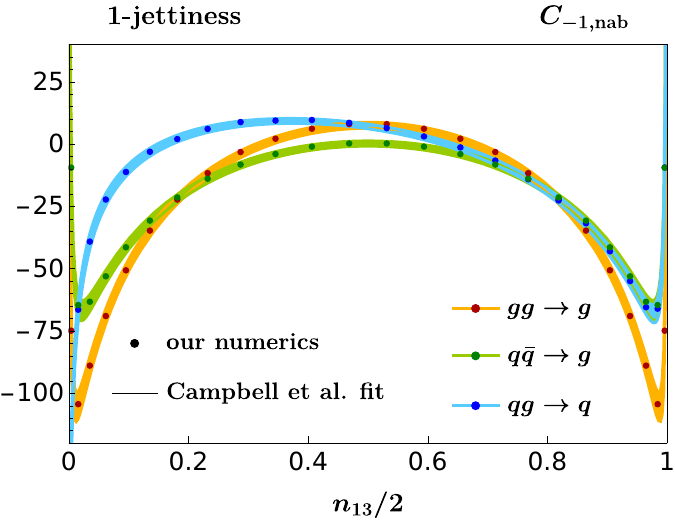}
				\includegraphics[width=0.49\textwidth]{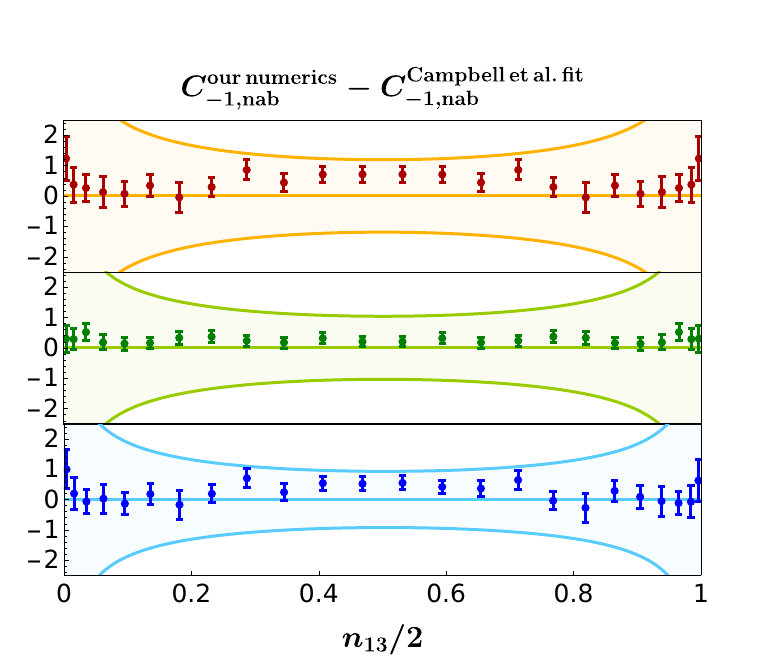}
	}
	\caption{Non-logarithmic terms of the renormalised 1-jettiness soft function in distribution space for all partonic channels. The dots represent our numerical results, and the bands show the fit functions from~\cite{Campbell:2017hsw}, for which we have added the uncertainties from the fit parameters in quadrature. The right panel displays the difference between our numbers and these fits.}
	\label{fig:1jet:comparison}
\end{figure}

In~\cite{Campbell:2017hsw} the 1-jettiness results are presented for the physical channels $gg\to g$, $q \bar q \to g$ and $qg\to q$. For a three-parton amplitude, colour conservation implies
\begin{align}\label{eq:casimir}
\bfT_i\cdot \bfT_j &= \frac{1}{2}(\bfT_k^2-\bfT_i^2-\bfT_j^2) \quad \quad {\rm with} \quad \quad {i\neq j \neq k} \in \{1,2,3\}\,,
\end{align}
and all products of colour generators can hence be expressed through Casimir operators, \mbox{$\bfT_i^2=C_i$}, with $C_q=C_{\bar q}=C_F=4/3$ and $C_g=C_A=3$ (we furthermore set $T_F=1/2$ and $n_f=5$ in our numerical analysis). The authors of~\cite{Campbell:2017hsw} computed 21 points in the variable \mbox{$y_{13}=n_{13}/2$}, and they fitted these points to an ad-hoc functional form that is supposed to approximate the $C_{-1,nab}$ coefficients over the entire kinematic range. Our numbers are compared to these fit functions in Fig.~\ref{fig:1jet:comparison}, which shows a very good agreement for all partonic channels. As can be read off from the right panel in this figure, which displays the difference between our numbers and the results of~\cite{Campbell:2017hsw}, the agreement is well within the uncertainty bands (at $95\%$ confidence level) of the provided fit functions. More specifically, our numbers seem to agree slightly better in the endpoint regions than in the bulk of the distributions, which could be explained by the fact that the fits of~\cite{Campbell:2017hsw} are dominated by the endpoint regions. We also note that the uncertainties of~\cite{Campbell:2017hsw} may be slightly overestimated in this figure, since the authors did not publish the correlation between the fit parameters. In contrast we provide the raw data of our calculation in the ancillary electronic file.

The plots in Fig.~\ref{fig:1jet:comparison} furthermore show that the $C_{-1,nab}$ coefficients diverge at both endpoints, i.e.~in the limit in which the final-state jet becomes collinear to one of the beam directions. This divergent behaviour is parametrised in~\cite{Campbell:2017hsw} by cubic logarithms in the kinematic variables, and we will analyse in Sec.~\ref{sec:method-of-regions} more closely if this is the correct ansatz. 

As a final remark, we note that the 1-jettiness results presented in this section have been obtained from a more general 2-jettiness setup, in which we have made the two jet directions collinear to each other. As an identical setup was used for the 2-jettiness calculation, the agreement found in this section also provides an indirect check for our 2-jettiness results that we are going to present in the following section.

\subsection{2-jettiness}

\begin{figure}[t!]
	\centering{
		\includegraphics[width=0.325\textwidth]{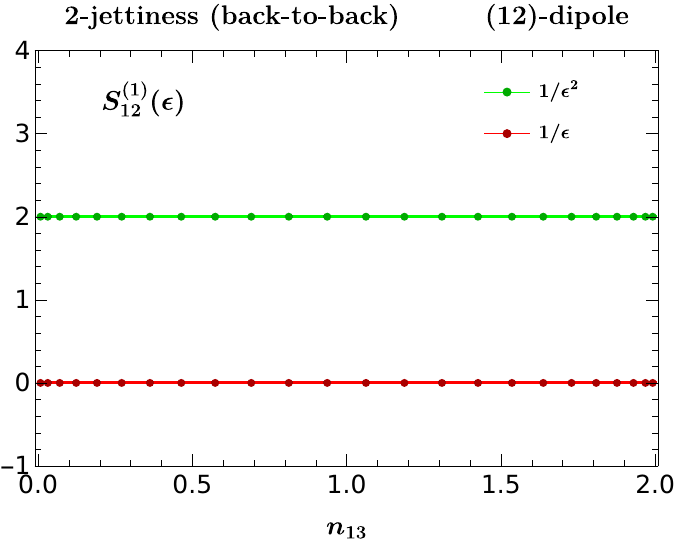}
		\includegraphics[width=0.325\textwidth]{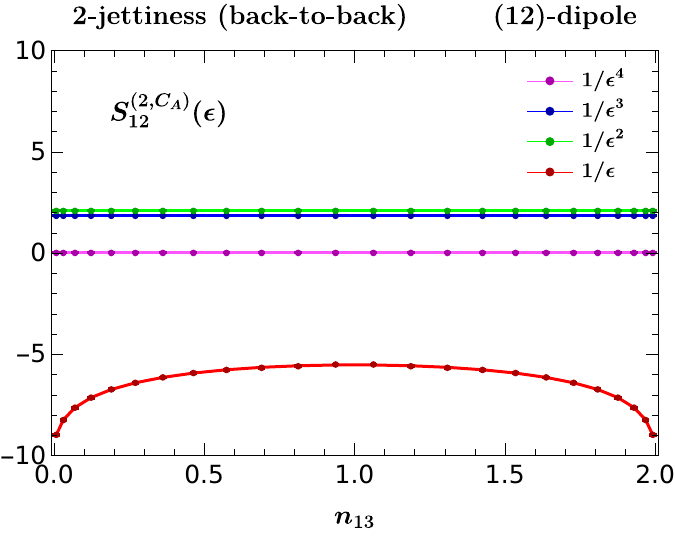}
		\includegraphics[width=0.325\textwidth]{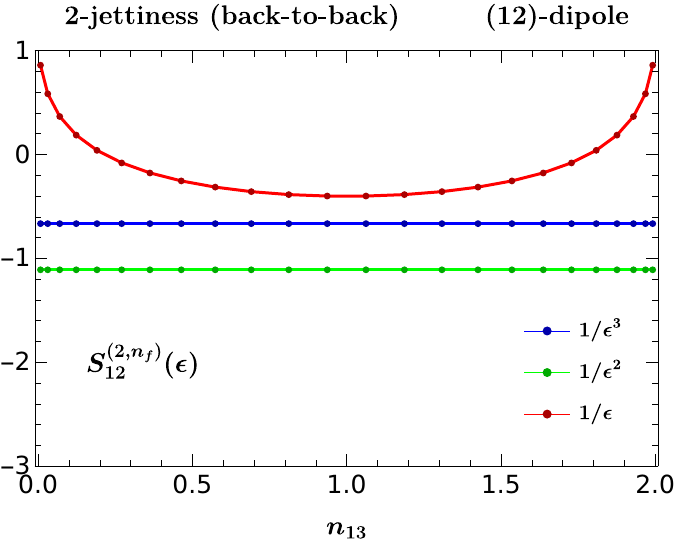}
	}
	\caption{Divergence structure of the bare 2-jettiness soft function in the back-to-back configuration. The plots show the NLO dipole $S_{12}^{(1)}(\eps) $ (left) and the two colour structures of the NNLO dipole $S_{12}^{(2,C_A)}(\eps)= S_{12}^{(2,{\rm Re})}(\eps)+ S_{12}^{(2,gg)}(\eps)$ (middle) and $S_{12}^{(2,n_f)}(\eps)= S_{12}^{(2,q\bar{q})}(\eps)$ (right) as a function of $n_{13}$. The dots represent our numerical results and the solid lines show the RG constraints.}
	\label{fig:dipole:divergences:2jet:b2b}
\end{figure}

The 2-jettiness soft function can again be inferred at NLO from the integral representations in~\cite{Jouttenus:2011wh}, while it has only been computed so far for specific kinematic configurations (consisting of about $20$ phase-space points) at NNLO in~\cite{Bell:2018mkk,Jin:2019dho}. In contrast we sample the entire phase space of the hard emitters for the first time in this paper.

To do so, we parametrise the external geometry of the soft function by two polar angles $\theta_{13}, \theta_{14}\in[0,\pi]$ and one azimuthal angle $\varphi_4\in[0,2\pi]$ in the form,
\begin{align}
n_1^\mu &= (1,0,0,1)\,,
&&&
n_3^\mu &= (1,\sin \theta_{13},0,\cos  \theta_{13})\,,
\nonumber\\
n_2^\mu &= (1,0,0,-1)\,,
&&&
n_4^\mu &= (1,\sin \theta_{14}\cos \varphi_{4},\sin \theta_{14}\sin \varphi_{4},\cos  \theta_{14})\,,
\label{eq:2jet:ni}
\end{align}
where the first two vectors refer to the beam directions, and the remaining ones to the jet directions. Similar to the 1-jettiness case, we then sample these angles in steps of $\pi/25$, which yields a total of $24\times24\times50=28,800$ phase-space points. We note that 24 of these points describe exactly collinear jets that belong to the 1-jettiness rather than the 2-jettiness soft function. As the 2-to-1 jettiness transition is not necessarily smooth (see the discussion in Sec.~\ref{sec:method-of-regions}), we explicitly removed these numbers in the 2-jettiness grids provided in the ancillary electronic file.

\begin{figure}[t!]
	\centering{
		\includegraphics[width=0.375\textwidth]{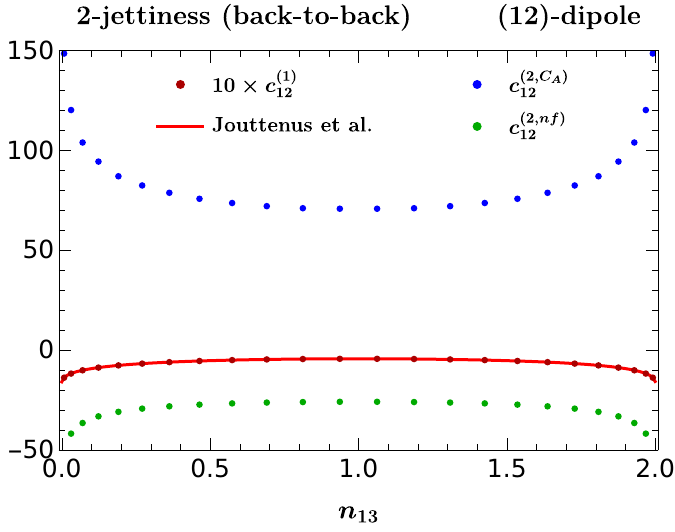}
		\hspace{10mm}
		\includegraphics[width=0.375\textwidth]{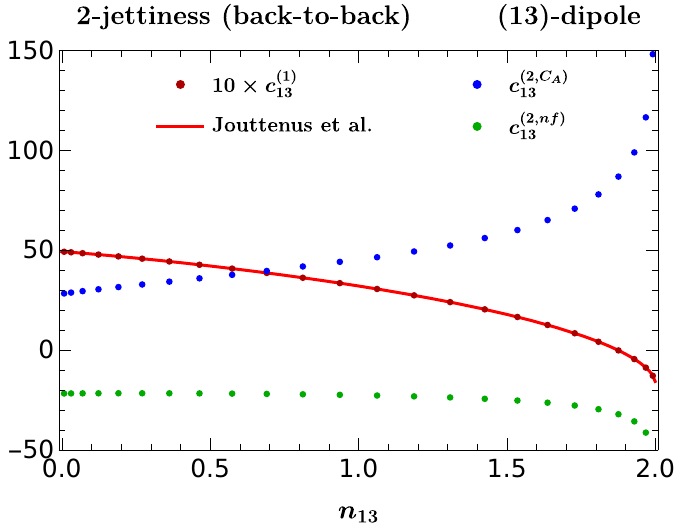}
	}
	\caption{The same as in Fig.~\ref{fig:dipole:finite:1jet} for the two independent dipole contributions to the \mbox{2-jettiness} soft function in the back-to-back configuration.}
	\label{fig:dipole:finite:2jet:b2b}
\end{figure}

The 2-jettiness soft function consists of six dipole and 24 tripole contributions, and the tripole sum yields a non-vanishing correction to the soft function in this case. In the previous section, we noted that the 1-jettiness soft function exhibits a certain symmetry, since it cannot distinguish between the two beam directions. Likewise the 2-jettiness soft function obeys similar symmetry relations, which we use to reduce the computing time somewhat. More specifically, the beam-beam dipole obeys the relations
\begin{align}
S^{(k)}_{12}(n_{13},n_{14},n_{34}) = S^{(k)}_{21}(n_{23},n_{24},n_{34}) = S^{(k)}_{12}(2-n_{13},2-n_{14},n_{34})\,,
\end{align}
and likewise for the jet-jet dipole (34). The mixed beam-jet dipoles, on the other hand, map onto their conjugate partners, e.g.
\begin{align}
S^{(k)}_{13}(n_{13},n_{14},n_{34}) = S^{(k)}_{23}(n_{23},n_{24},n_{34}) = S^{(k)}_{23}(2-n_{13},2-n_{14},n_{34})\,,
\end{align}
and similar relations can be derived for the tripoles $S^{(2,{\rm Im})}_{ilj}(n_{13},n_{14},n_{34})$ as well. In addition, there is a symmetry associated with the azimuthal angle $\varphi_4$, which parametrises the deviation of the second jet from the $(xz)$-plane that is spanned by the two beams and the first jet in our parametrisation. The second jet, obviously, cannot distinguish between the two directions that are orthogonal to this plane, and the scalar products $n_{ij}$ indeed only depend on $\cos \varphi_{4}$, which is symmetric under the exchange $\varphi_{4}\to 2\pi - \varphi_{4}$. Taken together, this implies that the 2-jettiness soft function only needs to be sampled in the range $\theta_{13}\in[0,\pi]$, $\theta_{14}\in[0,\pi/2]$ and $\varphi_4\in[0,\pi]$, which reduces the computing time by a factor of four.

Due to the high dimensionality of the phase space, it becomes difficult to display the \mbox{2-jettiness} results visually. In the following we present some projections of the phase space for illustration, but we reemphasise that we provide results for arbitrary kinematics in the attached electronic file.

\begin{figure}[t!]
	\centering{
		\includegraphics[width=0.375\textwidth]{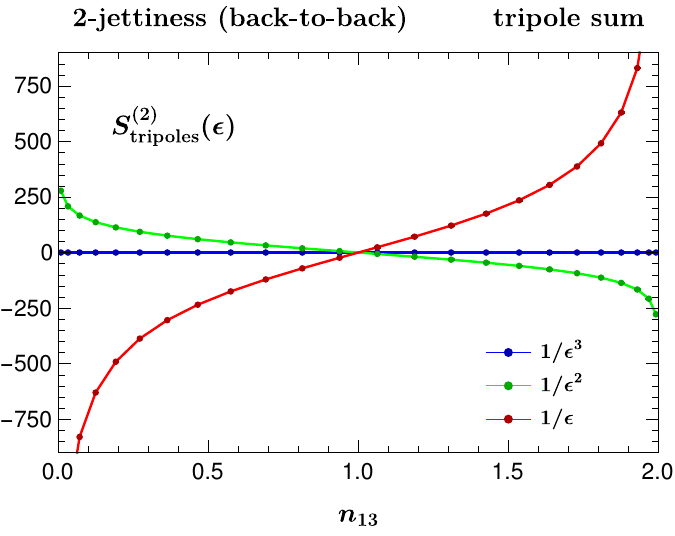}
		\hspace{10mm}
		\includegraphics[width=0.375\textwidth]{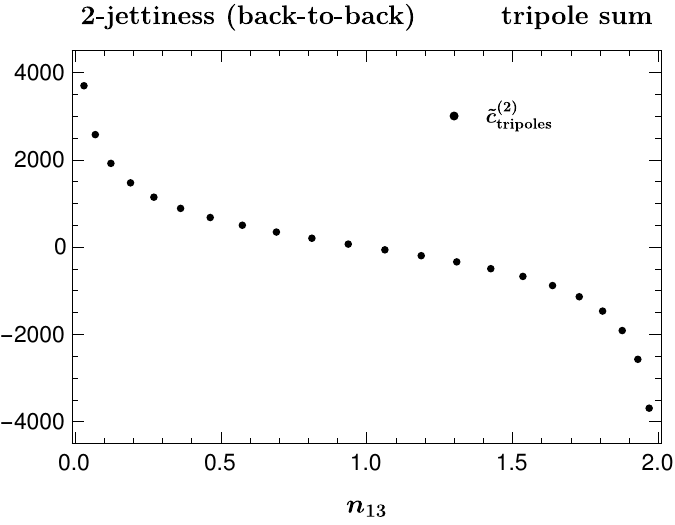}
	}
	\caption{
		Tripole contribution to the 2-jettiness soft function in the back-to-back configuration. Left: Divergences of the bare soft function according to the definition~\eqref{eq:2jet:baretripolesum}. Our numerical results (dots) are compared to the RG prediction (solid lines).  Right: Non-logarithmic term of the renormalised soft function according to the definition~\eqref{eq:2jet:rentripolesum:ctilde}.}
	\label{fig:tripole:2jet:b2b}
\end{figure}

\paragraph{Back-to-back jets:}
First of all, we focus on the situation in which the two jets are produced back-to-back, which is the configuration that was studied before in~\cite{Bell:2018mkk,Jin:2019dho}. In this setup, which corresponds to $\theta_{14}=\pi-\theta_{13}$ and $\varphi_4=\pi$ in our notation, the kinematic invariants become $n_{12}=n_{34}=2$, $n_{13}=n_{24}=1-\cos \theta_{13}$ and $n_{14}=n_{23}=1+\cos \theta_{13}=2-n_{13}$. There are, moreover, only two independent dipoles in this case -- we choose (12) and (13) -- which are functions of a single variable $n_{13}$.

\begin{figure}[t!]
	\centering{
		\includegraphics[width=0.375\textwidth]{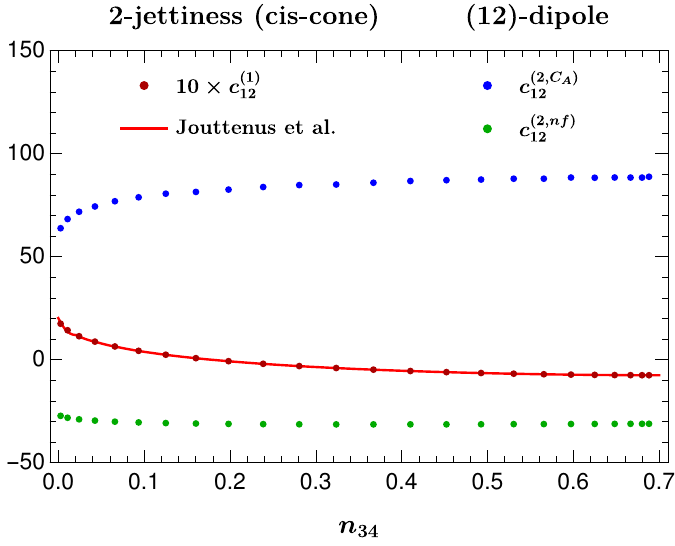}
		\hspace{10mm}
		\includegraphics[width=0.375\textwidth]{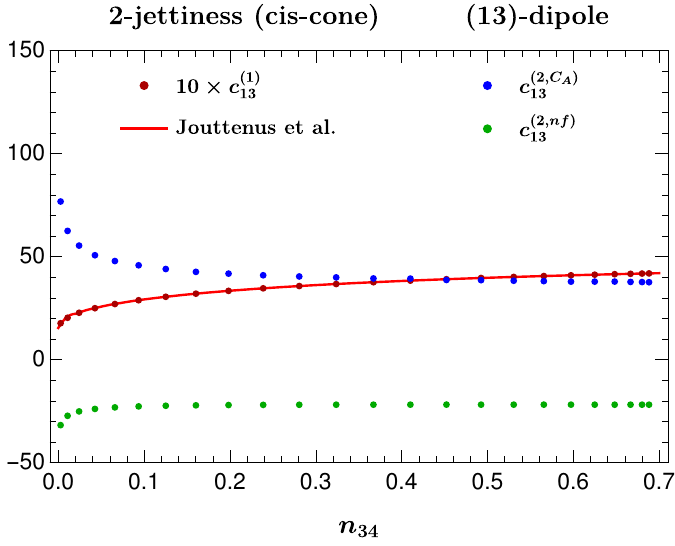}\\[1.5em]
		\includegraphics[width=0.375\textwidth]{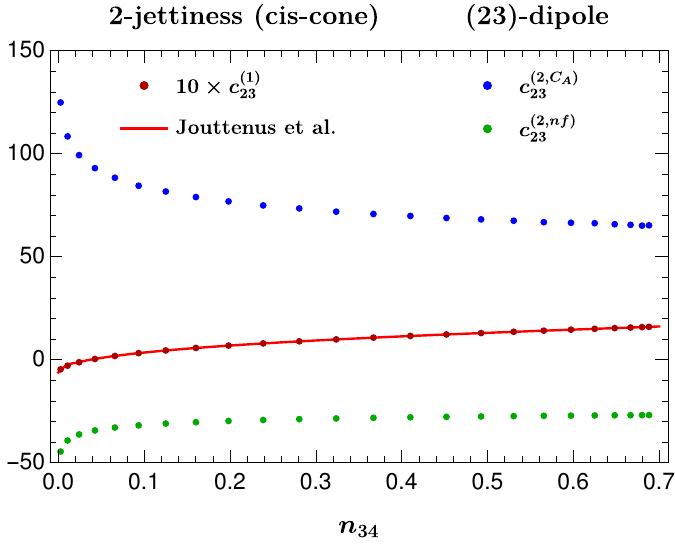}
		\hspace{10mm}
		\includegraphics[width=0.375\textwidth]{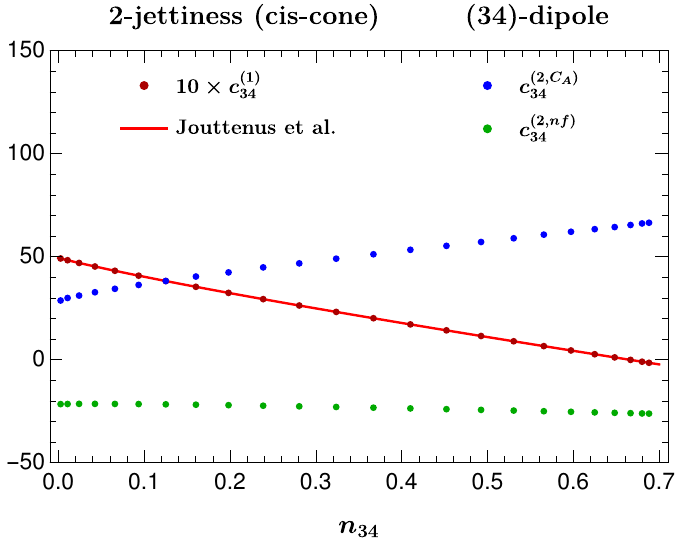}
	}
	\caption{The same as in Fig.~\ref{fig:dipole:finite:1jet} for the four independent dipole contributions to the \mbox{2-jettiness} soft function in the cis-cone configuration.}
	\label{fig:dipole:finite:2jet:cone}
\end{figure}

Similar to the previous section, we first compare the divergence structure of the bare soft function to the RG prediction~\eqref{eq:RGE:baresoft}. The corresponding plots for the (13)-dipole were already shown in~\cite{Bell:2018mkk}, and our results for the (12)-dipole are displayed in Fig.~\ref{fig:dipole:divergences:2jet:b2b}.  In each case we find perfect agreement with the RG prediction within the numerical uncertainties.

We next turn to the non-logarithmic coefficients $c_{ij}^{(1)}$ and $c_{ij}^{(2)}=T_F n_f \,c_{ij}^{(2,n_f)}+ C_A \, c_{ij}^{(2,C_A)}$ of the renormalised soft function that were defined in~\eqref{eq:RGE:softsolution}. As illustrated in Fig.~\ref{fig:dipole:finite:2jet:b2b}, our NLO results (red dots) again agree with the calculation in~\cite{Jouttenus:2011wh}, after transforming the latter into Laplace space (red lines). Our NNLO results for the coefficients $c_{ij}^{(2,n_f)}$ and $c_{ij}^{(2,C_A)}$ are displayed by the green and blue dots in this figure, respectively, and their numerical uncertainties are again too small to be visible in the plots. Whereas our results for the (13)-dipole were already presented in~\cite{Bell:2018mkk}, our numbers for the (12)-dipole can be compared to the calculation in~\cite{Jin:2019dho}, but since no detailed numerical results are provided in that paper, we can only conclude here that our numbers agree qualitatively with these results.

We finally consider the tripole contribution to the 2-jettiness soft function in the back-to-back configuration. On the level of the bare soft function, we use colour conservation to rewrite the tripole sum in the form
\begin{align}
\sum_{i\neq l\neq j} (\lambda_{il} - \lambda_{ip}-\lambda_{lp})\,
f_{ABC}\; \bfT_{i}^A \;\bfT_{l}^B \;\bfT_{j}^C  \,
\Big( \frac{n_{ij}}{2} \Big)^{2\eps} \,
S_{ilj}^{(2,{\rm Im})}(\eps)  &\equiv
f_{ABC}\; \bfT_{1}^A \;\bfT_{2}^B \;\bfT_{3}^C  \;
S_{\rm tripoles}^{(2)}(\eps)\,,
\label{eq:2jet:baretripolesum}
\end{align}
and the divergences of this contribution are compared against the RG prediction~\eqref{eq:RGE:baresoft} in the left panel of Fig.~\ref{fig:tripole:2jet:b2b}. In the right panel of this figure, we display the corresponding finite term of the renormalised soft function in the form
\begin{align}
2\pi \sum_{i\neq j\neq k} f_{ABC}\; \bfT_{i}^A \;\bfT_{j}^B \;\bfT_{k}^C\;
\tilde c_{ijk}^{(2)}
\equiv
f_{ABC}\; \bfT_{1}^A \;\bfT_{2}^B \;\bfT_{3}^C\;
\tilde c_{\rm tripoles}^{(2)}\,,
\label{eq:2jet:rentripolesum:ctilde}
\end{align}
and we recall that the $\tilde c_{ijk}^{(2)}$ coefficients refer to a representation in which the RG logarithms $L_\mu$ are manifest, see~\eqref{eq:c2ijknotilde}.
As can be seen in the figure, our numbers for the divergences of the tripole contribution (coloured dots) match the RG prediction (coloured  lines), whereas our prediction for the finite term $\tilde c_{\rm tripoles}^{(2)}$ (black dots) can be compared against the corresponding term in~\cite{Jin:2019dho}. While it is difficult to read off precise numbers from the plots provided in that paper, it seems that our results disagree with these numbers roughly by a factor of two. We will come back to this discrepancy in Sec.~\ref{sec:regions:discussion} below. We also note that the symmetry we discussed earlier -- related to the exchange of the two beam directions -- manifests in the renormalised tripole coefficient defined through \eqref{eq:2jet:rentripolesum:ctilde} as an \emph{anti-symmetry} under $n_{13} \to n_{23}=2-n_{13}$ and (in general) $n_{14} \to n_{24}=2-n_{14}$ as can clearly be seen in the figure.

\paragraph{Azimuthal dependence:}
We next consider a configuration in which the two jets lie on the same nappe of a cone with half angle $\theta_{13}=\theta_{14}=\pi/5$, and we probe the azimuthal dependence of the soft function by varying the azimuthal separation of the two jets $\varphi_{4}$. In this case, we have $n_{12}=2$, $n_{13}=n_{14}=(3 - \sqrt{5})/4$, $n_{23}=n_{24}=(5 + \sqrt{5})/4$ and \mbox{$n_{34}=(5 - \sqrt{5})/4\,\sin^2 (\varphi_{4}/2)$}. There are, moreover, four independent dipoles in this ``cis-cone'' configuration , viz.~(12), (13), (23) and (34), which are now functions of $n_{34}\in\{0,0.691\}$.

\begin{figure}[t!]
	\centering{
		\includegraphics[height=0.23\textheight]{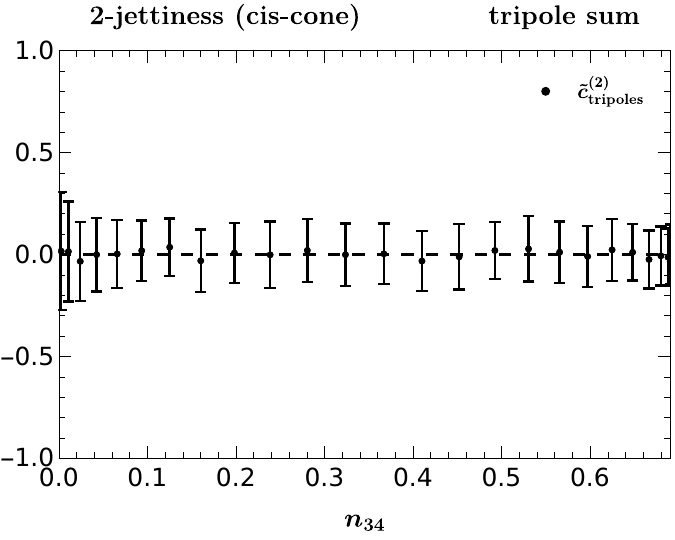}
 \hspace{1cm}
 \includegraphics[height=0.23\textheight]{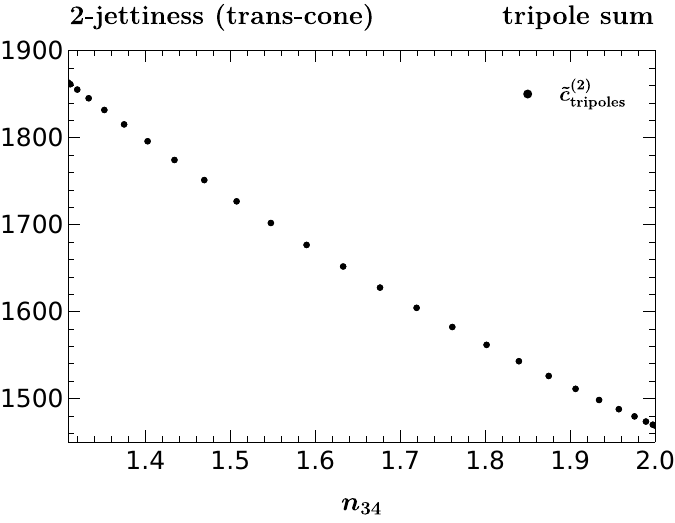}
	}
	\caption{Tripole contribution to the renormalised soft function in the convention~\eqref{eq:2jet:rentripolesum:ctilde} for cis-cone events (left) and trans-cone events (right) as a function of the invariant $n_{34}$.}
	\label{fig:tripole:finite:2jet:cone}
\end{figure}

We refrain from showing the divergence structure here, and display instead directly the dipole coefficients $c_{ij}^{(1)}$ and $c_{ij}^{(2)}$ of the renormalised soft function in Fig.~\ref{fig:dipole:finite:2jet:cone}. As can be seen from this figure, our NLO results (red dots) again agree with the calculation from~\cite{Jouttenus:2011wh} (red lines), whereas our NNLO results that are indicated by the green and blue dots are new. The corresponding term for the tripole sum defined in \eqref{eq:2jet:rentripolesum:ctilde} is shown in the left panel of Fig.~\ref{fig:tripole:finite:2jet:cone}, which reveals an interesting feature: The tripole contribution in fact vanishes for all cis-cone configurations, since the two jets at equal scattering angles are functionally indistinguishable, and the tripole contribution is antisymmetric under the exchange of two colour generators. Note that this is not the case for ``trans-cone'' configurations of jets on different nappes of the cone as demonstrated for comparison in the right panel of Fig.~\ref{fig:tripole:finite:2jet:cone} (note that $n_{34}\in\{1.309,2\}$ in this case because of the modified kinematics). Accordingly this can be viewed as another check of our numerics: While the results for individual tripoles are non-zero, the tripole sum vanishes for cis-cone events, and any errors missed by our uncertainty estimate should therefore show up as a deviation from zero that is larger than the error bars. Fig.~\ref{fig:tripole:finite:2jet:cone} clearly shows that this is not the case. Moreover, as our error estimate shown in this figure combines the uncertainties of individual tripoles in quadrature, it is also worth pointing out that the fluctuations of the central values around zero are smaller than the average error estimate for individual tripoles, which is typically around 0.1 for these events, implying that our uncertainty estimate for individual tripoles is also reliable.

\paragraph{Planar events:}
We finally show a two-dimensional projection of the phase space by considering events with two jets at different scattering angles $\theta_{13}$ and $\theta_{14}$ that are confined to the same $(xz)$-halfplane, i.e.~we set the azimuthal angle $\varphi_4=0$ in this setup. The kinematic invariants then become $n_{12}=2$, $n_{13}=1-\cos \theta_{13}$, $n_{14}=1-\cos \theta_{14}$, $n_{23}=2-n_{13}$, $n_{24}=2-n_{14}$, and $n_{34}=n_{13} + n_{14} - n_{13} \,n_{14} - \sqrt{n_{13} (2 - n_{13}) \,n_{14} (2 - n_{14})}$. There are furthermore now three independent dipoles -- we choose (12), (13) and (34)  -- which are functions of two variables $n_{13}$ and $n_{14}$.

\begin{figure}[t!]
	\centering{
		\includegraphics[height=0.19\textheight]{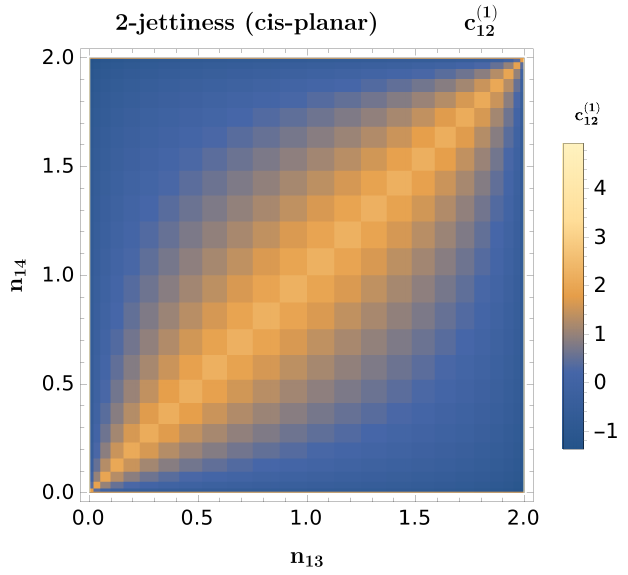}
		\includegraphics[height=0.19\textheight]{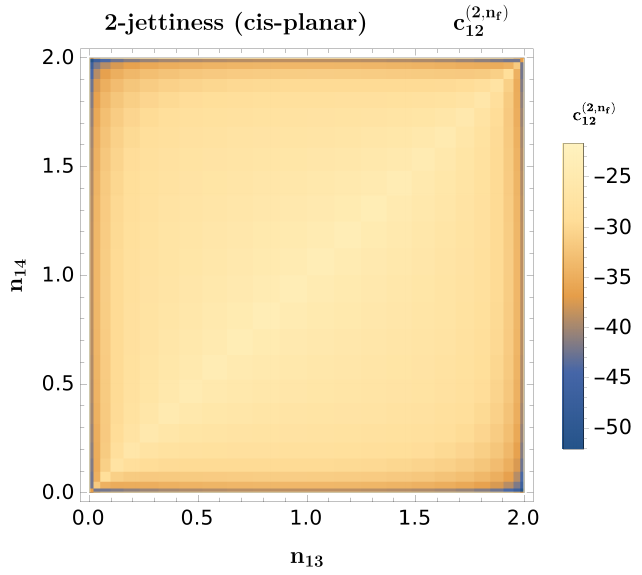}
		\includegraphics[height=0.19\textheight]{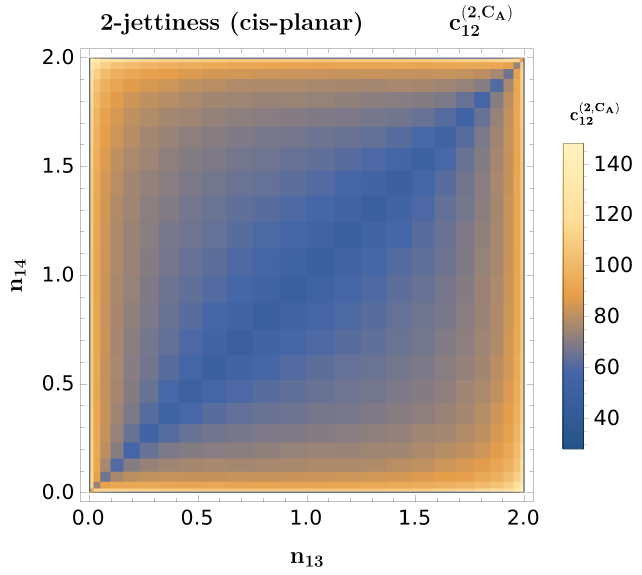}\\[1em]
		\includegraphics[height=0.19\textheight]{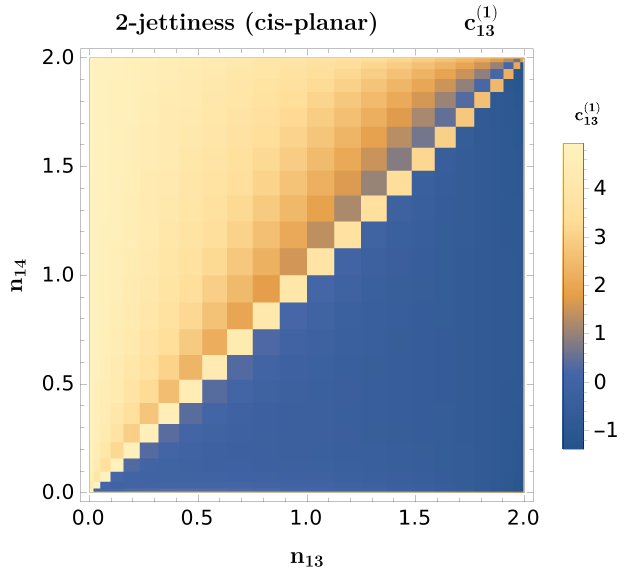}
		\includegraphics[height=0.19\textheight]{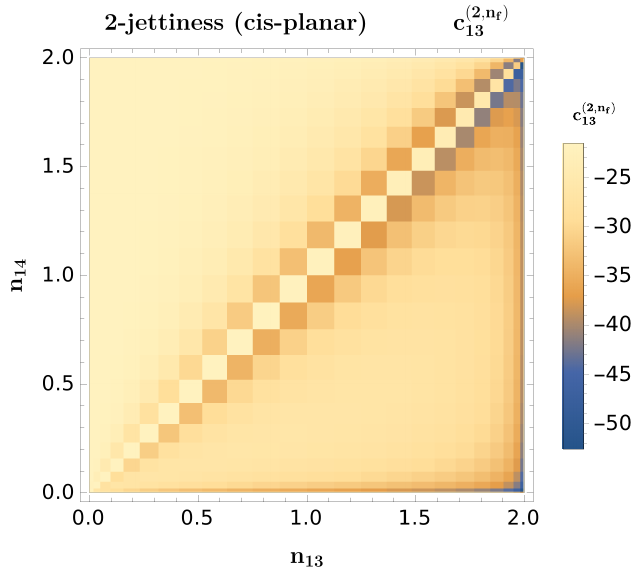}
		\includegraphics[height=0.19\textheight]{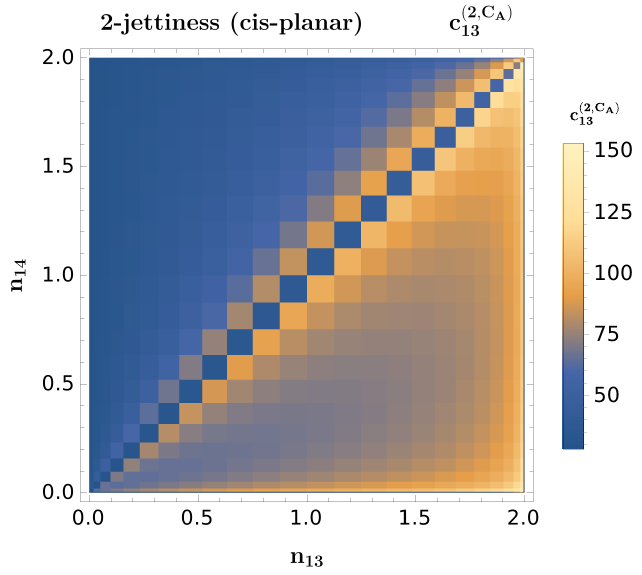}\\[1em]
		\includegraphics[height=0.19\textheight]{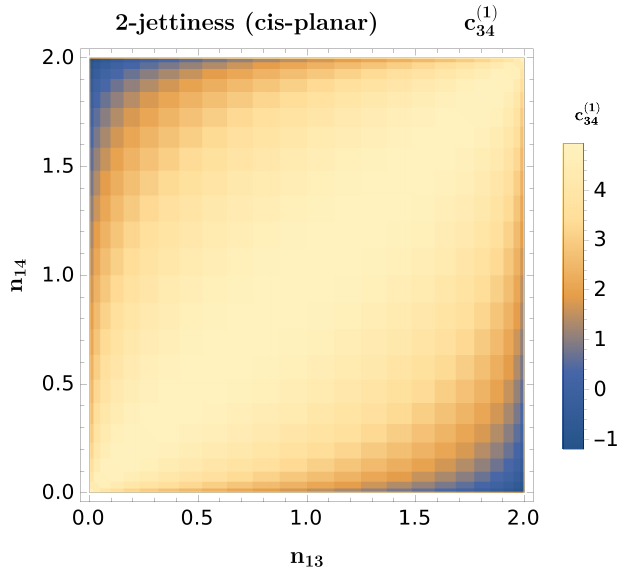}
		\includegraphics[height=0.19\textheight]{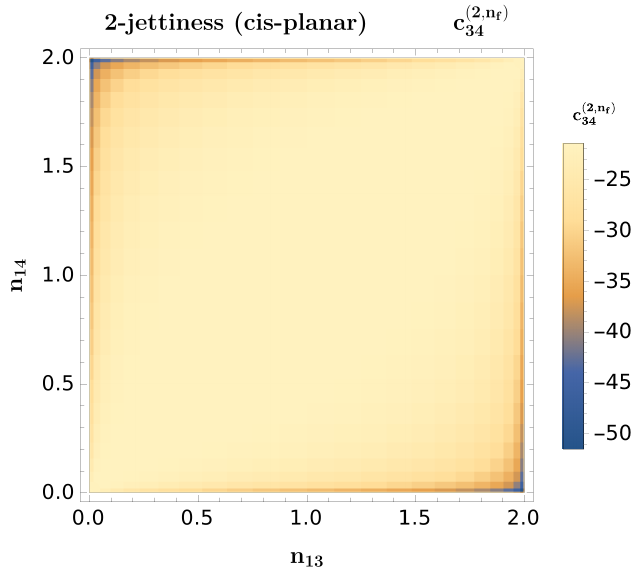}
		\includegraphics[height=0.19\textheight]{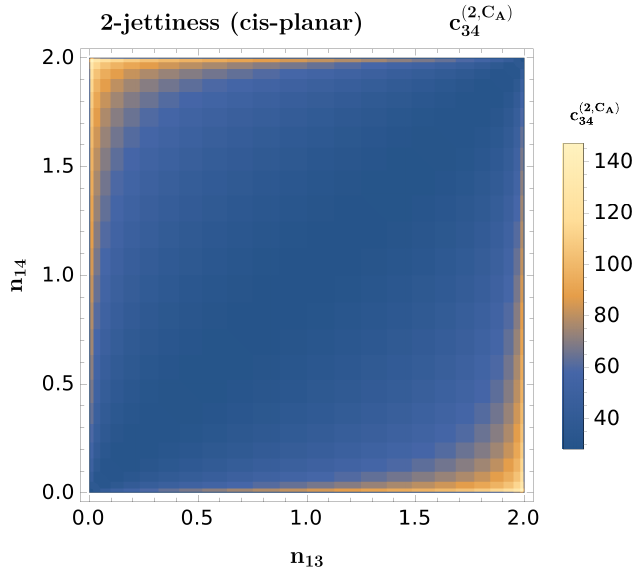}
	}
	\caption{Dipole contributions to the 2-jettiness soft function in the planar configuration  described in the text. The plots show the non-logarithmic terms $c_{ij}^{(1)}$ (left column) and the two colour structures $c_{ij}^{(2,n_f)}$ (middle column) and $c_{ij}^{(2,C_A)}$ (right column) of the renormalised soft function~\eqref{eq:RGE:softsolution} for the $(12)$-dipole (top row), $(13)$-dipole (middle row) and the $(34)$-dipole (bottom row) as a function of the two kinematic variables $n_{13}$ and $n_{14}$.}
	\label{fig:dipole:finite:2jet:planar}
\end{figure}

In Fig.~\ref{fig:dipole:finite:2jet:planar} we show these two-dimensional projections for the NLO dipole coefficients $c_{ij}^{(1)}$ (left column) and the two colour structures $c_{ij}^{(2,n_f)}$ and $c_{ij}^{(2,C_A)}$ (middle and right column), which together make up the NNLO dipole coefficient $c_{ij}^{(2)}=T_F n_f \,c_{ij}^{(2,n_f)}+ C_A \, c_{ij}^{(2,C_A)}$. Similar to the one-dimensional projections shown before, the plots display the raw data of our calculation -- i.e.~$26\times26 = 676$ pixels -- without any type of interpolation in between. Our NLO results, of course, again confirm the calculation from~\cite{Jouttenus:2011wh}, whereas the NNLO results are new. Note, in particular, that the diagonal in these plots -- corresponding to two jets that are exactly collinear to each other -- are \emph{de facto} 1-jettiness numbers, which are excluded in the attached electronic grids. We nevertheless find it instructive to include these numbers in the plots, since it shows very clearly that for the (12)-dipole (top row) and the (34)-dipole (bottom row) the respective 1-jettiness numbers are approached smoothly, whereas the (13)-dipole (middle row) is discontinuous in this limit. We will investigate the origin of this behaviour in Sec.~\ref{sec:method-of-regions} in detail.

\begin{figure}[t!]
	\centering{
		\includegraphics[height=0.2\textheight]{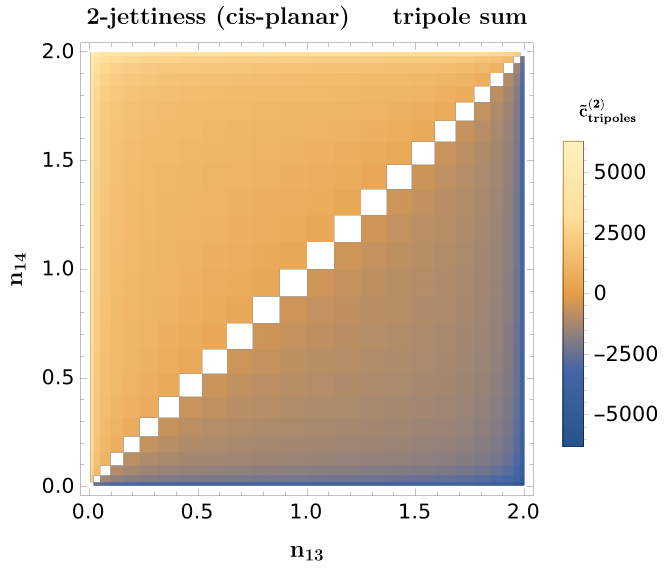}
  \hspace{2cm}
		\includegraphics[height=0.2\textheight]{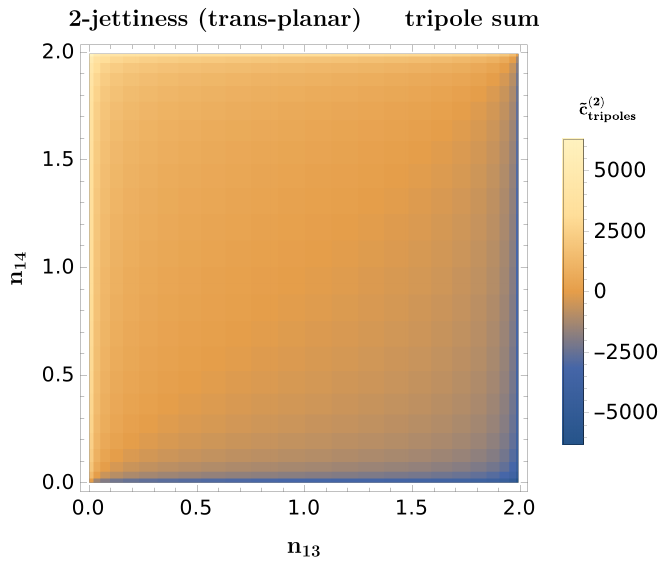}
	}
	\caption{Tripole contribution to the renormalised soft function in the convention~\eqref{eq:2jet:rentripolesum:ctilde} for cis-planar (left) and trans-planar (right) events as a function of the invariants $n_{13}$ and $n_{14}$.}
	\label{fig:tripole:finite:2jet:planar}
\end{figure}

The corresponding result for the renormalised tripole sum is shown in the left panel of  Fig.~\ref{fig:tripole:finite:2jet:planar} in the convention \eqref{eq:2jet:rentripolesum:ctilde}. As the values for the tripole sum are meaningless for the 1-jettiness -- since they multiply a vanishing colour structure in this case -- we exclude the tiles on the diagonal and on the edges in this figure, unlike we did for the dipoles in Fig.~\ref{fig:dipole:finite:2jet:planar}. For completeness we also show the relevant plot for trans-planar events, i.e.~events with the jets on opposite sides of the detector ($\varphi_{4}=\pi$), in the right panel of Fig.~\ref{fig:tripole:finite:2jet:planar}.  Here the entries on the diagonal exist and they all evaluate to zero, for the same invariance-under-exchange reason as for cone-type events.

\subsection{3-jettiness}

\begin{table}[t!]
	\center
	\setlength{\extrarowheight}{4pt}
	\scalebox{0.96}{\begin{tabular}{|c|c|c|c||c|c|c|c|}
			\hline \hline
			$ij$ & $c_{ij}^{(1)}$  &  $c_{ij}^{(2,n_f)}$ & $c_{ij}^{(2,C_A)}$  &
			$ij$ & $c_{ij}^{(1)}$  &  $c_{ij}^{(2,n_f)}$ & $c_{ij}^{(2,C_A)}$  \\[6pt]
			\hline \hline
			$12$  & $-3.3450(5) $ & $-36.244(9)$  & $116.20(16)$ &
			$24$ & $1.5850(3)$ & $-26.815(7)$ & $67.505(81)$
			\\[6pt]
			\hline
			$13$  & $4.1353(1)$& $-21.732(5)$  & $37.631(34)$ &
			$25$ & $4.7549(0)$ & $-21.561(5)$ & $30.735(37)$
			\\[6pt]
			\hline
			$14$  & $1.2979(2)$& $-25.192(6)$ & $63.663(63)$ &
			$34$ & $0.5340(3)$
   & $-25.861(6)$ & $69.243(74)$
			\\[6pt]
			\hline
			$15$  & $-2.4610(5)$& $-35.256(9)$  & $106.99(12)$ &
			$35$ & $-2.2155(5)$& $-34.796(8)$ & $105.97(13)$
			\\[6pt]
			\hline
			$23$  & $-1.2260(3)$& $-32.872(8)$ & $96.975(100)$ &
			$45$ & $0.6110(4)$
   & $-28.251(7)$ & $74.363(87)$
			\\[6pt]
			\hline
	\end{tabular}}
	\caption{Dipole contributions to the renormalised 3-jettiness soft function for the benchmark point defined in \eqref{eq:3jet:benchmark}.}
	\label{tab:3jet:dipoles:benchmark}
\end{table}

We finally consider the 3-jettiness soft function, which has not yet been computed to NNLO before. Similar to the variables that we introduced in the previous section, we parametrise the direction of the third jet by one polar angle $\theta_{15}\in[0,\pi]$ and one azimuthal angle $\varphi_5\in[0,2\pi]$ in the form
\begin{align}
n_5^\mu &= (1,\sin \theta_{15}\cos \varphi_{5},\sin \theta_{15}\sin \varphi_{5},\cos  \theta_{15})\,,
\end{align}
which --  as long as we continued to sample all angles in steps of $\pi/25$ -- would yield a total of around 45 million phase-space points. While this number could be somewhat reduced by exploiting the symmetry associated with the two beam directions, the complexity of the \mbox{3-jettiness} soft function also rises for another reason, since the number of independent dipole and tripole contributions grows to 10 and 60, respectively. A systematic scan of the full 3-jettiness phase space therefore seems out of reach to date.\footnote{To gauge this statement, we remark that the evaluation of a single 3-jettiness phase-space point lasts around 15-20 minutes on current machines with our default integrator settings. }

We could, of course, try to solve this problem by either reducing the precision of the numerical integrator significantly, or by interpolating the 3-jettiness grids from a much smaller number of phase-space points. While we do not attempt to follow such an approach here, we instead present a proof-of-concept calculation by providing explicit results (with uncertainties) for a single 3-jettiness benchmark point. The motivation for this exercise is two-fold: First, we want to illustrate that our framework can be used for any number of jets N and, second, our numbers represent a useful point of comparison for future calculations of the 3-jettiness soft function. Moreover, we remark that the 3-jettiness soft function will also be studied for a different reason in the context of a method-of-regions analysis in the following section.

The benchmark point we consider in this section refers to a configuration, in which the three jets are widely separated with
\begin{align}
\theta_{13}=\frac{3\pi}{10}\,,
&&
\theta_{14}=\frac{6\pi}{10}\,,
&&
\varphi_{4}=\frac{3\pi}{5}\,,
&&
\theta_{15}=\frac{9\pi}{10}\,,
&&
\varphi_{5}=\frac{6\pi}{5}\,.
\label{eq:3jet:benchmark}
\end{align}
Starting with the dipole contributions, we first observe that the divergences of the bare soft function agree with the RG prediction at an accuracy that is comparable to the 2-jettiness calculation. Our numbers for the finite terms of the renormalised soft function~\eqref{eq:RGE:softsolution} are summarised in Tab.~\ref{tab:3jet:dipoles:benchmark}, and at NLO they can once more be compared against the calculation of~\cite{Jouttenus:2011wh}. Our NNLO results, on the other hand, are new and they provide a useful point of reference for future calculations of the 3-jettiness soft function.

The tripole contribution is even more interesting in the sense that four independent colour structures can be identified in the tripole sum, compared to just one for the 2-jettiness.  While there exists a freedom in defining these terms -- depending on the colour generator that is resolved using colour conservation --  we choose the one of the first jet (with index 3) here for reasons that will become clearer in the following section. Anyhow, the result can, of course, easily be mapped onto any choice of colour basis. Specifically, we now write the corresponding finite term of the renormalised soft function as
\begin{align}
\label{eq:3jet:tripoles}
&2\pi \sum_{i\neq j\neq k} f_{ABC}\; \bfT_{i}^A \;\bfT_{j}^B \;\bfT_{k}^C\;
\tilde c_{ijk}^{(2)}
\equiv
f_{ABC}\; \bfT_{1}^A \;\bfT_{2}^B \;\bfT_{4}^C\;
\tilde c_{\rm tripoles}^{(2,124)}
+f_{ABC}\; \bfT_{1}^A \;\bfT_{2}^B \;\bfT_{5}^C\;
\tilde c_{\rm tripoles}^{(2,125)}
\nonumber \\
&\qquad +f_{ABC}\; \bfT_{1}^A \;\bfT_{4}^B \;\bfT_{5}^C\;
\tilde c_{\rm tripoles}^{(2,145)}
+f_{ABC}\; \bfT_{2}^A \;\bfT_{4}^B \;\bfT_{5}^C\;
\tilde c_{\rm tripoles}^{(2,245)}\,,
\end{align}
and the corresponding numbers for the considered benchmark point are collected in Tab.~\ref{tab:3jet:tripoles:benchmark}.

\begin{table}[t!]
	\center
	\setlength{\extrarowheight}{5pt}
	\scalebox{1}{\begin{tabular}{|c|c|c|c|}
			\hline \hline
			$\tilde c_{\rm tripoles}^{(2,124)}$ &
			$\tilde c_{\rm tripoles}^{(2,125)}$ &
			$\tilde c_{\rm tripoles}^{(2,145)}$ &
			$\tilde c_{\rm tripoles}^{(2,245)}$  \\[6pt]
			\hline \hline
			  ~$-683.23(4)$~ & ~$-2203.5(0)$~ &  $-6.3247(389)$ & 
            $-0.8304(393)$
			\\[6pt]
			\hline
	\end{tabular}}
	\caption{The same as in Tab.~\ref{tab:3jet:dipoles:benchmark} for the four independent tripole contributions defined in~\eqref{eq:3jet:tripoles}.}
		\label{tab:3jet:tripoles:benchmark}
\end{table}

\section{Method-of-regions analysis}
\label{sec:method-of-regions}

In the previous section we observed interesting features in corners of phase space where two of the reference vectors $n_i^\mu$ become collinear to each other, i.e.~when the number of jets N is reduced by one (or more). Consider e.g.~the dipole coefficients of the renormalised \mbox{1-jettiness} soft function from Fig.~\ref{fig:dipole:finite:1jet} in the limit $n_{13}\to 0$, i.e.~when the jet folds into a beam.
Of course, the underlying N-jettiness factorisation theorem assumes that the directions $n_i^\mu$ are all distinct and well separated, but we would nevertheless naively expect that the numerical values of the 1-jettiness coefficients should correspond exactly to the 0-jettiness numbers for $n_{13}=0$, since an event with a very forward jet is virtually identical to an event with only beams, and no jet at all. We thus would expect that the soft-function values for the beam-beam dipole (12) of the 1-jettiness should approach the values for the corresponding 0-jettiness dipole, which are $(4.93,-21.7,28.2)$, referring to (NLO, NNLO $T_F n_f$, NNLO $C_A$) or the (red, green, blue) curves of Fig.~\ref{fig:dipole:finite:1jet}. But as can clearly be seen by inspecting
the left panel of this figure, this is not the case: the NNLO numbers diverge, while the NLO value approaches a completely different (finite) number. By contrast the (13) dipole formed by the two collapsing beam and jet directions \emph{does} approach the 0-jettiness numbers, as visible in the middle panel of Fig.~\ref{fig:dipole:finite:1jet}. This is surprising, since this dipole has no counterpart in the 0-jettiness case.

These features, which may also arise in the bulk of the phase space as illustrated in Fig.~\ref{fig:dipole:finite:2jet:planar}, where the diagonal reflects the configuration in which the two final-state jets become collinear to each other, are clearly puzzling and call for an explanation. To this end, we will perform a dedicated method-of-regions analysis to compute the coefficients that control the asymptotic behaviour of the N-jettiness coefficients in these limits analytically. While this may seem a purely academic exercise, we stress that our analysis will provide important cross checks for the NNLO calculation, and it may help to guide fits to the multi-dimensional phase-space grids. In fact, in the 1-jettiness analysis of~\cite{Campbell:2017hsw}, the authors performed such fits, and we will study the quality of these fits in the endpoint region in some detail in this section. Moreover, controlling the endpoint behaviour can also be of direct phenomenological importance when the soft function is boosted to another reference frame~\cite{Alioli:2023rxx}.

\subsection{Qualitative analysis}
\label{sec:regions:qualitative}

Before we start the method-of-regions analysis, we will classify the different patterns that arise when two of the reference vectors $n_i^\mu$ become collinear to each other more clearly. For this purpose, we consider the 3-jettiness soft function, which we already studied in the previous section. We choose the 3-jettiness here because it approaches the 2-jettiness, which by itself has a non-trivial tripole structure, in the limit of two jets becoming collinear to each other. If instead we had started from the 2-jettiness, the tripole sum would vanish trivially in the 1-jettiness limit because of colour conservation.

More specifically, we consider a 3-jet configuration that is described by the angles
\begin{align}
\theta_{13}=\frac{2\pi}{5}\,,
&&
\theta_{14}=\frac{3\pi}{5}\,,
&&
\varphi_{4}=\frac{3\pi}{5}\,,
&&
\varphi_{5}=\frac{3\pi}{5}\,,
\label{eq:3To2kinematics}
\end{align}
and we vary the scattering angle of the third jet in the region $\theta_{15}\in[0.6001\pi,0.61\pi]$. As the second and third jet lie in a plane with $\varphi_{4}=\varphi_{5}$, this allows us to study configurations in which the two jets have an angular separation between $0.02\degree$ and $1.8\degree$, which is sufficient for the qualitative analysis in this section.

\begin{figure}[t!]
 	\centering{
 		\includegraphics[height=0.23\textheight]{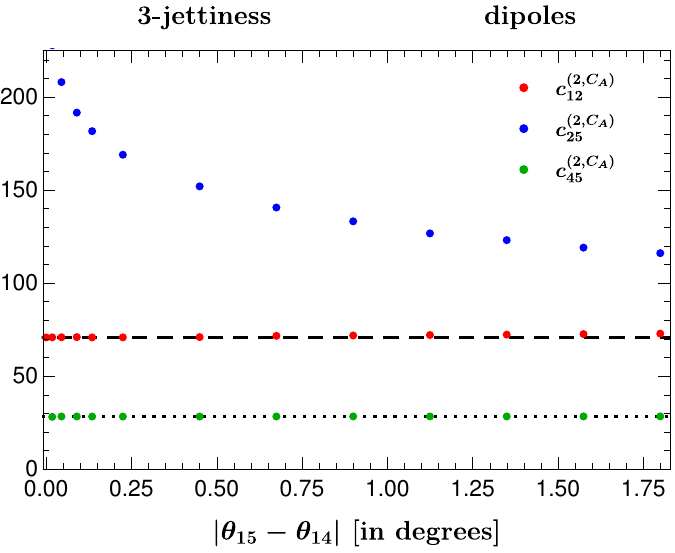}
 		\hspace{10mm}
 		\includegraphics[height=0.23\textheight]{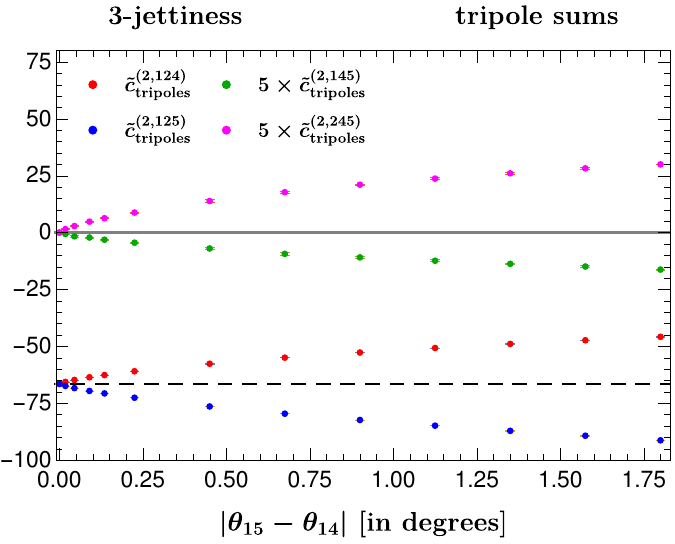}
 	}
 	\caption{
 		Asymptotic behaviour of the 3-jettiness soft function for various dipole (left) and tripole (right) contributions in the limit when two jets become collinear to each other. The dashed lines show the respective 2-jettiness numbers, and the dotted line marks the 0-jettiness value.}
 	\label{fig:regions:qualitative}
 \end{figure}

We first consider the dipole coefficients of the renormalised N-jettiness soft function, which we also plotted in Fig.~\ref{fig:dipole:finite:1jet} for the 1-jettiness soft function, and in Figs.~\ref{fig:dipole:finite:2jet:b2b}, \ref{fig:dipole:finite:2jet:cone} and~\ref{fig:dipole:finite:2jet:planar} for the various 2-jettiness scenarios. In all of these plots we found that the NLO coefficients $c_{ij}^{(1)}$ have a smooth behaviour in the N to (N-1)-jet transition, although as argued before the asymptotic value may not correspond to the one that one may have naively expected. We will clarify the origin of this offset in the following section, and concentrate here on the NNLO coefficients $c_{ij}^{(2,n_f)}$ and $c_{ij}^{(2,C_A)}$ instead. In particular, we saw in the previous section that these coefficients may or may not diverge in the collinear limit. More precisely, we find three characteristic patterns (recall that directions $n_4^\mu$ and $n_5^\mu$ become collinear in our setup):
\begin{itemize}
\item
The dipoles that do \emph{not} involve \emph{either} of the directions that become collinear -- i.e.~dipoles $c_{12}^{(2)}$, $c_{13}^{(2)}$ and $c_{23}^{(2)}$ in our setup -- approach the (N-1)-jettiness limit smoothly.
\item
The dipoles that involve \emph{one} of the collinear directions -- i.e.~dipoles $c_{14}^{(2)}$, $c_{15}^{(2)}$, $c_{24}^{(2)}$, $c_{25}^{(2)}$, $c_{34}^{(2)}$ and $c_{35}^{(2)}$ -- diverge.
\item
The dipole that is spanned by \emph{both} of the collinear directions -- i.e.~dipole $c_{45}^{(2)}$ -- approaches a constant, despite being physically meaningless in this limit.
\end{itemize}
This behaviour is illustrated in the left panel of Fig.~\ref{fig:regions:qualitative}, which displays the NNLO coefficients $c_{ij}^{(2,C_A)}$ for one representative of each class as a function of the angular separation of the two jets. In particular, we observe that the (12)-dipole (red dots) transitions smoothly into the \mbox{2-jettiness} value, whereas the (25)-dipole (blue dots) is divergent in this limit. Moreover, the (45)-dipole (green dots) also approaches a constant, which is however not the 2-jettiness number. We will see below that this ``pathological'' dipole approaches the \emph{0-jettiness} value instead. The remaining dipoles show the same qualitative behaviour, and the same is true for the $c_{ij}^{(2,n_f)}$ coefficients.

We next turn to the tripole contribution. As we argued in the previous sections, it is not possible to study individual tripoles on the renormalised level, and we instead have to consider the four independent tripole sums of the 3-jettiness soft function. For the following discussion, the specific colour basis in \eqref{eq:3jet:tripoles} is a particularly convenient choice, as it makes the dependence on the two collinear directions $n_4^\mu$ and $n_5^\mu$ explicit. The tripole contribution to the 2-jettiness soft function, on the other hand, only contains a single term as in \eqref{eq:2jet:rentripolesum:ctilde}. If we resolve the same jet (with index 3) using colour conservation, the colour generators of the 2-jettiness soft function can be written in the form $\bfT_{1}^A \;\bfT_{2}^B \;\bfT_{X}^C$, where $\bfT_{X}^C=\bfT_{4}^C+\bfT_{5}^C$ is the colour charge of the merged jet in the collinear limit. In terms of these tripole sums, we then observe the following patterns:
\begin{itemize}
\item
The tripole sums that involve \emph{one} of the collinear directions -- i.e.~$\tilde c_{\rm tripoles}^{(2,124)}$ and $\tilde c_{\rm tripoles}^{(2,125)}$ in our setup -- approach the (N-1)-jettiness limit smoothly. The very fact that the two corresponding sums approach the \emph{same} constant, in particular, allows their colour generators to ``fuse'' into the one of the merged jet.
\item
The tripole sums that involve \emph{both} of the collinear directions -- i.e.~$\tilde c_{\rm tripoles}^{(2,145)}$ and $\tilde c_{\rm tripoles}^{(2,245)}$ -- vanish, as one may have naively expected since there exists no analogue for them in the (N-1)-jettiness calculation.
\item
Moreover, the tripole sums that do \emph{not} involve \emph{either} of the collinear directions -- which only exist for N\,$\geq4$ -- also approach the (N-1)-jettiness limit smoothly.
\end{itemize}
This behaviour is illustrated for the four independent tripole sums of the 3-jettiness soft function in the right panel of Fig.~\ref{fig:regions:qualitative}. Finally, we stress that the tripole contribution is process-dependent and the above pattern holds only for two collinear jet directions, while it is somewhat different if one considers the limit in which the jet becomes collinear to one of the beam directions. We will come back to this latter point in Sec.~\ref{sec:regions:discussion} and~\ref{app:method-of-regions::2jet-tripoles}.

\subsection{NLO dipoles}
\label{sec:regions:NLO}

We will now put these findings onto a solid footing using a dedicated method-of-regions analysis of the limit in which two of the reference vectors $n_i^\mu$ become collinear to each other. To this end, we start with the NLO dipole contribution that we studied in detail in Sec.~\ref{sec:NLO}. Although the NLO dipoles $c_{ij}^{(1)}$ are not divergent, we argued above that their limiting behaviour may not be trivial. The goal of our analysis consists in understanding the origin of this feature and in computing the offset analytically.

\begin{figure}
	\centering
	\includegraphics[width=0.7\columnwidth]{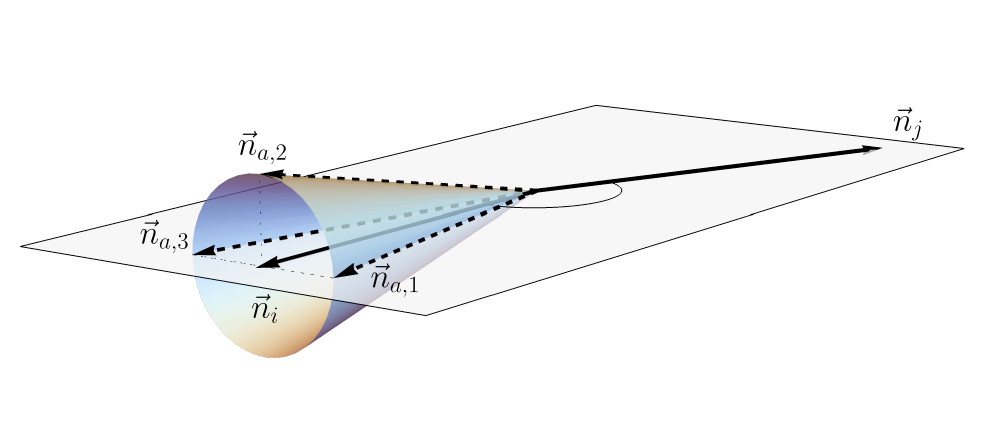}
	\caption{Geometric arrangement for the method-of-regions analysis. The two dipole directions $n_i^\mu=(1,\vec{n}_i)$ and $n_j^\mu=(1,\vec{n}_j)$ are widely separated, and the additional direction $n_a^\mu=(1,\vec{n}_a)$ lies on a cone with a small opening angle of $\mathcal{O}(\sqrt{\delta})$ with respect to the first dipole direction.}
	\label{fig:regions:cone}
\end{figure}

Specifically, we first consider the most interesting configuration -- i.e.~a dipole that involves one of the collinear directions -- and we will come back to the other scenarios at the end of this section. In order to focus on the essential ingredients, it will furthermore be convenient to analyse a subsystem first that is spanned by two dipole directions $n_i^\mu$ and $n_j^\mu$, and the direction of an additional jet $n_a^\mu$ that is almost collinear to $n_i^\mu$. The geometric arrangement of this configuration is illustrated in Fig.~\ref{fig:regions:cone}. Although this setup may resemble a variant of the \mbox{1-jettiness} calculation (with non-back-to-back beams), we stress that it should rather be considered as a building block of the N-jettiness calculation, since it allows us to include further (generic) reference directions $n_b^\mu$ in a straightforward manner as we will see below.

We start by parametrising the reference vectors of this subsystem in the form
\begin{align}
n_i^\mu&=(1,0,0,1)\,,
\nonumber\\
n_j^\mu&=(1,0,\sqrt{n_{ij} (2-n_{ij})},1-n_{ij})\,,
\nonumber\\
n_a^\mu&=(1,2 \sqrt{\delta(1-\delta)} \,s_a, 2\sqrt{\delta(1-\delta)}\,c_a, 1-2\delta),
\label{eq:regions:kinematics}
\end{align}
and we are interested here in the limit $\delta\to 0$. For $n_{ij}=\mathcal{O}(1)$, this corresponds to the situation of two widely separated dipole directions, whereas the non-dipole direction $n_{a}^\mu$ with $n_{ai}=2\delta$ and $n_{aj}=\mathcal{O}(1)$ is almost collinear to the direction $n_i^\mu$. More precisely, the jet lies on a cone with an opening angle of $\mathcal{O}(\sqrt{\delta})$, and its position on this cone is parametrised by its azimuthal angle $\varphi_a$ via its sine $s_a$ and cosine $c_a$ in \eqref{eq:regions:kinematics}. Sample configurations on this cone are shown for $\varphi_a=0$ ($\vec{n}_{a,1}$), $\varphi_a=\pi/2$ ($\vec{n}_{a,2}$) and $\varphi_a=\pi$ ($\vec{n}_{a,3}$) in Fig.~\ref{fig:regions:cone}.

The starting point of our analysis is the master formula \eqref{eq:NLO:master} for the NLO dipole contribution $S_{ij}^{(1)}(\eps)$. In particular, this formula contains integrations over two angular variables that enter the calculation via the projection of the jet direction $n_a^\mu$ onto the basis vectors $e_{\perp_1}^{\mu}$ and $e_{\perp_2}^{\mu}$ in the transverse space. Following the construction described in Sec.~\ref{sec:parametrisations}, we obtain $n_{a\perp_1}= 2\delta \sqrt{\frac{2-n_{ij}}{n_{ij}}}-2 \sqrt{\delta(1-\delta)} \,c_a$ and $n_{a\perp_2}= -2 \sqrt{\delta(1-\delta)} \,s_a$. For the current analysis, however, it will be more convenient to rotate the transverse basis such that the first basis vector $e_{\perp_1}^{\prime \,\mu}$ aligns with $n_{a\perp}^\mu$, which has the advantage that the dependence on the angular variable $\theta_{k_2}$ drops out in \eqref{eq:single-emission:ZX}. In other words, we have $n_{a\perp_2}^\prime=0$ as well as
\begin{align}
n_{a\perp_1}^\prime&=
\sqrt{n_{a\perp_1}^2+n_{a\perp_2}^2}
= 2 \sqrt{\delta} - 2 {\textstyle\sqrt{\frac{2-n_{ij}}{n_{ij}}}} \, c_a \,\delta
+\mathcal{O}(\delta^{3/2})
\label{eq:regions:nJperp}
\end{align}
in the new basis. Note that the leading term in the limit $\delta\to 0$ is independent of both the azimuthal orientation of the jet and the dipole geometry $n_{ij}$.

As long as we focus on the 1-jettiness subsystem, the dependence on the angular variable $\theta_{k_2}$ drops out in the measurement function, and the corresponding integration can thus be trivially performed. It is, in fact, more convenient to start from the representation \eqref{eq:NLO:intermediate} of the NLO dipole contribution, which yields
\begin{align}
S_{ij}^{(1)}(\eps)  &= -\frac{e^{-\gamma_E\eps}}{4^{\eps}\pi}\,
\Gamma(-\eps)\,
\int_0^1 \! dy_k \;\, y_k^{-1+\eps}\,
\int_{-1}^1 \!dc_{k_1} \,s_{k_1}^{-1-2\eps} \,
\Big\{f_A(y_k,\theta_{k_1})^{2\eps}
+ f_B(y_k,\theta_{k_1})^{2\eps}\Big\}\,,
\label{eq:regions:dipoles:master}
\end{align}
with
\begin{align}
f_A(y_k,\theta_{k_1}) &= \min \bigg( 1, \frac{1}{y_k},
\frac{n_{aj}}{n_{ij}} + \frac{n_{ai}}{n_{ij}} \, \frac{1}{y_k}
+ \sqrt{\frac{2}{n_{ij}}}\, \frac{1}{\sqrt{y_k}} \;
n_{a\perp_1}^\prime \, c_{k_1} \bigg)\,,
\nonumber\\
f_B(y_k,\theta_{k_1}) &= \min \bigg( 1, \frac{1}{y_k},
\frac{n_{ai}}{n_{ij}} + \frac{n_{aj}}{n_{ij}} \, \frac{1}{y_k}
+ \sqrt{\frac{2}{n_{ij}}}\, \frac{1}{\sqrt{y_k}} \;
n_{a\perp_1}^\prime \, c_{k_1} \bigg)\,.
\label{eq:regions:dipoles:measure}
\end{align}
This formula is the starting point of our method-of-regions~\cite{Beneke:1997zp} analysis. Specifically, we need to understand which regions contribute in the limit $\delta\to 0$, given that $n_{ij}=\mathcal{O}(1)$, $n_{ai}=\mathcal{O}(\delta)$, $n_{aj}=\mathcal{O}(1)$ and $n_{a\perp_1}^\prime=\mathcal{O}(\sqrt{\delta})$ as argued above. It is, in fact, an easy exercise to show that there are only two scalings that do not lead to scaleless integrals:
\begin{itemize}
  \item
	$y_k=\mathcal{O}(1)$ and $c_{k_1}=\mathcal{O}(1)$, i.e.~the gluon is emitted at generic rapidities and azimuthal angles. We call this region the \emph{base region}.
  \item $y_k=\mathcal{O}(\delta)$ and $c_{k_1}=\mathcal{O}(1)$, i.e.~the gluon is emitted into the forward direction and it may therefore resolve the difference between the $n_{a}^\mu$ and $n_{i}^\mu$ directions. We call this region the \emph{correction region}.
\end{itemize}
At leading power in $\delta\ll 1$, the measurement function in the base region becomes
\begin{align}
f_A^{\rm base}(y_k,\theta_{k_1}) &\,\simeq\, \min \bigg( 1, \frac{1}{y_k},1 \bigg)=1\,,
\nonumber\\
f_B^{\rm base}(y_k,\theta_{k_1}) &\,\simeq\, \min \bigg( 1, \frac{1}{y_k}, \frac{1}{y_k}\bigg)=1\,,
\end{align}
which is precisely the measurement function for 0-jettiness.

The offset we observed in previous sections must therefore be related to the correction region. In this case the integration over the variable $y_k$ extends to infinity, and the measurement function simplifies to
\begin{align}
f_A^{\rm corr}(y_k,\theta_{k_1}) &\,=\, \min \bigg(
\underbrace{1\rule[-3.7mm]{1mm}{0mm}}_{\sim 1}\,,
\underbrace{\frac{1}{y_k}}_{\sim 1/\delta}\,,
\underbrace{\frac{n_{ij}+\mathcal{O}(\sqrt{\delta})}{n_{ij}}}_{\sim 1}
+ \underbrace{\frac{2\delta}{n_{ij}} \, \frac{1}{y_k}}_{\sim 1}
+ \underbrace{\sqrt{\frac{2}{n_{ij}}}\, \frac{2\sqrt{\delta}+\mathcal{O}(\delta)}{\sqrt{y_k}} \;  c_{k_1}}_{\sim 1} \bigg)
\nonumber\\
&\,\simeq\, \min \bigg(
1,1
+ \frac{2\delta}{n_{ij}} \, \frac{1}{y_k}
+ \sqrt{\frac{2}{n_{ij}}}\, \frac{2\sqrt{\delta}}{\sqrt{y_k}} \;  c_{k_1} \bigg)\,,
\nonumber\\
f_B^{\rm corr}(y_k,\theta_{k_1}) &\,=\, \min \bigg(
\underbrace{1\rule[-3.7mm]{1mm}{0mm}}_{\sim 1}\,,
\underbrace{\frac{1}{y_k}}_{\sim 1/\delta}\,,
\underbrace{\frac{2\delta}{n_{ij}}}_{\sim \delta}
+ \underbrace{\frac{n_{ij}+\mathcal{O}(\sqrt{\delta})}{n_{ij}} \, \frac{1}{y_k}}_{\sim1/\delta}
+ \underbrace{\sqrt{\frac{2}{n_{ij}}}\, \frac{2\sqrt{\delta}+\mathcal{O}(\delta)}{\sqrt{y_k}} \;
c_{k_1}}_{\sim 1} \bigg)
\nonumber\\
&\,\simeq\, 1
\,.
\label{eq:regions:measure:corr}
\end{align}
Region B then yields a scaleless integral, whereas the minimisation procedure is non-trivial in region A, since the gluon can resolve the difference between the $n_{a}^\mu$ and $n_{i}^\mu$ directions as anticipated above. The subsequent calculation is straightforward and leads to the integral
\begin{align}
S_{ij}^{(1,\rm{corr})}(\eps)  &= -\frac{e^{-\gamma_E\eps}}{4^{\eps}\pi}\,
\Gamma(-\eps)\bigg(\frac{2\delta}{n_{ij}}\bigg)^\eps \!
\int_{-1}^0 \!\!dc_{k_1} \,s_{k_1}^{-1-2\eps} \!
\int_0^{4c_{k_1}^2} \!\! dx \; x^{-1-\eps}\,
\Big\{ (1+x+2\sqrt{x} \,c_{k_1})^{2\eps} \! -1\Big\}
\end{align}
where we substituted $y_k=\frac{2\delta}{n_{ij}}\frac{1}{x}$ and the non-trivial phase-space boundaries arise from the minimisation procedure in \eqref{eq:regions:measure:corr}. This integral evaluates to
\begin{align}
S_{ij}^{(1,\rm{corr})}(\eps)  &=
\bigg(\frac{2\delta}{n_{ij}}\bigg)^\eps\,
\bigg\{ -\frac{\pi^2}{3} - 4 \zeta_3 \,\eps
-\frac{41 \pi^4}{180}\eps^2 + \mathcal{O}(\eps^3)\bigg\}\,,
\label{eq:regions:NLO:correction}
\end{align}
where we kept higher-order terms in the dimensional regulator that may be required for NNLO calculations. On the level of the renormalised NLO coefficient $c_{ij}^{(1)}$, we thus find that the \mbox{1-jettiness} result approaches the sum of the 0-jettiness value ($\pi^2/2\approx4.935$) that we found in the base region and the correction term ($-\pi^2/3\approx-3.290$) such that $c_{ij}^{(1)}=\pi^2/6\approx 1.645$. This is indeed what is observed in the left panel of Fig.~\ref{fig:dipole:finite:1jet} in the limit $n_{13}\to 0$ (note that this coefficient is multiplied by a factor 10 in that figure).

We recall that this discussion refers to the class of dipoles that involve one of the collinear directions. Before we come to the other classes, let us add some remarks:
\begin{itemize}
  \item
	First of all, it is now an easy task to add further generic directions $n_b^\mu$ with \mbox{$n_{bi}=\mathcal{O}(1)$} and $n_{bj}=\mathcal{O}(1)$ to this setup.
 In the base region, this adds more terms to the measurement function that are independent of $\delta$, and we thus recover the (N-1)-jettiness reference value in the general case. Physically, this means that a gluon at generic rapidities cannot resolve the difference between the two collinear directions, but it sees all the remaining directions in the event. In the correction region, on the other hand, the additional terms in the measurement function are of $\mathcal{O}(1/\delta)$, and they are therefore irrelevant for the minimisation procedure. This corresponds to the situation that the gluon at forward rapidities is blind to all the non-collinear directions. The upshot of this discussion is that the correction term in \eqref{eq:regions:NLO:correction} is \emph{universal} in the sense that it applies to the \mbox{N-jettiness} case as well. This can be verified e.g.~in the second panel of Fig.~\ref{fig:dipole:finite:2jet:cone}, where the corresponding 1-jettiness value for the (13)-dipole is $4.758$, but the curve approaches the value $4.758-3.290\approx1.468$ in the limit $n_{34}\to 0$ (similarly for the (23)-dipole with $2.621-3.290\approx-0.669$).
\item
	Although we did not study next-to-leading power corrections in the method-of-regions analysis systematically, we can make a conjecture here, which is based on the observation that next-to-leading power terms to the measurement function are encoded in $n_{a\perp_1}^\prime$ from \eqref{eq:regions:nJperp} and $n_{aj}= n_{ij} - 2 \sqrt{\delta} \sqrt{n_{ij}(2-n_{ij})} \, c_a +\mathcal{O}(\delta)$. We are thus led to expect that power corrections scale as $\mathcal{O}(\sqrt{\delta})$ for generic configurations, but they are presumably of higher orders for back-to-back dipoles (with $n_{ij}=2$) and for maximally non-planar configurations labelled by $\vec{n}_{a,2}$ in Fig.~\ref{fig:regions:cone} (with $\varphi_a=\pi/2$ and thus $c_a=0$). This behaviour is indeed borne out by the convergence of our numerical results, as we will illustrate explicitly in Sec.~\ref{sec:regions:discussion}.
\end{itemize}

To close this section, we briefly comment on the other two classes of dipoles. If the jet $n_a^\mu$ is not close to the directions forming the dipole $n_i^\mu$ and $n_j^\mu$, but to some other jet $n_b^\mu$, all invariants $n_{ij}$, $n_{ai}$, $n_{aj}$ and $n_{a\perp_1}^\prime$ are of $\mathcal{O}(1)$, whereas the invariant $n_{ab}=\mathcal{O}(\delta)$ is small. As this scalar product does not appear in the measurement function -- see \eqref{eq:single-emission:ZX} -- we trivially obtain the (N-1)-jettiness number in the collinear limit. To put it differently, the phase-space region in the vicinity of the two collinear directions has a negligibly small weight, but this region was enhanced by a collinear singularity in the previous case. For a dipole that does not involve either of the collinear directions, there is no such enhancement and the (N-1)-jettiness limit is therefore smoothly recovered. This can be read off e.g.~from the first panel of Fig.~\ref{fig:dipole:finite:2jet:cone}, where the corresponding 1-jettiness value $2.086$ is approached smoothly for the (12)-dipole in the limit $n_{34}\to 0$.

Finally, for the pathological dipole that is spanned by the directions which themselves are collinear the kinematic invariant $n_{ij}$ is of $\mathcal{O}(\delta)$, whereas all the other invariants are of $\mathcal{O}(1)$. As the inverse of $n_{ij}$ controls the contributions to the measurement function from all terms except for the dipole directions -- see again \eqref{eq:single-emission:ZX} -- the minimisation will always pick one of the dipole contributions in this case. In the base region, this implies that the \emph{0-jettiness} number is recovered, irrespective of the value of N, whereas the correction region yields a scaleless integral. The 0-jettiness number $4.935$ is indeed approached for the (34)-dipole in the limit $n_{34}\to 0$ in the last panel of Fig.~\ref{fig:dipole:finite:2jet:cone}.

\subsection{NNLO dipoles}
\label{sec:regions:NNLO:dipoles}

The extension to the NNLO dipole contribution is comparatively straightforward. First of all, our analysis of the two simple classes of dipoles (without collinear direction or the pathological case with two collinear directions) immediately carries over to NNLO, since the measurement function has the same overall structure in this case. We can therefore concentrate here on the interesting class with one collinear direction, focusing again on the subsystem from Fig.~\ref{fig:regions:cone} to derive a universal correction term to the (N-1)-jettiness reference number. In contrast to the NLO dipoles, however, we will find that the NNLO dipoles diverge in the collinear limit, and we will compute the coefficient that controls this divergence explicitly.

According to the discussion in Sec.~\ref{sec:NNLO}, the NNLO dipole contribution consists of two pieces: mixed real-virtual and double real-emission corrections. The structure of the former is identical to the NLO dipole contribution, and we can immediately follow the steps from the previous section to derive the corresponding correction term,
\begin{align}
S_{ij}^{(2,\text{Re},\rm{corr})}(\eps)  &=
\bigg(\frac{2\delta}{n_{ij}}\bigg)^{2\eps}\,
\bigg\{ \frac{2\pi^2}{3\eps^2} + \frac{8 \zeta_3}{\eps}
+\frac{76 \pi^4}{45} + \mathcal{O}(\eps)\bigg\}\,.
\end{align}
For the double real-emission contribution, on the other hand, the general principles of our analysis also apply. In particular, we again find two regions which are distinguished by the scaling of the rapidity-type variable $y$, whereas all remaining integration variables that parametrise the two-particle phase space are of $\mathcal{O}(1)$. The scaling $y=\mathcal{O}(1)$ in the base region then yields the (N-1)-jettiness number, and the non-trivial correction term is encoded in the second region with $y=\mathcal{O}(\delta)$. The evaluation of the correction term is, however, more involved in this case due to the higher-dimensional integrations and the fact that the measurement function consists of two terms, each of which comes with its own minimisation prescription, see \eqref{eq:M2:Njettiness}. Without going into further details here, we note that the integrals can be brought into a form that can be evaluated numerically with the public version of \texttt{SoftSERVE}. Using this strategy, we find
\begin{align}
S_{ij}^{(2,q\bar q,\rm{corr})}(\eps)  &=
\bigg(\frac{2\delta}{n_{ij}}\bigg)^{2\eps}\,
\bigg\{ \frac{4\pi^2}{9\eps}+16.4445(2)+\mathcal{O}(\eps)\bigg\}\,,
\nonumber\\
S_{ij}^{(2,gg,\rm{corr})}(\eps)  &=
\bigg(\frac{2\delta}{n_{ij}}\bigg)^{2\eps}\,
\bigg\{ - \frac{2\pi^2}{3\eps^2} - \bigg( \frac{11\pi^2}{9} + 8 \zeta_3 \bigg) \frac{1}{\eps}
-190.166(2) + \mathcal{O}(\eps)\bigg\}\,,
\end{align}
where we restored the (known) pole terms analytically. This then translates into a correction to the renormalised NNLO dipole coefficients of the form
\begin{align}
c_{ij}^{(2,n_f,\rm{corr})}  &=
\frac{4 \pi^2}{9}  \ln \Big( \frac{2 \delta}{n_{ij}}\Big) + 10.0335(2)\,,
\nonumber\\
c_{ij}^{(2,C_A,\rm{corr})}  &=
-\frac{11 \pi^2}{9}  \ln \Big( \frac{2 \delta}{n_{ij}}\Big) -8.023(2)\,,
\label{eq:regions:NNLO:dipoles:cij}
\end{align}
which is divergent in the limit $\delta\to 0$ as anticipated. In particular, this explains the asymptotic behaviour of the (13) and (23)-dipoles in Fig.~\ref{fig:dipole:finite:2jet:cone}, for which the green (blue) dots tend to $-\infty$ ($+\infty$) in the limit $n_{34}=2\delta\to 0$. The (12)-dipole, on the other hand, smoothly approaches the corresponding 1-jettiness number $-26.18$ ($57.04$) for the green (blue) dots, and the (34)-dipole does so for the 0-jettiness result $-21.70$ ($28.25$).

\subsection{NNLO tripoles}
\label{sec:regions:tripoles}

\begin{figure}[t]
	\begin{subfigure}[b]{0.19\textwidth}
		\centering
		\begin{tikzpicture}
		\draw[line width=1pt,Cyan,-stealth](0,0)--(1,0)
		node[anchor= south]{$~$};
		\draw[line width=1pt,Cyan,-stealth](0,0)--(-1,0)
		node[anchor= south]{$~$};
		\draw[line width=1pt,LimeGreen,-stealth](0,0)--(0,-1)
		node[anchor=west]{$~$};
		\draw[line width=1pt,black,-stealth](0,0)--(.1,0.9)
		node[anchor=south west]{$~$};
		\draw[line width=1pt,black,-stealth](0,0)--(-.1,0.9)
		node[anchor=south east]{$~$};
		\end{tikzpicture}
		\caption*{(I)}
		\label{fig:I}
	\end{subfigure}
	\hfill
	\begin{subfigure}[b]{0.19\textwidth}
		\centering
		\begin{tikzpicture}
		\draw[line width=1pt,Cyan,-stealth](0,0)--(1,0)
		node[anchor= south]{$~$};
		\draw[line width=1pt,Cyan,-stealth](0,0)--(-1,0)
		node[anchor= south]{$~$};
		\draw[line width=1pt,black,-stealth](0,0)--(0,-1)
		node[anchor=west]{$~$};
		\draw[line width=1pt,LimeGreen,-stealth](0,0)--(.1,0.9)
		node[anchor=west]{$~$};
		\draw[line width=1pt,black,-stealth](0,0)--(-.1,0.9)
		node[anchor=south east]{$~$};
		\end{tikzpicture}
		\caption*{(IIa)}
		\label{fig:IIa}
	\end{subfigure}
	\hfill
	\begin{subfigure}[b]{0.19\textwidth}
		\centering
		\begin{tikzpicture}
		\draw[line width=1pt,Cyan,-stealth](0,0)--(1,0)
		node[anchor= south]{$\vec{n}_j$};
		\draw[line width=1pt,black,-stealth](0,0)--(-1,0)
		node[anchor= south]{$~$};
		\draw[line width=1pt,LimeGreen,-stealth](0,0)--(0,-1)
		node[anchor=west]{$~$};
		\draw[line width=1pt,Cyan,-stealth](0,0)--(.1,0.9)
		node[anchor=west]{$\vec{n}_i$};
		\draw[line width=1pt,black,-stealth](0,0)--(-.1,0.9)
		node[anchor=south east]{$~$};
		\end{tikzpicture}
		\caption*{(IIb1)}
		\label{fig:IIb}
	\end{subfigure}
	\hfill
	\begin{subfigure}[b]{0.19\textwidth}
		\centering
		\begin{tikzpicture}
		\draw[line width=1pt,Cyan,-stealth](0,0)--(1,0)
		node[anchor= south]{$\vec{n}_j$};
		\draw[line width=1pt,black,-stealth](0,0)--(-1,0)
		node[anchor= south]{$~$};
		\draw[line width=1pt,black,-stealth](0,0)--(0,-1)
		node[anchor=west]{$~$};
		\draw[line width=1pt,Cyan,-stealth](0,0)--(.1,0.9)
		node[anchor=west]{$\vec{n}_i$};
		\draw[line width=1pt,LimeGreen,-stealth](0,0)--(-.1,0.9)
		node[anchor=east]{$~$};
		\end{tikzpicture}
		\caption*{(IIIa1)}
		\label{fig:IIIa}
	\end{subfigure}
	\hfill
	\begin{subfigure}[b]{0.19\textwidth}
		\centering
		\begin{tikzpicture}
		\draw[line width=1pt,LimeGreen,-stealth](0,0)--(1,0)
		node[anchor= south]{$~$};
		\draw[line width=1pt,black,-stealth](0,0)--(-1,0)
		node[anchor= south]{$~$};
		\draw[line width=1pt,black,-stealth](0,0)--(0,-1)
		node[anchor=west]{$~$};
		\draw[line width=1pt,Cyan,-stealth](0,0)--(.1,0.9)
		node[anchor=west]{$~$};
		\draw[line width=1pt,Cyan,-stealth](0,0)--(-.1,0.9)
		node[anchor=east]{$~$};
		\end{tikzpicture}
		\caption*{(IIIb)}
		\label{fig:IIIb}
	\end{subfigure}
	\caption{Different classes of tripole contributions. In each class there are two directions that are almost collinear to each other (pointing upwards). The decomposition vectors $n_i^\mu$ and $n_j^\mu$ are shown in blue, the tripole vector $n_l^\mu$ in green, and further generic directions in black. The classes (IIb2) and (IIIa2) are obtained by interchanging $n_i^\mu$ and $n_j^\mu$ in (IIb1) and (IIIa1), respectively.}
	\label{fig:regions:tripoles}
\end{figure}
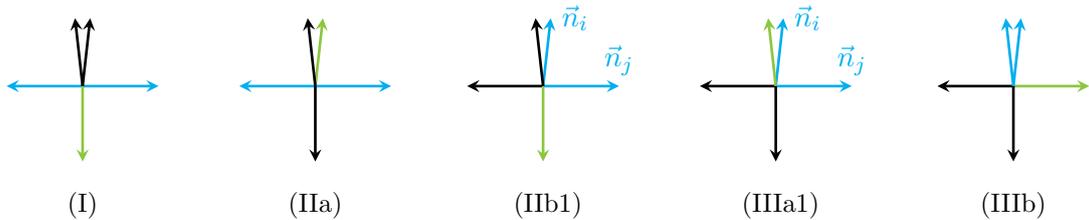

Although the master formula for individual tripoles $S_{ilj}^{(2,{\rm Im})}(\eps)$ in \eqref{eq:NNLO:RV:tripoles:master} is structurally similar to the one of the NLO dipoles, the analysis of the tripole contribution turns out to be significantly more involved. This is primarily related to the fact that the tripoles are not symmetric under the exchange of any pair of indices, and we therefore have to distinguish more cases, depending on whether one, two or none of the indices belong to the collinear directions. As we argued in \mbox{Sec.~\ref{sec:renormalisation}}, moreover, the tripoles do not renormalise individually, and instead have to be summed up to understand the different patterns that were shown for the 3-jettiness soft function in the right panel of Fig.~\ref{fig:regions:qualitative}.

In order to set up the notation that will be used throughout this section, we recall that the first two indices $(i,l)$ belong to the directions that enter the relevant loop integral, whereas the indices $(i,j)$ refer to the directions that are used for the Sudakov decomposition and hence the phase-space parametrisation. We will highlight these properties by calling $n_i^\mu$ and $n_j^\mu$ the \emph{decomposition vectors}, and $n_l^\mu$ the \emph{tripole vector} in this section. As long as we focus on individual tripoles, we thus need to analyse the object $S_{ilj}^{(2,{\rm Im})}(\eps)$ for various cases of $n_i^\mu$, $n_l^\mu$ and $n_j^\mu$ being involved in the collinear directions (or not), while keeping the positions of the indices fixed.

Specifically, there are seven different tripole classes that are illustrated in Fig.~\ref{fig:regions:tripoles}:
\begin{itemize}
	\item[(I)]
	None of the decomposition and tripole vectors belongs to the collinear directions.
	\item[(IIa)]
	The tripole vector $n_l^\mu$ is one of the collinear directions, and the decomposition vectors are chosen from the remaining directions.
	\item[(IIb1)]
	The decomposition vector $n_i^\mu$ is one of the collinear directions, and the second decomposition vector and the tripole vector are chosen from the remaining directions.
	\item[(IIb2)]
	Similar to the previous case with the roles of $n_i^\mu$ and $n_j^\mu$ interchanged.
	\item[(IIIa1)]
	The decomposition vector $n_i^\mu$ and the tripole vector $n_l^\mu$ form the collinear pair, and the second decomposition vector is among the remaining directions.
	\item[(IIIa2)]
	Similar to the previous case with the roles of $n_i^\mu$ and $n_j^\mu$ interchanged.
	\item[(IIIb)]
	The decomposition vectors $n_i^\mu$ and $n_j^\mu$ form the collinear pair, and the tripole vector is among the remaining directions.
\end{itemize}
The simplest case for which all these classes exist is the 3-jettiness, which has five external directions. In this case the 60 different 3-jettiness tripoles split into  6 (class I), 12 (class IIa), 12 (class IIb1), 12 (class IIb2), 6 (class IIIa1), 6 (class IIIa2) and 6 (class IIIb). We will now analyse each of these classes in turn:

\paragraph{Class I:}
The discussion is particularly simple for the tripoles that involve two generic collinear directions $n_a^\mu$ and $n_b^\mu$, i.e.~$n_{ab}=\mathcal{O}(\delta)$. As the phase-space region in the vicinity of these directions is not enhanced by a collinear divergence, we trivially recover the (N-1)-jettiness number in this case. This is similar to what we have seen for the dipoles that do not involve either of the collinear directions in the previous section.

\paragraph{Class IIa:}
The same argument holds for the tripoles that involve the tripole vector $n_l^\mu$  and a generic vector $n_a^\mu$ among the collinear directions, i.e.~$n_{al}=\mathcal{O}(\delta)$. As the decomposition is based on the vectors $n_i^\mu$ and $n_j^\mu$, all scalar products in the factor $Z_X(y_k, \theta_{k_1}, \theta_{k_2})$ from \eqref{eq:single-emission:ZX} involve at least one of the $(i,j)$ labels, and the scalar product $n_{al}$ therefore does not appear explicitly in the calculation. We thus again recover the (N-1)-jettiness number.

\paragraph{Class IIb1:}
The situation is more interesting for the tripoles that involve at least one of the decomposition vectors among the collinear directions. Similar to the dipole analysis, the collinear divergence may then enhance the contribution from forward emissions, which induces a correction term to the naive (N-1)-jettiness result.

The starting point of the tripole analysis is the formula \eqref{eq:NNLO:RV:tripoles:master}, which yields
\begin{align}
\label{eq:regions:tripoles:master}
&S_{ilj}^{(2,{\rm Im})}(\eps)  =  \frac{4\,e^{-2\gamma_E\eps}}{4^{\eps}\pi}\, \frac{1}{\eps^2} \,
\frac{\Gamma^3(1-\eps)\Gamma^2(1+ \eps)\Gamma(-4\eps)\Gamma(-\eps)\sin(\pi\eps) }{\Gamma(1-2 \eps)\Gamma(-2\eps)}\,
\int_0^1 \! dy_k \,
\int_{-1}^1 \!dc_{k_1} \,s_{k_1}^{-1-2\eps} \,
\nonumber \\
&\;
\times
\bigg\{y_k^{-1+2\eps}\, \bigg( \frac{n_{il}}{n_{ij} \,y_k\, Z_l(y_k,\theta_{k_1})} \bigg)^\eps
f_A(y_k,\theta_{k_1})^{4\eps}
+ y_k^{-1+3\eps}\, \bigg( \frac{n_{il}}{n_{ij} \,Z_l(1/y_k,\theta_{k_1})} \bigg)^\eps
f_B(y_k,\theta_{k_1})^{4\eps}\bigg\}
\end{align}
after integrating out the second angular variable, which is possible because of the particular basis choice in the transverse space that we introduced earlier. As anticipated, this expression has a similar structure as the NLO dipole formula from \eqref{eq:regions:dipoles:master}, except for a factor
\begin{align}
Z_l(y_k, \theta_{k_1}) &=
 \frac{n_{lj}}{n_{ij}} + \frac{n_{il}}{n_{ij}} \, \frac{1}{y_k}
+ \sqrt{\frac{2}{n_{ij}}}\, \frac{1}{\sqrt{y_k}} \, n_{l\perp_1}^\prime c_{k_1}\,,
\end{align}
outside the measurement function, which is not affected by the minimisation procedure. The measurement function, on the other hand, is again given by \eqref{eq:regions:dipoles:measure}.

The evaluation of the tripole contribution then follows along the same lines as for the NLO dipoles, and we in particular find the same contributing regions. In the base region with $y_k=\mathcal{O}(1)$, we then again recover the (N-1)-jettiness number, and the region with $y_k=\mathcal{O}(\delta)$ yields a non-trivial correction term. Specifically, the tripoles of class IIb1 assume the counting $n_{ai}=\mathcal{O}(\delta)$ and $n_{a\perp_1}^\prime=\mathcal{O}(\sqrt{\delta})$ for one of the other hard emitters $n_a^\mu$, whereas all remaining scalar products are of $\mathcal{O}(1)$. In the correction region the measurement function then reduces to \eqref{eq:regions:measure:corr}, whereas the additional factors in \eqref{eq:regions:tripoles:master} become
\begin{align}
\frac{n_{il}}{n_{ij} \,y_k\, Z_l(y_k,\theta_{k_1})} &\,\simeq\,1\,, &
\frac{n_{il}}{n_{ij} \,Z_l(1/y_k,\theta_{k_1})} &\,\simeq\, \frac{n_{il}}{n_{lj}}\,.
\end{align}
As we have seen above for the NLO dipoles, the non-trivial correction term then emerges in region A, whereas region B yields a scaleless integral. Following the subsequent steps that were discussed in Sec.~\ref{sec:regions:NLO} leads to the result,
\begin{align}
S_{ilj,\rm{(IIb1)}}^{(2,{\rm Im,\rm{corr}})}(\eps)   &= \pi\,
\bigg(\frac{2\delta}{n_{ij}}\bigg)^{2\eps}\,
\bigg\{ \frac{4\pi^2}{3\eps} + 16 \zeta_3 + \mathcal{O}(\eps)\bigg\}\,.
\label{eq:regions:tripoles:IIb1}
\end{align}
Note that the correction is again \emph{universal} and that it depends on the tripole kinematics only via the factor $n_{ij}$, which in fact cancels against the prefactor in \eqref{eq:NNLO:RV}.

\paragraph{Class IIb2:}
The calculation proceeds similarly for the tripoles of class IIb2, as only the role of the decomposition vectors $n_i^\mu$ and $n_j^\mu$ needs to be interchanged. As a result, the non-trivial correction term to the (N-1)-jettiness number now arises in region B, and since \eqref{eq:regions:tripoles:master} is not symmetric in the two regions, the result takes a slightly different form. Explicitly, we now find
\begin{align}
S_{ilj,\rm{(IIb2)}}^{(2,{\rm Im,\rm{corr}})}(\eps)   &= \pi\,
\bigg(\frac{2\delta}{n_{ij}}\bigg)^{2\eps}\,
\bigg(\frac{2\delta \,n_{il}}{n_{ij}\,n_{lj}}\bigg)^{\eps}\,
\bigg\{ \frac{4\pi^2}{3\eps} + 32 \zeta_3 + \mathcal{O}(\eps)\bigg\}\,,
\label{eq:regions:tripoles:IIb2}
\end{align}
which has a slightly more involved dependence on the tripole kinematics that will become important below when we discuss the tripole sum.

\paragraph{Remaining classes:}
The tripoles of the remaining classes share the common characteristic that they do not have an (N-1)-jettiness analogue, because they involve \emph{both} collinear directions among their indices. While their evaluation proceeds along the same lines as in the previous cases, we refrain from discussing them here (some of them will be computed explicitly in~\ref{app:method-of-regions::2jet-tripoles}), and we will instead only focus on their kinematic dependence. Explicitly, we find that the tripoles of class (IIIa1) are symmetric under $n_i^\mu \leftrightarrow n_l^\mu$ exchange in the collinear limit, whereas the same is true for the tripoles of class (IIIa2) under $n_j^\mu \leftrightarrow n_l^\mu$ exchange. Finally, the tripoles of class (IIIb) are the analogue of the pathological dipoles that we discussed earlier, and they are found to give a universal contribution that is independent of the tripole kinematics. 
\bigskip

Having understood the contributions from individual tripoles, we can now combine them to construct the bare tripole sum. As our goal consists in understanding the patterns that were shown in the right panel of Fig.~\ref{fig:regions:qualitative}, we consider the same setup here, i.e.~we focus on the 3-jettiness in the limit in which the two jets with reference vectors $n_4^\mu$ and $n_5^\mu$ become collinear to each other. Choosing furthermore the same colour basis as in \eqref{eq:3jet:tripoles}, we obtain
\begin{align}
& \sum_{i\neq l\neq j} (\lambda_{il} - \lambda_{ip}-\lambda_{lp})\,
f_{ABC}\; \bfT_{i}^A \;\bfT_{l}^B \;\bfT_{j}^C  \;
\tilde{S}_{ilj}
\nonumber \\
&\quad 
=(-\I{\tilde{S}_{1 2 3}} + \IIb{\tilde{S}_{1 2 4}} - \I{\tilde{S}_{1 3 2}} + \IIb{\tilde{S}_{1 3 4}} + \IIa{\tilde{S}_{1 4 2}} - \IIa{\tilde{S}_{1 4 3}} + \I{\tilde{S}_{2 1 3}} - \IIb{\tilde{S}_{2 1 4}} 
+ \I{\tilde{S}_{2 3 1}} -  \IIb{\tilde{S}_{2 3 4}} 
\nonumber \\ 
&\qquad\;\,
- \IIa{\tilde{S}_{2 4 1}} + \IIa{\tilde{S}_{2 4 3}} +  \I{\tilde{S}_{3 1 2}} - \IIb{\tilde{S}_{3 1 4}} - \I{\tilde{S}_{3 2 1}} +  \IIb{\tilde{S}_{3 2 4}} 
+ \IIa{\tilde{S}_{3 4 1}} - \IIa{\tilde{S}_{3 4 2}} -  \IIb{\tilde{S}_{4 1 2}} + \IIb{\tilde{S}_{4 1 3}} 
\nonumber\\ 
&\qquad\;\, 
+ \IIb{\tilde{S}_{4 2 1}} -  \IIb{\tilde{S}_{4 2 3}} - \IIb{\tilde{S}_{4 3 1}} + \IIb{\tilde{S}_{4 3 2}})
\;f_{ABC}\;\bfT_1^A\; \bfT_2^B\;\bfT_4^C
\nonumber\\ &
\quad +
(-\I{\tilde{S}_{1 2 3}} + \IIb{\tilde{S}_{1 2 5}} - \I{\tilde{S}_{1 3 2}} + \IIb{\tilde{S}_{1 3 5}} + \IIa{\tilde{S}_{1 5 2}} - \IIa{\tilde{S}_{1 5 3}} +  \I{\tilde{S}_{2 1 3}} - \IIb{\tilde{S}_{2 1 5} }
+ \I{\tilde{S}_{2 3 1}} - \IIb{\tilde{S}_{2 3 5}} 
\nonumber\\ 
&\qquad\;\,
- \IIa{\tilde{S}_{2 5 1}} + \IIa{\tilde{S}_{2 5 3}} +  \I{\tilde{S}_{3 1 2}} - \IIb{\tilde{S}_{3 1 5}} - \I{\tilde{S}_{3 2 1}} +  \IIb{\tilde{S}_{3 2 5} }
+ \IIa{\tilde{S}_{3 5 1}} - \IIa{\tilde{S}_{3 5 2}} - \IIb{\tilde{S}_{5 1 2}} + \IIb{\tilde{S}_{5 1 3}} 
\nonumber\\ 
&\qquad\;\, 
+ \IIb{\tilde{S}_{5 2 1}} -  \IIb{\tilde{S}_{5 2 3}} - \IIb{\tilde{S}_{5 3 1}} + \IIb{\tilde{S}_{5 3 2}})
\;f_{ABC}\;\bfT_1^A\; \bfT_2^B\;\bfT_5^C
\nonumber\\ & 
\quad +
(- \IIb{\tilde{S}_{1 3 4}} + \IIb{\tilde{S}_{1 3 5}} + \IIa{\tilde{S}_{1 4 3}} -  \IIIa{\tilde{S}_{1 4 5}} - \IIa{\tilde{S}_{1 5 3}} + \IIIa{\tilde{S}_{1 5 4}} +  \IIb{\tilde{S}_{3 1 4}} - \IIb{\tilde{S}_{3 1 5}} - \IIa{\tilde{S}_{3 4 1}} +  \IIIa{\tilde{S}_{3 4 5}} \nonumber \\ 
&\qquad\;\,
+ \IIa{\tilde{S}_{3 5 1}} - \IIIa{\tilde{S}_{3 5 4}} -  \IIb{\tilde{S}_{4 1 3}} + \IIIb{\tilde{S}_{4 1 5}} + \IIb{\tilde{S}_{4 3 1}} -  \IIIb{\tilde{S}_{4 3 5}} - \IIIa{\tilde{S}_{4 5 1}} + \IIIa{\tilde{S}_{4 5 3}} +  \IIb{\tilde{S}_{5 1 3}} - \IIIb{\tilde{S}_{5 1 4}} 
\nonumber\\ 
&\qquad\;\, 
- \IIb{\tilde{S}_{5 3 1}} +  \IIIb{\tilde{S}_{5 3 4}} + \IIIa{\tilde{S}_{5 4 1}} - \IIIa{\tilde{S}_{5 4 3}})
\;f_{ABC}\;\bfT_1^A\; \bfT_4^B\;\bfT_5^C
\nonumber\\ &
\quad +
(- \IIb{\tilde{S}_{2 3 4}} + \IIb{\tilde{S}_{2 3 5}} + \IIa{\tilde{S}_{2 4 3}} -  \IIIa{\tilde{S}_{2 4 5}} - \IIa{\tilde{S}_{2 5 3}} + \IIIa{\tilde{S}_{2 5 4}} +  \IIb{\tilde{S}_{3 2 4}} - \IIb{\tilde{S}_{3 2 5} } - \IIa{\tilde{S}_{3 4 2}} +  \IIIa{\tilde{S}_{3 4 5}} 
\nonumber\\ &
\qquad\;\, 
+ \IIa{\tilde{S}_{3 5 2}} - \IIIa{\tilde{S}_{3 5 4}} -  \IIb{\tilde{S}_{4 2 3}} + \IIIb{\tilde{S}_{4 2 5}} + \IIb{\tilde{S}_{4 3 2}} -  \IIIb{\tilde{S}_{4 3 5} }- \IIIa{\tilde{S}_{4 5 2}} + \IIIa{\tilde{S}_{4 5 3}} + \IIb{\tilde{S}_{5 2 3}} - \IIIb{\tilde{S}_{5 2 4}} 
\nonumber\\ 
&\qquad\;\, 
- \IIb{\tilde{S}_{5 3 2}} + \IIIb{\tilde{S}_{5 3 4}} + \IIIa{\tilde{S}_{5 4 2}} - \IIIa{\tilde{S}_{5 4 3}})
\;f_{ABC}\;\bfT_2^A\; \bfT_4^B\;\bfT_5^C\,.
\label{eq:regions:tripolesum}
\end{align}
where we introduced the short-hand notation $\tilde{S}_{ilj}=(n_{ij}/2)^{2\eps}\,S_{ilj}^{(2,{\rm Im})}(\eps)$, recombining the tripoles with the kinematic factor from \eqref{eq:NNLO:RV}, and we indicated the different tripole classes using a colour coding. Specifically, we highlight class I (\I{dark green}), class IIa (\IIa{light green}), class IIb (\IIb{orange}), class IIIa (\IIIa{red}) and class IIIb (\IIIb{blue}), and we refrain from splitting classes IIb and IIIa further to keep the number of colours small. The respective subclasses can easily be inferred from the order of indices.

Before we start interpreting this formula, let us remind ourselves of the patterns that we observed in Fig.~\ref{fig:regions:qualitative}. There we saw that the coefficients of the first two colour structures merge smoothly into the 2-jettiness number, whereas the ones of the last two colour structures vanish in the collinear limit. Although this observation was made in Sec.~\ref{sec:regions:qualitative} on the level of renormalised coefficients, we will now see that the same pattern arises in the bare tripole sum in \eqref{eq:regions:tripolesum}. 

We start with the first colour structure, which is supposed to have a smooth 2-jettiness limit. The 2-jettiness itself has 24 tripoles that combine exactly in the form specified in \eqref{eq:regions:tripolesum}. We furthermore note that the contributions from classes IIIa (red) and IIIb (blue) are absent in this case, and we thus only need to combine tripoles from the classes that we discussed explicitly above. There we saw that the base region always reproduces the (N-1)-jettiness number, and we hence only need to show that the correction terms from \eqref{eq:regions:tripoles:IIb1} and \eqref{eq:regions:tripoles:IIb2} vanish in the sum. It is thus sufficient to focus on the orange terms in the first colour structure of \eqref{eq:regions:tripolesum}.  For the tripoles of class IIb1 (of type $4lj$) the kinematic $n_{4j}$-dependence in \eqref{eq:regions:tripoles:IIb1}  cancels against the one that is implicit in the $\tilde{S}_{4lj}$ notation. The respective correction terms are therefore independent of the tripole kinematics, and it is easy to verify that they cancel in the sum of the six IIb1 tripoles, since they contribute with an equal number of plus and minus signs.  The situation is similar for the IIb2 tripoles (of type $il4$), except that the $\tilde{S}_{il4}$ have a residual kinematic dependence due to the additional $(n_{il}/n_{i4}n_{l4})^\eps$ factor in \eqref{eq:regions:tripoles:IIb2}. This factor is, however, symmetric under $i \leftrightarrow l$ exchange, and the correction terms therefore now cancel pairwise in the sum. The same argument holds, of course, for the second colour structure, whose coefficient also approaches the 2-jettiness number.

We next consider the third colour structure in \eqref{eq:regions:tripolesum}, which we expect to vanish in the collinear limit. Starting with the IIa tripoles (light green), we immediately see that they cancel pairwise in the sum, since the $n_4^\mu$ and $n_5^\mu$ directions cannot be distinguished in the collinear limit. The same is true for the IIb tripoles (orange), for which the pairwise cancellation holds for both the base and the correction region. As we argued above, the IIIa tripoles (red) are either symmetric in the first two indices (IIIa1) or in the last two indices (IIIa2), and in either case we can identify tripole pairs that cancel exactly. Finally, the tripoles of class IIIb (blue) only depend on the kinematics via the $n_{ij}$ factor implicit in $\tilde{S}_{ilj}$, and they are thus symmetric under the exchange of the two collinear directions. Once again, we observe that these tripoles cancel out in pairs, and we can thus conclude that the coefficient of the third colour structure in \eqref{eq:regions:tripolesum} indeed vanishes in the collinear limit. Similar arguments hold for the fourth colour structure as well.

While this discussion refers to the bare tripole sum, we show in~\ref{app:method-of-regions:3jet-tripoles} that it carries over to the renormalised coefficients as long as the $\tilde c_{ijk}^{(2)}$ notation is used in the sum rather than the $c_{ijk}^{(2)}$. We also note that the analysis is slightly different if one considers the limit when one of the jets is made collinear to the one of beam directions, because of the process-dependent sign factors in \eqref{eq:regions:tripolesum}. We refrain from discussing these configurations explicitly for the 3-jettiness here, and will instead come back to this point in the following section.

\subsection{Discussion}
\label{sec:regions:discussion}

Having gained a better understanding of the asymptotic behaviour of the soft-function coefficients in the limit when two of the reference directions $n_i^\mu$ become collinear to each other, we will now re-examine some of the plots we have shown in Sec.~\ref{sec:results}. Specifically, we consider the 1-to-0 jettiness transition (Fig.~\ref{fig:1jet:comparison}), as well as some aspects of the 2-to-1 jettiness (Fig.~\ref{fig:tripole:finite:2jet:planar}) and the 2-to-0 jettiness transition (Fig.~\ref{fig:tripole:2jet:b2b}).

\begin{figure*}[t!]
\centering{
 \includegraphics[width=0.32\textwidth]{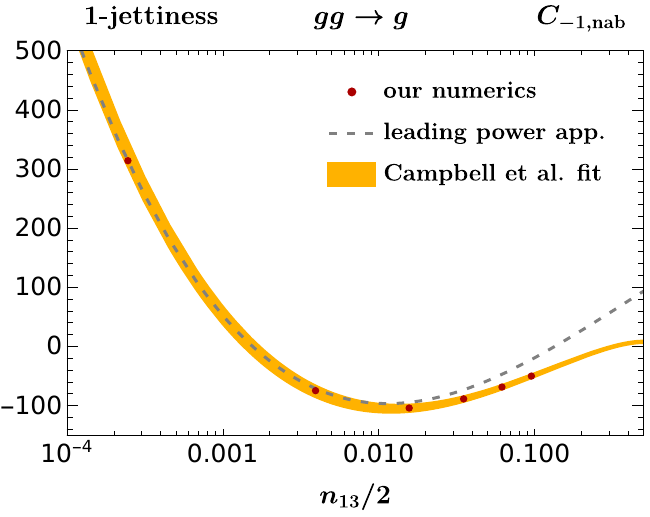}\hspace{1mm}
 \includegraphics[width=0.32\textwidth]{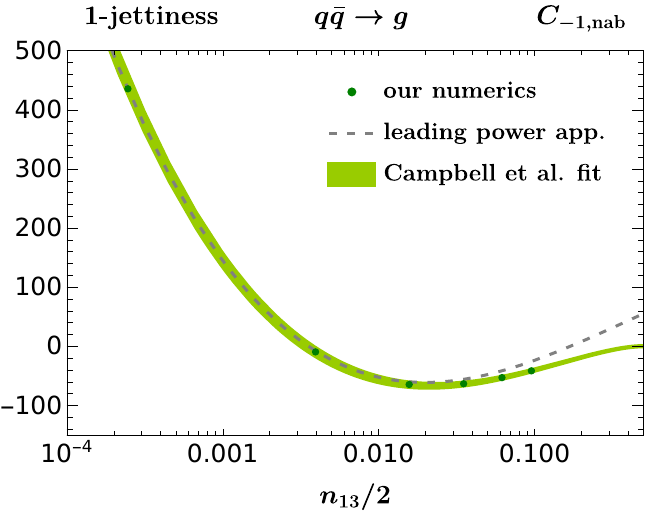}\hspace{1mm}
 \includegraphics[width=0.32\textwidth]{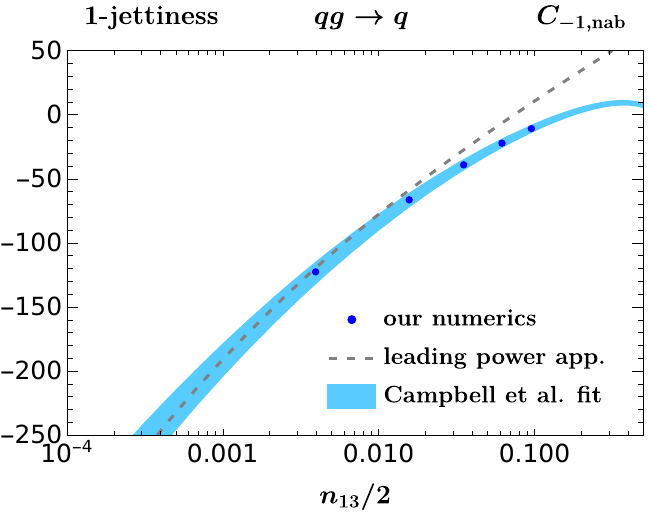}\vspace{2mm}
 \includegraphics[width=0.32\textwidth]{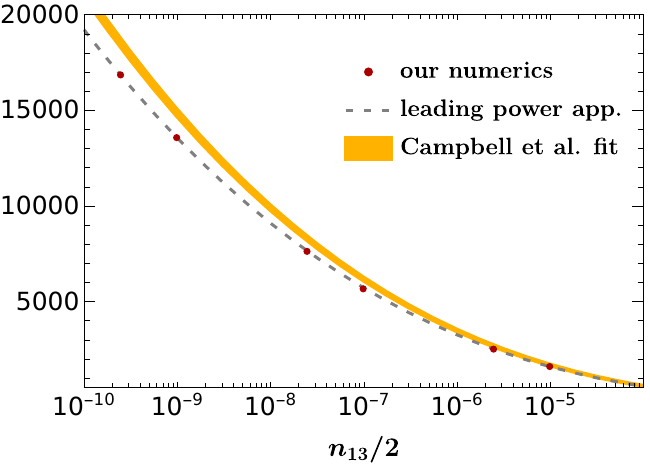}\hspace{1mm}
 \includegraphics[width=0.32\textwidth]{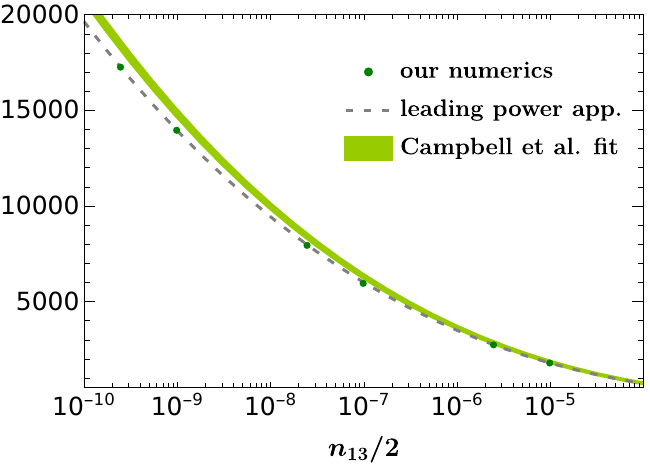}\hspace{1mm}
 \includegraphics[width=0.32\textwidth]{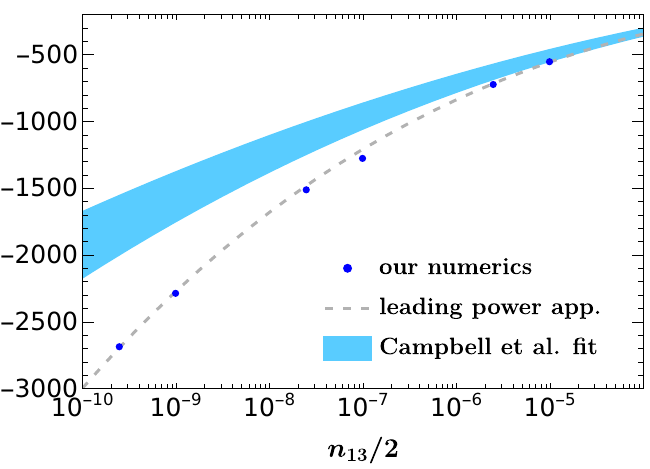}
 } 
\caption{Comparison of the $C_{-1,nab}$ coefficients of the 1-jettiness soft function in distribution space. The plots refer to the $gg\to g$ (left), $q\bar q\to g$ (middle) and $qg\to q$ channel (right) for two different regions of $n_{13}$ (top and bottom). The dots represent our numerical results, the bands the fits from~\cite{Campbell:2017hsw}, and the dashed lines the leading-power approximation from \eqref{eq:1JN:Cm1:LP}.}
\label{fig:1jet:comparison:Cm1nab:LP}
\end{figure*}

\paragraph{1-jettiness:} 
We start by analysing the endpoint region $n_{13}\to 0$ of the 1-jettiness dipole coefficients in Fig.~\ref{fig:1jet:comparison} for all partonic channels. As mentioned earlier, the authors of \cite{Campbell:2017hsw} fitted their numerical results to an ad-hoc functional form consisting of logarithms of the kinematic invariants up to the third power. In contrast to that, we can now derive the exact form of the coefficients in the limit $n_{13}\ll 1$ at leading power. Explicitly, we find
\begin{align}
 C_{-1,nab}^{gg\to g} &=
 -\frac{23}{12}\,L^3+ \left(\frac{3\pi^2}{4} -\frac{151}{12}\right) L^2 + \left(\frac{63\zeta_3}{2}  + \frac{23\pi^2}{4}-\frac{233}{9}\right)L + 142.689(9) + \,\ldots
 \nonumber\\
 C_{-1,nab}^{q\bar q\to g} &=
 -\frac{23}{12}\,L^3+ \left(\frac{3\pi^2}{4} -\frac{151}{12}\right) L^2 + \left(\frac{63\zeta_3}{2}  + \frac{391\pi^2}{108}-\frac{233}{9}\right)L + 90.1233(4) + \,\ldots
 \nonumber\\
 C_{-1,nab}^{qg\to q} &=
 \frac{23}{108}\,L^3+\left(-\frac{\pi^2}{12}+\frac{151}{108}\right)L^2+\left(-\frac{7\zeta_3}{2}+\frac{391}{108}\pi^2+\frac{233}{81}\right)L + 89.2769(9) + \,\ldots
 \label{eq:1JN:Cm1:LP}
\end{align}
where $L=\ln(\frac{n_{13}}{2})$ and the dots indicate terms of $\mathcal{O}(\sqrt{n_{13}})$. These expressions are displayed together with the fit functions from~\cite{Campbell:2017hsw} as well as our own data in Fig.~\ref{fig:1jet:comparison:Cm1nab:LP}. On the one hand, the upper plots illustrate that our numerical results (dots) agree with the fits (bands) for values of $n_{13}\gtrsim 10^{-4}$, whereas the leading-power approximation (dashed) asymptotes to these results for $n_{13}\lesssim 10^{-2}$. In the lower plots, on the other hand, we can observe that the fits start to deviate from the leading-power expressions for $n_{13}\lesssim 10^{-5}$. This disagreement is particularly pronounced for the $qg\to q$ channel (right), since the coefficient of the leading logarithm in \eqref{eq:1JN:Cm1:LP} is particularly small in this case and therefore hard to fit accurately. The lower plots, moreover, show some data points that are not included in our electronic grids, and which we especially derived for this comparison using quadruple precision. As these results perfectly align with the leading-power prediction, we conclude that (i) our method-of-regions analysis for the dipole coefficients works and (ii) the uncertainties of our numerical approach are well under control even in the deep collinear region. We finally remark that the deviation between our results and the ones from~\cite{Campbell:2017hsw} for very small values of $n_{13}$ can safely be considered irrelevant in the hadronic center-of-mass frame, but it may well become noticeable for soft-function definitions in highly boosted frames~\cite{Alioli:2023rxx}.

\begin{figure}[t!]
	\centering{
		\includegraphics[height=0.17\textheight]{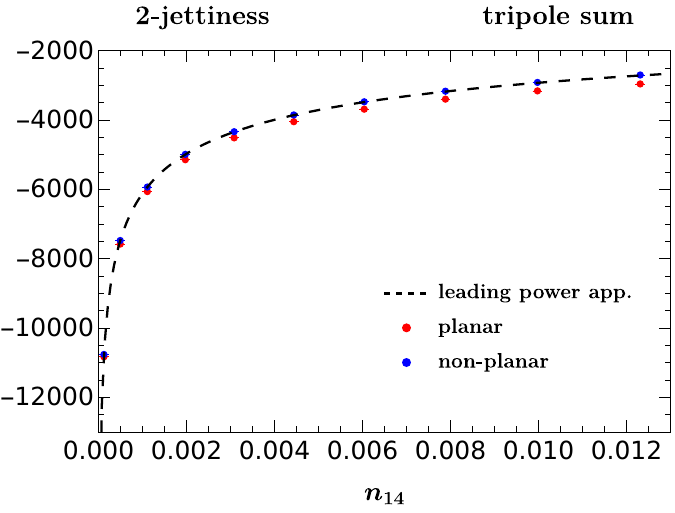}
		\includegraphics[height=0.17\textheight]{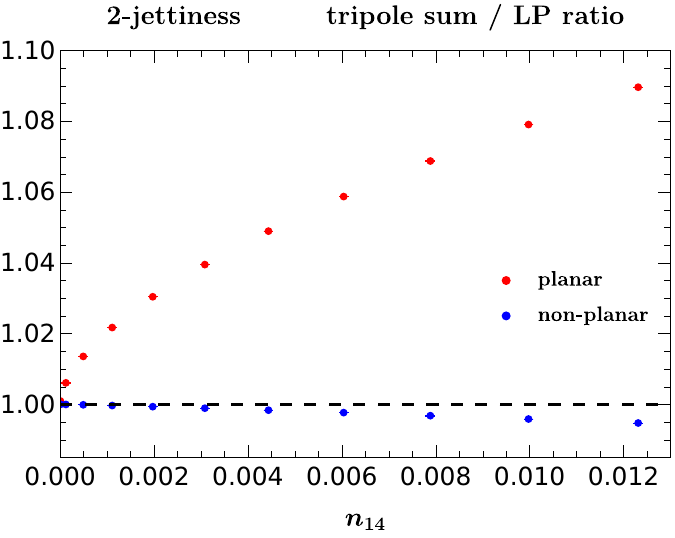}
		\includegraphics[height=0.17\textheight]{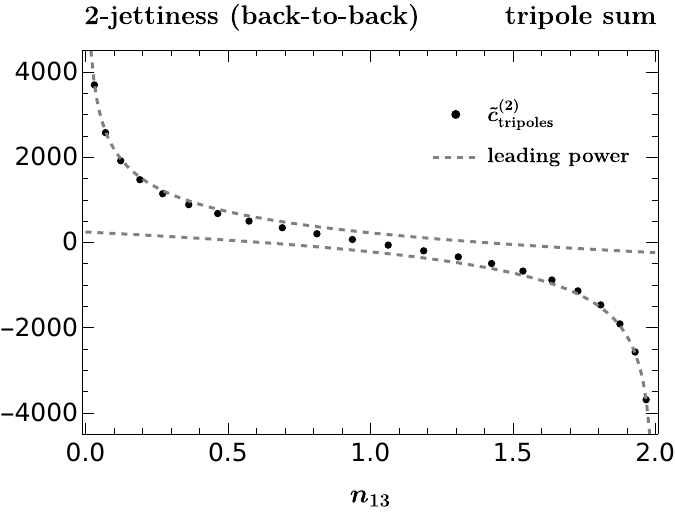}
	}
	\caption{Tripole contribution to the renormalised soft function in the convention~\eqref{eq:2jet:rentripolesum:ctilde} for the 2-to-1 jettiness transition (left and middle) and the 2-to-0 jettiness transition (right). The dashed lines show the respective leading-power (LP) approximation derived in our method-of-regions analysis, and the dots are our numerical results. For further information, see text.}
	\label{fig:2to1jetplot-beamjet}
\end{figure}

\paragraph{2-jettiness (planar events):} 
As we have already demonstrated that we control the collinear limits of the dipole contributions, we next consider the asymptotic behaviour of the 2-jettiness tripole sum. The limit in which two jets become collinear to each other has, moreover, already been addressed in the context of the 3-jettiness soft function in Sec.~\ref{sec:regions:tripoles}. We therefore consider the alternative configuration here, in which one of the jets becomes collinear to one of the beam directions. As we will see in the following, the jet-beam collinear limit shows very different features than its jet-jet counterpart, which can be traced to the different $\lambda_{AB}$ factors in the tripole sum.

To be specific we start from a planar event with $\theta_{13}=10\pi/25$ and $\varphi_4=0$ in the notation of \eqref{eq:2jet:ni}. Our goal consists in studying the behaviour of the tripole sum in the limit $\theta_{14}\to 0$, i.e.~we consider a one-dimensional projection of the left plot in Fig.~\ref{fig:tripole:finite:2jet:planar} for $n_{13}\approx 0.691 $ and $n_{14}\to 0$. The plot, in fact, already suggests that this limit is not smooth, contrary to what we have found for the jet-jet collinear limit in the previous section.

We refer to~\ref{app:method-of-regions::2jet-tripoles} for the technical details of this analysis, where we show that the renormalised 2-jettiness tripole coefficient defined in \eqref{eq:2jet:rentripolesum:ctilde} can be written at leading power in $n_{14}\ll 1$ in the form 
\begin{align}
\tilde c_{\rm tripoles}^{(2)} &= 
 \frac{8\pi}{3} L^3 + \frac{32 \pi^3}{3} L + \Delta\tilde c_{\rm tripoles}^{(2)}(n_{13}) + 
 \mathcal{O}(\sqrt{n_{14}}) \,,
 \label{eq:regions:2jet:tripoles:deltac}
\end{align}
with $L=\ln(\frac{n_{14}}{2})$. In contrast to the jet-jet collinear limit, we thus find that the tripole coefficient is logarithmically divergent as $n_{14}\to 0$, and the non-logarithmic terms $\Delta\tilde c_{\rm tripoles}^{(2)}(n_{13})$ in this equation are, moreover, found to be non-universal, i.e.~they have a different structure for each value of N. Specifically for the 2-jettiness case, our results for these coefficients can also be found in the ancillary electronic file. 

In the left panel of Fig.~\ref{fig:2to1jetplot-beamjet} we show the leading-power approximation (dashed line) together with a few data points (in red) that were calculated with quadruple precision.  Similar to the dipole pattern discussed earlier, the data points nicely asymptote to the leading-power curves as $n_{14}\to 0$, which becomes even more apparent in the middle panel of Fig.~\ref{fig:2to1jetplot-beamjet}, which shows the same data normalised to the leading-power result. We also use this opportunity to visualise another aspect that we briefly mentioned in Sec.~\ref{sec:regions:NLO}, i.e.~that power corrections are expected to be smaller for maximally non-planar configurations. To this end, we display some blue data points in the same figures that were obtained in a setup with $\theta_{13}=10\pi/25$ and $\varphi_4=\pi/2$. The jet with index 4 is thus in a maximally non-planar configuration indicated by $\vec{n}_{a,2}$ in Fig.~\ref{fig:regions:cone}, and we again consider the limit that it becomes collinear to the first beam direction. As can clearly be observed in the first two panels of Fig.~\ref{fig:2to1jetplot-beamjet}, the convergence to the leading-power curves is indeed much faster for non-planar events.

\paragraph{2-jettiness (back-to-back jets):} 
Interestingly, the planar configuration also serves as a starting point to examine the endpoint behaviour of the 2-jettiness tripole coefficient for back-to-back jets shown in Fig.~\ref{fig:tripole:2jet:b2b}, where we found a discrepancy with an earlier calculation~\cite{Jin:2019dho}. We, in fact, explicitly verified that the method-of-regions analysis for merging the second jet onto the first beam direction -- as described previously -- and subsequently merging the other (planar) jet onto the second beam direction gives the same result as collapsing two back-to-back jets onto the respective beam directions simultaneously. We can therefore directly use the same equation \eqref{eq:regions:2jet:tripoles:beamjet} -- which underlies \eqref{eq:regions:2jet:tripoles:deltac} as explained in the appendix -- to study the limit $n_{13}\to 2$ of Fig.~\ref{fig:tripole:2jet:b2b}, whereas the behaviour for $n_{13}\to 0$ then follows by the asymmetry of the tripole coefficient. Explicitly, we find that the leading-power approximation in the limit $n_{13}\to 2$ becomes
\begin{align}
\tilde c_{\rm tripoles}^{(2)} &= \pi \bigg\{
 \frac{16}{3} L^3 + \frac{64 \pi^2}{3} L + 64\zeta_3 \bigg\} + 
 \mathcal{O}(\sqrt{n_{14}}) \,,
\end{align}
where again $L=\ln(\frac{n_{14}}{2})$ and the variable $n_{13}=2-n_{14}$ is used in the plot. The same data that is shown in Fig.~\ref{fig:tripole:2jet:b2b} is then displayed in the right panel of Fig.~\ref{fig:2to1jetplot-beamjet} together with the leading-power approximations (dashed) at both endpoints.  The resulting agreement clearly confirms our numerical results and disfavours the ones from~\cite{Jin:2019dho} (which are roughly by a factor of two larger).

\section{Conclusion}
\label{sec:conclusion}

The N-jettiness event shape is an essential reference observable for precision studies at hadron colliders. In this work we have presented a systematic framework for the calculation of the N-jettiness soft function at next-to-next-to leading order (NNLO) in the QCD expansion. While our method is valid for any number N of jets,  we have presented numerical results for \mbox{N~$=1,2,3$} in this work. Specifically, our numbers for the 1-jettiness soft function confirm existing results in the literature~\cite{Boughezal:2015eha,Campbell:2017hsw}, whereas our 2-jettiness numbers  -- consisting of about 30,000 points in the phase space of the hard emitters -- are new, and represent the main result of our study. In addition, we for the first time computed the 3-jettiness soft function for a few benchmark points, which provides a useful point of comparison for future calculations. Our complete numerical results can be found in an ancillary file to this publication. 

Our calculation is based on an extension of the \texttt{SoftSERVE} framework~\cite{Bell:2018vaa,Bell:2018oqa,Bell:2020yzz} -- originally devised for back-to-back dipole configurations -- to the case of arbitrarily many hard emitters in any kinematic orientation. In its present form, the framework can  be applied to any SCET-1 observable obeying the non-abelian exponentiation theorem, and fulfilling certain conditions that are similar to those required in the original \texttt{SoftSERVE} approach~\cite{Bell:2018oqa}. More specifically, our novel N-jet extension primarily consists of the capability to accommodate non-back-to-back dipole configurations, as well as support for non-trivial tripole contributions that become relevant for N~$\geq 2$. The latter, in particular, are a perpetual source of confusion in the literature, and we hope that our comprehensive discussion -- covering in particular their properties under renormalisation -- will help to clarify their structure.

We supplemented our numerical results with a detailed analytic study of the asymptotic behaviour of the soft-function coefficients in the limit of two of the reference vectors $n_i^\mu$ becoming collinear to each other. Interestingly, we found that this limit is not always smooth, but can instead lead to discontinuities or even logarithmic divergences. Using method-of-regions techniques we were able to clarify the origin of this behaviour, and we could prove that it shows some universal properties -- meaning that it is independent of the number of jets N -- in most situations (with the only exception being the tripole coefficient of the jet-beam collinear limit). These findings serve as an independent validation of our numerical results, and they may also prove useful for constructing fits to the numerical grids, as configurations with closely adjacent hard emitters can, for example, become relevant whenever the soft function is evaluated in a highly boosted frame.

With high-precision results for the 2-jettiness NNLO soft function provided, the capabilities of our approach are not yet exhausted. The calculation of the 3-jettiness soft function is in principle already feasible, but requires a substantial amount of computing time due to the extremely large phase space that needs to be sampled. The extension to SCET-2 observables as well as observables that violate the non-Abelian exponentiation theorem can also be addressed in our framework, as has been demonstrated for back-to-back Wilson-line configurations in~\cite{Bell:2018vaa,Bell:2018oqa,Bell:2020yzz}. Further generalisations concerning soft functions that involve non-lightlike Wilson lines or more general non-global measurements are also possible. We believe that such an approach -- together with similar progress for the automated calculation of collinear functions~\cite{Bell:2021dpb,Bell:2022nrj} -- will prove useful to improve the theoretical predictions for many key observables at the LHC and beyond.

\section*{Acknowledgement}
We are grateful to Frank Tackmann for discussions regarding the results in~\cite{Jouttenus:2011wh}. The research of G.B.~was supported by the Deutsche Forschungsgemeinschaft (DFG, German Research Foundation) under grant 396021762 - TRR 257. R.R.~is supported by the Royal Society through Grant URF\textbackslash R1\textbackslash 201500, and acknowledges prior funding through the ERC grant ERC-STG-2015-677323 and the NWO projectruimte 680-91-12, as well as the SNSF grant CRSII2\textunderscore160814. The work of B.D.~has partially been funded by the Helmholtz Association Grant W2/W3-116 and the European Research Council (ERC) under the European Union's Horizon 2020 research and innovation programme (Grant agreement No. 101002090 COLORFREE). We thank the Erwin-Schrödinger International Institute for Mathematics and Physics at the University of Vienna for partial support during the Programme ``Quantum Field Theory at the Frontiers of the Strong Interactions'', July 31 - September 1, 2023.

\begin{appendix}

\section{Angular parametrisation}
\label{app:angles}

In this appendix we present a systematic method to parametrise the angular phase-space integrations in the double real-emission contribution. We first remark that all four-vectors in this appendix refer to the boosted back-to-back frame, in which the transverse components $k_\perp^\mu$ and $l_\perp^\mu$ are purely spatial, but in contrast to the main text we drop the primes on these vectors to keep the notation simple. Adopting the usual Cartesian basis for the vectors $\vec{e}_{\perp n}$ that were introduced in Sec.~\ref{sec:parametrisations}, the $(d-2)$-dimensional transverse components of the external lightlike vectors $n_X^\mu$ take the form $\vec{n}_{X,\perp}=(-n_{X\perp_1},-n_{X\perp_2},0,\ldots,0)$, since these vectors live in the physical two-dimensional transverse space that is spanned by $\vec{e}_{\perp_1}$ and $\vec{e}_{\perp_2}$. The transverse momenta of the emitted soft partons $k_\perp^\mu$ and $l_\perp^\mu$ are, on the other hand, not restricted to the physical subspace, and we parametrise these in general $(d-2)$-dimensional spherical coordinates according to \eqref{eq:spherical-coordinates},
\begin{align}
\vec{k}_{\perp}\equiv 
k_T \,\begin{pmatrix}
c_{k_1}\\
s_{k_1}\,c_{k_2}\\ 
\vdots \\ 
s_{k_1}\,\dots\,s_{k_{d-4}}\,c_{k_{d-3}} \\ 
s_{k_1}\,\dots\,s_{k_{d-4}}\,s_{k_{d-3}}
\end{pmatrix} 
, \qquad
\vec{l}_{\perp}\equiv 
l_T \,\begin{pmatrix}
c_{l_1}\\
s_{l_1}\,c_{l_2}\\ 
\vdots \\ 
s_{l_1}\,\dots\,s_{l_{d-4}}\,c_{l_{d-3}} \\ 
s_{l_1}\,\dots\,s_{l_{d-4}}\,s_{l_{d-3}}
\end{pmatrix} 
.
\end{align} 
As discussed in the main text, we want to parametrise the phase-space integrals in terms of the angle $\theta_{kl}$ between the vectors $\vec{k}_{\perp}$ and $\vec{l}_{\perp}$, since this angle controls the singular limit in which the two emissions become collinear to each other. In the given parametrisation, the cosine of this angle becomes
\begin{align}
c_{kl} = c_{k_1} c_{l_1} + 
s_{k_1}\,c_{k_2} s_{l_1}\,c_{l_2} + \ldots 
+ s_{k_1}\,\dots\,s_{k_{d-3}} s_{l_1}\,\dots\,s_{l_{d-3}}\,,
\end{align}
and it is hence a non-trivial function of all angular variables.

A generic phase-space integral involves, on the other hand, a scalar integrand that depends only on the scalar products between the vectors $\vec{k}_{\perp}$, $\vec{l}_{\perp}$ and $\vec{n}_{X,\perp}$,
\begin{align}
\label{eq:generic-phase-space-integral}
\int {d}^{d-2}   \,l_{\perp}\,  \int {d}^{d-2}  \, k_{\perp}\;
 F\big(\vec{k}_\perp \cdot \vec{l}_\perp \, , \,\{ \vec{k}_\perp \cdot  \vec{n}_{X,\perp} \} \, ,\,\{ \vec{l}_\perp \cdot \vec{n}_{X,\perp}\}\big)\,.
\end{align}
It is therefore possible to rotate the $(d-2)$-dimensional coordinate system in a way that exposes the dependence on $\theta_{kl}$, and which shows that there are, in fact, only five non-trivial angles at this order. To do so, we first perform a rotation that aligns the vector $\vec{l}_{\perp}$ with the first basis vector. This is achieved by successive rotations in two-dimensional subspaces that are encoded in the rotation matrix
\begin{align}
\mathcal{R}_1(\{\theta_{l_i}\}) = 
\begin{pmatrix}
c_{l_{1}} & s_{l_{1}} & 0&0\\
-s_{l_{1}} & c_{l_{1}} & 0 &0\\
0 & 0 & 1&0 \\
0 & 0 & 0&\mathbf{1}_{d-5,d-5}
\end{pmatrix} 
\begin{pmatrix}
1 & 0 & 0 & 0 \\
0 & c_{l_{2}} & s_{l_{2}} & 0\\
0 & -s_{l_{2}} & c_{l_{2}} & 0 \\
0 & 0 & 0&\mathbf{1}_{d-5,d-5}
\end{pmatrix} 
\dots
\begin{pmatrix}
\mathbf{1}_{d-5,d-5} & 0 & 0 & 0 \\
0 & 1 & 0 & 0\\
0 & 0  &c_{l_{d-3}} & s_{l_{d-3}}\\
0 &0 &-s_{l_{d-3}} & c_{l_{d-3}}
\end{pmatrix} 
,
\end{align} 
such that $\vec{l}_\perp^{\,\prime} = \mathcal{R}_1(\{\theta_{l_i}\}) \cdot \vec{l}_\perp = l_T
(1,0, \ldots,0 )^T$ as desired. Note that the prime has a different meaning in this appendix than in the main text, since it refers here to a rotated coordinate system within the back-to-back frame.

We next write the rotated transverse momentum of the second emission in the form 
\begin{align}
\vec{k}_\perp^{\,\prime}  = \mathcal{R}_1(\{\theta_{l_i}\})\cdot \vec{k}_\perp \equiv 
k_T \,\begin{pmatrix}
c_{kl}\\
s_{kl}\,c_{5}\\ 
\vdots \\ 
s_{kl}\,\dots\,s_{d-1}\,c_{d} \\ 
s_{kl}\,\dots\,s_{d-1}\,s_{d}
\end{pmatrix} 
,
\end{align}
where we traded the angles  $\{\theta_{k_1},\ldots,\theta_{k_{d-3}}\}$ for a new set of angles $\{\theta_{kl},\theta_{5},\ldots, \theta_{d}\}$. While the relation between the two sets of angular variables is complicated, it is immediately clear that the first angle in the new set is the desired angle with
\begin{align}
\vec{k}_{\perp} \cdot \vec{l}_{\perp} =
\vec{k}_{\perp}^{\,\prime}  \cdot \vec{l}_{\perp}^{\,\prime}  =
k_T \,l_T \,c_{kl}\,.
\label{eq:kperplperp}
\end{align} 
We then perform a second rotation that brings the vector $\vec{k}_{\perp}^{\,\prime}$ into the plane that is spanned by the first two basis vectors. The corresponding rotation matrix is now given by
\begin{align}
\mathcal{R}_2(\{\theta_{i}\}) = 
\begin{pmatrix}
1 & 0 & 0 & 0 \\
0 & c_{5} & s_{5} & 0\\
0 & -s_{5} & c_{5} & 0 \\
0 & 0 & 0&\mathbf{1}_{d-5,d-5}
\end{pmatrix} 
\dots
\begin{pmatrix}
\mathbf{1}_{d-5,d-5} & 0 & 0 & 0 \\
0  &c_{d-1} & s_{d-1} & 0\\
0 &-s_{d-1} & c_{d-1} &0\\
0 & 0 & 0 & 1
\end{pmatrix} 
\begin{pmatrix}
\mathbf{1}_{d-5,d-5} & 0 & 0 & 0 \\
0 & 1 & 0 & 0\\
0 & 0  &c_{d} & s_{d}\\
0 &0 &-s_{d} & c_{d}
\end{pmatrix} 
,
\end{align} 
which yields $\vec{k}_\perp^{\,\prime\prime} = \mathcal{R}_2(\{\theta_{i}\}) \cdot \vec{k}_\perp^{\,\prime} = k_T
(c_ {kl},s_ {kl},0, \ldots,0 )^T$, whereas the transverse momentum of the first emission
$\vec{l}_\perp^{\,\prime\prime} = \mathcal{R}_2(\{\theta_{i}\}) \cdot \vec{l}_\perp^{\,\prime} = l_T
(1,0, \ldots,0 )^T$ is invariant under this rotation. 

Under the two successive rotations, the external lightlike reference vectors $\vec{n}_{X,\perp}$ then transform into
\begin{align}
\vec{n}_{X,\perp}^{\,\prime\prime}  = \mathcal{R}_2(\{\theta_{i}\}) \cdot \mathcal{R}_1(\{\theta_{l_i}\})\cdot \vec{n}_{X,\perp} =
-\begin{pmatrix}
n_{X\perp_1} c_{l_1} + n_{X\perp_2} s_{l_1} c_{l_2}\\
- n_{X\perp_1} s_{l_1} c_5 + n_{X\perp_2} (c_{l_1} c_{l_2} c_5 - s_{l_2} s_5 c_6)\\ 
n_{X\perp_1} s_{l_1} s_5 - n_{X\perp_2} (c_{l_1} c_{l_2} s_5 + s_{l_2} c_5 c_6)\\
n_{X\perp_2}  s_{l_2} s_6 \\
0\\
\vdots \\ 
0\\
\end{pmatrix} 
,
\label{eq:nXperp}
\end{align}
and their scalar products that enter the phase-space integrals \eqref{eq:generic-phase-space-integral} become 
\begin{align}
 \vec{k}_\perp \cdot  \vec{n}_{X,\perp}&=
  \vec{k}_\perp^{\,\prime\prime} \cdot  \vec{n}_{X,\perp}^{\,\prime\prime}=
  - k_T \Big(  n_{X\perp_1} \, \lambda_3(\theta_{kl},\theta_5,\theta_{l_1}) + n_{X\perp_2} \, \lambda_4(\theta_{kl},\theta_5,\theta_6,\theta_{l_1},\theta_{l_2}) \Big)\,,
\nonumber\\
 \vec{l}_\perp \cdot  \vec{n}_{X,\perp}&=
\vec{l}_\perp^{\,\prime\prime} \cdot  \vec{n}_{X,\perp}^{\,\prime\prime}=
- l_T \Big( n_{X\perp_1} \, \lambda_3(0,0,\theta_{l_1}) + 
n_{X\perp_2}\, \lambda_4(0,0,0,\theta_{l_1},\theta_{l_2}) \Big)\,,
\label{eq:klperpnXperp}
\end{align} 
where we introduced the functions
\begin{align}
&\lambda_3(\theta_{kl},\theta_5,\theta_{l_1})  =
c_{kl}\, c_{l_1}-s_{kl}\,c_5 \,  s_{l_1}\,,
\nn \\
& \lambda_4(\theta_{kl},\theta_5,\theta_6,\theta_{l_1},\theta_{l_2}) =  
c_{kl} \, s_{l_1} \, c_{l_2} + s_{kl} (c_5\, c_{l_1}\, c_{l_2} - s_5\,c_6\, s_{l_2})\,.
\end{align}
As the Jacobians of the two variable transformations are trivial, the phase-space integrals can finally be written in the form
\begin{align}
& \int {d}^{d-2}   \,l_{\perp}\,  \int {d}^{d-2}  \, k_{\perp}\;
F\big(\vec{k}_\perp \cdot \vec{l}_\perp \, , \,\{ \vec{k}_\perp \cdot  \vec{n}_{X,\perp} \} \, ,\,\{ \vec{l}_\perp \cdot \vec{n}_{X,\perp}\}\big)
\nonumber\\
& \; = 
\int {d}^{d-2}   \,l_{\perp}^{\,\prime\prime}\,  \int {d}^{d-2}  \, k_{\perp}^{\,\prime\prime}\;
F\big(\vec{k}_\perp^{\,\prime\prime} \cdot \vec{l}_\perp^{\,\prime\prime} \, , \,\{ \vec{k}_\perp^{\,\prime\prime} \cdot  \vec{n}_{X,\perp}^{\,\prime\prime} \} \, ,\,\{ \vec{l}_\perp^{\,\prime\prime} \cdot \vec{n}_{X,\perp}^{\,\prime\prime}\}\big)
\nonumber\\
& \; = \Omega_{d-4} \; \Omega_{d-5} 
\int \!d l_T \, l_T^{d-3}  \! \int \!d k_T \, k_T^{d-3}  \!
\int \!d c_{l_1} \, s_{l_1}^{d-5}  \! \int \!d c_{l_2} \, s_{l_2}^{d-6} \!
\int \!d c_{kl} \, s_{kl}^{d-5}  \! \int \!d c_{5} \, s_{5}^{d-6}  \! 
\int \!d c_{6} \, s_{6}^{d-7}
\nonumber\\
& \qquad \times F\big(\vec{k}_\perp^{\,\prime\prime} \cdot \vec{l}_\perp^{\,\prime\prime} \, , \,\{ \vec{k}_\perp^{\,\prime\prime} \cdot  \vec{n}_{X,\perp}^{\,\prime\prime} \} \, ,\,\{ \vec{l}_\perp^{\,\prime\prime} \cdot \vec{n}_{X,\perp}^{\,\prime\prime}\}\big)\,,
\end{align}
where $\Omega_{n}=2 \pi^{n/2}/\Gamma(n/2)$ is the solid angle in $n$ dimensions. This representation shows that the phase-space integrals in the double real-emission contribution depend on five non-trivial angles $\{\theta_{kl},\theta_5,\theta_6,\theta_{l_1},\theta_{l_2}\}$ that parametrise the various scalar products according to \eqref{eq:kperplperp} and \eqref{eq:klperpnXperp}. Specifically, the chosen set of integration variables contains the desired angle $\theta_{kl}$, two angles $\theta_{l_1}$ and $\theta_{l_2}$ that are introduced in analogy to the single-emission case from Sec.~\ref{sec:parametrisations}, and two auxiliary angles $\theta_5$ and $\theta_6$. Although the  interpretation of the latter angles is strictly speaking not needed, we can read off from \eqref{eq:nXperp} that $\theta_{5}$ is the angle between the $(\vec{l}_\perp,\vec{k}_\perp)$-plane and the $(\vec{l}_\perp,\vec{e}_{\perp_1})$-plane, while $\theta_{6}$ parametrises the misalignment of the three-dimensional $(\vec{l}_\perp,\vec{k}_\perp,\vec{e}_{\perp_1})$ and $(\vec{l}_\perp,\vec{k}_\perp,\vec{e}_{\perp_2})$ subspaces. The proposed method of applying a set of rotations that bring the momenta of the emitted partons into a canonical form can easily be generalised to a larger number of emissions. 

\section{Further details of the method-of-regions analysis}
\label{app:method-of-regions}

In this appendix we collect further details of the method-of-regions analysis that we presented in the main body of the text in Sec.~\ref{sec:method-of-regions}.

\subsection{Renormalisation of the 3-jettiness tripole contribution}
\label{app:method-of-regions:3jet-tripoles}

In Sec.~\ref{sec:regions:tripoles} we studied the bare 3-jettiness tripole contribution in the limit where the jets with reference vectors $n_4^\mu$ and $n_5^\mu$ become collinear to each other. In particular, we showed that the bare tripole sum has the same properties that we observed on the level of the renormalised coefficients in Fig.~\ref{fig:regions:qualitative}:  The coefficients of the $\bfT_1^A\; \bfT_2^B\;\bfT_4^C$ and $\bfT_1^A\; \bfT_2^B\;\bfT_5^C$ colour structures converge smoothly into the 2-jettiness number, whereas the coefficients of the $\bfT_1^A\; \bfT_4^B\;\bfT_5^C$ and $\bfT_2^A\; \bfT_4^B\;\bfT_5^C$ colour structures vanish in this limit. We will now show that this pattern carries over to the renormalised tripole coefficients.

For this discussion it will be crucial to use the $\tilde c_{ijk}^{(2)} $ notation that we introduced in \eqref{eq:c2ijktilde}. In this convention, the renormalised tripole contribution takes the form
\begin{align}
\label{eq:regions:3jet:tripoles:renormalised}
& 2\pi \sum_{i\neq j\neq k} f_{ABC}\; \bfT_{i}^A \;\bfT_{j}^B \;\bfT_{k}^C\;
\tilde c_{ijk}^{(2)}
\\
&\quad
= \pi   \sum_{i\neq l\neq j} (\lambda_{il} - \lambda_{ip}-\lambda_{jp})\,
f_{ABC}\; \bfT_{i}^A \;\bfT_{l}^B \;\bfT_{j}^C  \; \tilde  z_0^{ilj}
-2\pi  \,\Gamma_0  \sum_{i\neq j\neq k} f_{ABC}\; \bfT_{i}^A \;\bfT_{j}^B \;\bfT_{k}^C\;
 \lambda_{ij} R_{jk}\,,
\nonumber
\end{align}
where $\tilde  z_0^{ilj}$ is the $\mathcal{O}(\eps^0)$ coefficient of the bare $(ilj)$ tripole defined by
\begin{align}
\label{eq:tripole-bare-coeff}
\tilde{S}_{ilj}(\eps) =
\Big( \frac{n_{ij}}{2} \Big)^{2\eps} \, S_{ilj}^{(2,{\rm Im})}(\eps) = \pi\,
\bigg\{
\frac{\tilde z_3^{ilj}}{\eps^3}  + \frac{\tilde z_2^{ilj}}{\eps^2} + \frac{\tilde z_1^{ilj}}{\eps} + \tilde z_0^{ilj} + \mathcal{O}(\eps)\bigg\}\,,
\end{align}
and the second term in \eqref{eq:regions:3jet:tripoles:renormalised} originates from the interference of the bare NLO dipoles and the NLO counterterm contribution. Specifically, we have
\begin{align}
R_{jk}
= \frac{1}{3}\Gamma_0\tilde{L}_{jk}^3+\gSzero \tilde{L}_{jk}^2+ c_{jk}^{(1)} \tilde{L}_{jk} +\frac12 d_{jk}^{(1)} \,, 
\label{eq:regions:Rjk}
\end{align}
where $c_{jk}^{(1)}$ was introduced in \eqref{eq:RGE:softsolution} and $d_{jk}^{(1)}$ is the corresponding non-logarithmic term at $\mathcal{O}(\eps)$. As we have already shown in Sec.~\ref{sec:regions:tripoles}, the first term in \eqref{eq:regions:3jet:tripoles:renormalised} has the desired properties in the collinear limit, and we can therefore focus on the interference term in the following.

To this end, we resolve the colour generator of the first jet (with index 3) using colour conservation. The sum over the interference contributions then becomes
\begin{align}
& \sum_{i\neq j\neq k} f_{ABC}\; \bfT_{i}^A \;\bfT_{j}^B \;\bfT_{k}^C\;
\lambda_{ij} R_{jk}
\nonumber\\
&\quad 
=( {\color{ForestGreen}R_{13}} - {\color{Orange}R_{14}} - {\color{ForestGreen}R_{23}} + {\color{Orange}R_{24}} + {\color{ForestGreen}R_{31}} - {\color{ForestGreen}R_{32}} - {\color{Orange}R_{41}} + {\color{Orange}R_{42}} )
\;f_{ABC}\;\bfT_1^A\; \bfT_2^B\;\bfT_4^C
\nonumber\\ &
\quad +
( {\color{ForestGreen}R_{13}} - {\color{Orange}R_{15}} - {\color{ForestGreen}R_{23}} + {\color{Orange}R_{25}} + {\color{ForestGreen}R_{31}} - {\color{ForestGreen}R_{32}} - {\color{Orange}R_{51}} + {\color{Orange}R_{52}} )
\;f_{ABC}\;\bfT_1^A\; \bfT_2^B\;\bfT_5^C
\nonumber\\ & 
\quad +
( - {\color{Orange}R_{34}} + {\color{Orange}R_{35}} + {\color{Orange}R_{43}} - {\color{cyan}R_{45}} - {\color{Orange}R_{53}} + {\color{cyan}R_{54}} )
\;f_{ABC}\;\bfT_1^A\; \bfT_4^B\;\bfT_5^C
\nonumber\\ &
\quad +
( - {\color{Orange}R_{34}} + {\color{Orange}R_{35}} + {\color{Orange}R_{43}} - {\color{cyan}R_{45}} - {\color{Orange}R_{53}} + {\color{cyan}R_{54}} )
\;f_{ABC}\;\bfT_2^A\; \bfT_4^B\;\bfT_5^C\,,
\label{eq:regions:interference} 
\end{align}
where similar to \eqref{eq:regions:tripolesum} we used a colour coding to indicate the various dipole classes we studied in Sec.~\ref{sec:regions:NLO}. Specifically, the {\color{ForestGreen}green} contributions refer to dipoles that do not involve any of the collinear directions $n_4^\mu$ and $n_5^\mu$, and they simply yield the respective 2-jettiness values in the collinear limit. The {\color{Orange}orange} contributions, on the other hand, refer to  dipoles that involve one of the collinear directions among their indices, and they consist of a base and a correction term, of which the former is again equal to the 2-jettiness value, whereas the latter yields the correction in \eqref{eq:regions:NLO:correction}. Finally, the {\color{cyan}blue} terms, involving both collinear directions, represent the pathological dipoles that reduce to the 0-jettiness value in the collinear limit. 

As can immediately be seen in \eqref{eq:regions:interference}, the coefficients of the last two colour structures vanish as expected in the collinear limit, since the $R_{jk}$ in \eqref{eq:regions:Rjk} are symmetric under an exchange of the indices. This symmetry together with the fact that the $n_4^\mu$ and $n_5^\mu$ directions cannot be distinguished in the orange contributions allows us to rewrite the interference term as
\begin{align}
\sum_{i\neq j\neq k} f_{ABC}\; \bfT_{i}^A \;\bfT_{j}^B \;\bfT_{k}^C\;
\lambda_{ij} R_{jk}
=2 ( {\color{ForestGreen}R_{13}} - {\color{Orange}R_{14}} - {\color{ForestGreen}R_{23}} + {\color{Orange}R_{24}}  ) \;f_{ABC}\;\bfT_1^A\; \bfT_2^B\;(\bfT_4^C+\bfT_5^C)\,.
\label{eq:regions:interference:final} 
\end{align}
The green dipoles and the base contribution of the orange dipoles then precisely reduce to the 2-jettiness expressions that is multiplied here by the colour generator $\bfT_{X}^C=\bfT_{4}^C+\bfT_{5}^C$ of the merged jet. It therefore only remains to be shown that the correction contributions to the orange dipoles vanish in \eqref{eq:regions:interference:final}. 

The correction term \eqref{eq:regions:NLO:correction} does not modify the anomalous dimensions, but it contributes to $c_{jk}^{(1,\rm{corr})}=-\pi^2/3$ and $d_{jk}^{(1,\rm{corr})}=
-\pi^2/3 \ln \delta + 2\pi^2/3 \tilde L_{jk}- 4 \zeta_3$. Interestingly, the kinematic dependence cancels in the combination
\begin{align}
{\color{Orange}R_{jk}^{(\rm{corr})}}
=  c_{jk}^{(1,\rm{corr})} \tilde{L}_{jk} +\frac12 d_{jk}^{(1,\rm{corr})} 
= -\frac{\pi^2}{6} \ln \delta - 2 \zeta_3 \,,
\end{align}
and we stress that this would not have been the case if we had considered the renormalised tripole contribution on the level of the  $c_{ijk}^{(2)} $ rather than the $\tilde c_{ijk}^{(2)} $ coefficients. As the orange dipoles contribute in \eqref{eq:regions:interference:final} with opposite signs, the correction terms then cancel in the sum. 

To summarise, we have just shown that the coefficients of the first two colour structures in \eqref{eq:regions:interference} reduce to the 2-jettiness equivalent, whereas the ones of the last two colour structures vanish in the collinear limit. The observed behaviour therefore also applies to the renormalised tripole sum as long as it is formulated in terms of the $\tilde c_{ijk}^{(2)} $ coefficients.

\subsection{Collinear limit of the 2-jettiness tripole contribution}
\label{app:method-of-regions::2jet-tripoles}

In Sec.~\ref{sec:regions:discussion} we considered the 2-jettiness tripole contribution for a planar configuration with \mbox{$\theta_{13}=10\pi/25$}, $\varphi_4=0$, and we were interested in the limit when the second jet becomes collinear to the first beam direction, i.e.~$\theta_{14}\to 0$. Whereas we studied the jet-jet collinear limit for the 3-jettiness tripole contribution in Sec.~\ref{sec:regions:tripoles}, we will see now that the jet-beam collinear limit is more complicated, because of the different $\lambda_{AB}$ factors in the tripole sum. 

Our starting point is the bare 2-jettiness tripole contribution, for which we choose to resolve the colour generator of the second jet (with index 4) via colour conservation,
\begin{align}
& \sum_{i\neq l\neq j} (\lambda_{il} - \lambda_{ip}-\lambda_{lp})\,
f_{ABC}\; \bfT_{i}^A \;\bfT_{l}^B \;\bfT_{j}^C  \;
\tilde{S}_{ilj}
\nonumber \\
&\quad
=(\IIb{\tilde{S}_{1 2 3}} - \IIIb{\tilde{S}_{1 2 4}} + \IIb{\tilde{S}_{1 3 2}} - \IIIb{\tilde{S}_{1 3 4}} - \IIIa{\tilde{S}_{1 4 2}} + \IIIa{\tilde{S}_{1 4 3}} - \IIa{\tilde{S}_{2 1 3}} + \IIIa{\tilde{S}_{2 1 4}}
- \IIb{\tilde{S}_{2 3 1}} +  \IIb{\tilde{S}_{2 3 4}}
\nonumber \\
&\qquad\;\,
+ \IIIa{\tilde{S}_{2 4 1}} - \IIa{\tilde{S}_{2 4 3}} -  \IIa{\tilde{S}_{3 1 2}} + \IIIa{\tilde{S}_{3 1 4}} + \IIb{\tilde{S}_{3 2 1}} -  \IIb{\tilde{S}_{3 2 4}}
- \IIIa{\tilde{S}_{3 4 1}} + \IIa{\tilde{S}_{3 4 2}} +  \IIIa{\tilde{S}_{4 1 2}} - \IIIa{\tilde{S}_{4 1 3}}
\nonumber\\
&\qquad\;\,
- \IIIb{\tilde{S}_{4 2 1}} +  \IIb{\tilde{S}_{4 2 3}} + \IIIb{\tilde{S}_{4 3 1}} - \IIb{\tilde{S}_{4 3 2}})
\;f_{ABC}\;\bfT_1^A\; \bfT_2^B\;\bfT_3^C.
\label{eq:regions:2jet:tripolesum}
\end{align}
We recall that $\tilde{S}_{ilj}=(n_{ij}/2)^{2\eps}\,S_{ilj}^{(2,{\rm Im})}(\eps)$, and we furthermore used the same colour scheme as in Sec.~\ref{sec:regions:tripoles} with class IIa (\IIa{light green}), class IIb (\IIb{orange}), class IIIa (\IIIa{red}) and class IIIb (\IIIb{blue}) tripoles. From this expression, it is easy to see that some cancellations that occur in the limit when two jets are made collinear are not realised in this case: The class IIIb tripoles, for instance, contribute to \eqref{eq:regions:tripolesum} with an equal number of plus and minus signs, but this is not so in \eqref{eq:regions:2jet:tripolesum}. In fact, we find that besides the tripole classes that we discussed explicitly in Sec.~\ref{sec:regions:tripoles}, the ``remaining'' tripole classes IIIa2 and IIIb also contribute here. Of these, class IIIb is the tripole equivalent of the pathological dipole, which evaluates to
\begin{align}
S_{ilj,\rm{(IIIb)}}^{(2,{\rm Im})}(\eps)   &= \pi\,
\bigg\{ -\frac{5}{3\eps^3} -\frac{35\pi^2}{18\eps} - \frac{352\zeta_3}{9}  + \mathcal{O}(\eps)\bigg\}\,,
\end{align}
and which is independent of the kinematics as anticipated in Sec.~\ref{sec:regions:tripoles}. The structure of the class IIIa2 tripoles, on the other hand, is non-trivial since its base contribution does not have a tripole analogue in the 1-jettiness case. Instead it turns out that the base region can be expressed in terms of a  1-jettiness real-virtual dipole $S_{ij}^{(2,{\rm Re})}(\eps)$, whereas the correction region can be evaluated in analogy to the IIb1 tripoles described in Sec.~\ref{sec:regions:tripoles}. In total we find 
\begin{align}
S_{ilj,\rm{(IIIa2)}}^{(2,{\rm Im})}(\eps) &= 2\tan(\pi \eps) \,S_{ij}^{(2,{\rm Re})}(\eps)  
+ \pi\, \bigg(\frac{2\delta}{n_{ij}}\bigg)^{2\eps}\,
\bigg\{ \frac{1}{3\eps^3} + \frac{31 \pi^2}{18\eps} + \frac{272 \zeta_3}{9} + \mathcal{O}(\eps)\bigg\}\,,
\label{eq:regions:tripoles:2jet:IIIa2}
\end{align}
where the first (second) term reflects the contribution from the base (correction) region.

The NLO$\times$NLO interference contribution, on the other hand, can be derived along the same lines as in the previous section. Focusing again on the $\mathcal{O}(\eps^0)$ coefficient in the notation of \eqref{eq:regions:3jet:tripoles:renormalised}, we obtain
\begin{align}
\sum_{i\neq j\neq k} f_{ABC}\; \bfT_{i}^A \;\bfT_{j}^B \;\bfT_{k}^C\;
\lambda_{ij} R_{jk}
=2 ( {\color{ForestGreen}R_{23}} - {\color{Orange}R_{24}} - {\color{Orange}R_{13}} + {\color{cyan}R_{14}}  ) \;f_{ABC}\;\bfT_1^A\; \bfT_2^B\; \bfT_3^C\,,
\end{align}
where the same colour coding has been applied as in \eqref{eq:regions:interference}. In particular, we see that the pathological dipoles (in blue) do not cancel here.

Putting all contributions together, we can reconstruct the renormalised 2-jettiness tripole coefficient defined in \eqref{eq:2jet:rentripolesum:ctilde} at leading power in $n_{14}\ll 1$. As our goal consists in writing it as a modification applied to a 1-jettiness reference value, we denote the corresponding 1-jettiness coefficients -- defined in the previous section -- by $\tilde{z}^{ilj}_0$, $c_{ij}^{(1)}$, and $d_{ij}^{(1)}$ (they originate from the respective base contributions). Explicitly, we find
\begin{align}
\tilde c_{\rm tripoles}^{(2)} &= \pi\,
\bigg\{ \frac83 L^3 + \frac{32 \pi^2}{3} L + \frac83 \ln^3 \Big(\frac{n_{13}}{2}\Big) 
+ 8  c_{13}^{(1)} \ln \Big(\frac{n_{13}}{2}\Big) -  \frac83 \ln^3 \Big(\frac{2-n_{13}}{2}\Big)
\nonumber\\
&\qquad \quad
- 8  c_{23}^{(1)} \ln \Big(\frac{2-n_{13}}{2}\Big) + 2 \tilde  z_0^{123} - 2 \tilde  z_0^{213} 
 + \frac{4 \pi^2}{3} y_{3,{\rm RV}}^{12} + 4 y_{1,{\rm RV}}^{12}
 \nonumber\\
&\qquad \quad
+   8 d_{13}^{(1)}  - 8 d_{23}^{(1)}  + 8 d_{12}^{(1)}  
 + \frac{208 \zeta_3}{3} \bigg\} + \mathcal{O}(\sqrt{n_{14}}) \,,
\label{eq:regions:2jet:tripoles:beamjet}
\end{align}
where the first two terms represent logarithmically enhanced corrections with \mbox{$L=\ln(\frac{n_{14}}{2})$}, whereas the remaining logarithms are of $\mathcal{O}(1)$. Moreover, the $y_{k,{\rm RV}}^{ij}$ coefficients arise from the 1-jettiness real-virtual dipole contribution in \eqref{eq:regions:tripoles:2jet:IIIa2}, and they are defined by
\begin{align}
 S_{ij}^{(2,{\rm Re})}(\eps) =
\frac{y_{4,{\rm RV}}^{ij}}{\eps^4} + \frac{y_{3,{\rm RV}}^{ij}}{\eps^3}  + \frac{y_{2,{\rm RV}}^{ij}}{\eps^2} + \frac{y_{1,{\rm RV}}^{ij}}{\eps} + y_{0,{\rm RV}}^{ij} + \mathcal{O}(\eps)\,.
\end{align}
Merging other jet-beam combinations then arises from straightforward replacements: For the limit when the jet with index 4 is made collinear to the beam with index 2, for instance, the labels $1$ and $2$ in \eqref{eq:regions:2jet:tripoles:beamjet} need to be exchanged, including in the colour structure that multiplies the tripole coefficient in \eqref{eq:2jet:rentripolesum:ctilde}.

The above result has several remarkable features. First, we observe that the jet-beam collinear limit is not smooth in contrast to what we have found for the jet-jet collinear limit in Sec.~\ref{sec:regions:tripoles}. The renormalised 2-jettiness tripole coefficient is rather logarithmically divergent in this limit, similar to the NNLO dipoles that involve one collinear direction in~\eqref{eq:regions:NNLO:dipoles:cij}. Second, the pattern that arises in \eqref{eq:regions:2jet:tripoles:beamjet} in the jet-beam collinear limit is not universal, i.e.~the terms that are not logarithmically enhanced refer to very specific 1-jettiness contributions, and this structure would be different if we had considered e.g.~the 3-to-2 jettiness transition in the jet-beam collinear limit. This non-universality is, in fact, the main reason why we did not consider the jet-beam collinear limit in the main body of the text.

Nevertheless, since the 2-jettiness results are the main outcome of our study, we think that it is useful to publish the leading-power expressions in the jet-beam collinear limits in a compact form. We therefore rewrite the renormalised 2-jettiness tripole coefficient as
\begin{align}
\tilde c_{\rm tripoles}^{(2)} &= 
 \frac{8\pi}{3} L^3 + \frac{32 \pi^3}{3} L + \Delta\tilde c_{\rm tripoles}^{(2)}(n_{13}) + 
 \mathcal{O}(\sqrt{n_{14}}) \,,
\end{align}
and we provide a grid for the $\Delta\tilde c_{\rm tripoles}^{(2)}(n_{13})$ coefficients in the attached electronic file.

\end{appendix}

\bibliographystyle{JHEP}
\bibliography{Njet}

\end{document}